\documentclass[
reprint,
superscriptaddress,
showpacs,
amsmath, amssymb,
aps,
prl,
]{revtex4-1}%
\usepackage{graphicx}
\usepackage{chngcntr}
\usepackage{dcolumn}
\usepackage{bm}
\usepackage{hyperref}
\hypersetup{colorlinks=true, citecolor=blue, urlcolor=blue, linkcolor=blue}
\usepackage{breakurl}
\usepackage{amsfonts}
\usepackage{amsmath}
\usepackage{amssymb}
\usepackage{mathrsfs}
\usepackage{epsfig}
\usepackage{graphicx}
\usepackage{bm}
\begin{document}
	\title{Using $\mathcal{PT}$-symmetric Qubits to Break the Tradeoff Between Fidelity and the Degree of Quantum Entanglement}

	\author{B.-B. Liu}\thanks{The authors contributed equally to this work}
\affiliation{School of Physics and Laboratory of Zhongyuan Light, Zhengzhou University, Zhengzhou 450001, China}
\author{Shi-Lei Su\textsuperscript{*}}
\email{slsu@zzu.edu.cn}
\affiliation{School of Physics and Laboratory of Zhongyuan Light, Zhengzhou University, Zhengzhou 450001, China}
	\affiliation{Institute of Quantum Materials and Physics, Henan Academy of Sciences, Zhengzhou 450046, China}
	\author{Y.-L. Zuo}
	\affiliation{Key Laboratory of Low-Dimensional Quantum Structures and Quantum Control of Ministry of Education, Department of Physics and Synergetic Innovation Center for Quantum Effects and Applications, Hunan Normal University, Changsha 410081, China}
	\affiliation{School of Physics and Chemistry, Hunan First Normal University, Changsha 410205, China}
 \author{Qiongyi He}
 \affiliation{State Key Laboratory for Mesoscopic Physics, School of Physics, Frontiers Science Center for Nano-optoelectronics, Peking University, Beijing 100871, China}
	\author{Gang Chen}\email{chengang971@163.com}
	\affiliation{School of Physics and Laboratory of Zhongyuan Light, Zhengzhou University, Zhengzhou 450001, China}
	\author{F. Nori}
	\affiliation{Theoretical Quantum Physics Laboratory, RIKEN Cluster for Pioneering Research, Wako-shi, Saitama 351-0198, Japan}
    \affiliation{Quantum Information Physics Theory Research Team, RIKEN Center for Quantum Computing, Wako-shi, Saitama 351-0198, Japan}
    \affiliation{Department of Physics, University of Michigan, Ann Arbor, Michigan 48109-1040, USA}
	\author{H. Jing}\email{jinghui73@foxmail.com}
	\affiliation{Key Laboratory of Low-Dimensional Quantum Structures and Quantum Control of Ministry of Education, Department of Physics and Synergetic Innovation Center for Quantum Effects and Applications, Hunan Normal University, Changsha 410081, China}
\date{\today}	
\begin{abstract}
A noteworthy discovery is that the minimal evolution time is smaller for parity-time~($\mathcal{PT}$) symmetric systems compared to Hermitian setups. Moreover, there is a significant acceleration of two-qubit quantum entanglement preparation near the exceptional point (EP), or spectral coalescence, within such system. Nevertheless, an important problem often overlooked for quantum EP-based devices is their fidelity, 
greatly affected by the process of dissipation or post-selection, creating an inherent trade-off relation between the degree of entanglement and fidelity. Our study demonstrates that this limitation can be effectively overcome by harnessing an active $\mathcal{PT}$-symmetric system, which possesses balanced gain and loss, enabling maximal entanglement with rapid speed, high fidelity, and greater resilience to non-resonant errors. This new approach can efficiently prepare multi-qubit entanglement and use not only bipartite but also tripartite entanglement, as illustrative examples, even when the precise gain-loss balance is not strictly maintained. Our analytical findings are in excellent agreement with numerical simulations, confirming the potential of truly $\mathcal{PT}$-devices as a powerful tool for creating and engineering diverse quantum resources for applications in quantum information technology.
\end{abstract}
	
\maketitle
\emph{Introduction.---}
Quantum entanglement, the cornerstone of quantum information science and technology~\cite{RevModPhys.81.865}, plays a central role in large-scale quantum computing~\cite{Andrew,1097,Walther2005}, secure quantum communications~\cite{Bouwmeester1997,Ursin2007,doi:10.1126/science.1167209,Yin2012,Ma2012}, and high-performance quantum metrology~\cite{Roos2006,Riedel2010,Napolitano2011,RevModPhys}. To avoid detrimental effects of noises on the typically fragile quantumness, it is highly desirable to realize high-efficiency entanglement preparation~\cite{Gardiner2004QuantumNA,SCHLOSSHAUER20191,PhysRevLett.72.2508,RevModPhys.75.715}. An appealing strategy is to utilize the exceptional points (EPs), or spectral coalescence, of non-Hermitian systems to speed up quantum evolution processes~\cite{PhysRevLett.98.040403,PhysRevLett.101.230404} and accelerating entanglement preparation~\cite{PhysRevLett.131.100202}, together with directly achieving topologically protected robust entanglement operations around the EP~\cite{Tang2024Topologically}. Two widely adopted methods for implementing $\mathcal{PT}$-symmetric non-Hermitian Hamiltonians involve a dissipation process that neglects the jump operator~(passive $\mathcal{PT}$ symmetry) and its realization via dilated Hermitian systems~\cite{PhysRevLett.101.230404,doi:10.1126/science.aaw8205,Dogra2021}. However, a major challenge of this approach is the extremely low success rate for preparing high-degree of entanglement due to dissipation or post-selection~\cite{PhysRevLett.124.020501,Tang2016,doi:10.1126/sciadv.adk7616}.

Recent experiments have demonstrated active $\mathcal{PT}$-symmetric devices across various fields~\cite{doi:10.1126/science.1258479,Gao2015,Peng2014,Jahromi2017,Chang2014,Cao2022,Zhang2020}, including optics and photonics~\cite{El-Ganainy:07,Ruter2010,Liu2018,2019,doi:10.1126/science.aar7709,Feng2017,PhysRevLett.126.230402,10.1093/nsr/nwy011,Li2020,kremer2019}, acoustics~\cite{PhysRevX.4.031042,Shi2016}, and various lattices~\cite{Regensburger2012,PhysRevLett.117.123601,PhysRevA.81.063807}. These devices are characterized by their ability to achieve tunable gain and loss through direct technical means. In contrast to complex energy spectra exhibited by lossy EP systems, $\mathcal{PT}$-symmetric systems with balanced gain and loss, which remain invariant under the joint operator of parity and time reversal, possess entirely real spectra similar to Hermitian systems~\cite{PhysRevLett.80.5243,10.1063/1.532860,Bender_2007,ElGanainy2018,Mostafazadeh_2003,10.1063/1.1418246,1876991}, hence preserving conserved dynamics~\cite{PhysRevLett.100.103904,Ashida2017,Weimann2017,PhysRevX.4.031011,PhysRevLett.119.190401,Ge2016,PhysRevA.85.023802}, which is important to prepare high-fidelity entanglement and features superior performances in sensing applications~\cite{Fleury2015,Hodaei2017,110802,Chen2017,PhysRevLett.125.240506}, optical engineering~\cite{Peng2014,Chang2014,Jing2015,doi:10.1126/science.1258480,lllLiu2017}, and phonon-laser operations~\cite{PhysRevLett.113.053604,Zhang2018}, to name only a few~\cite{11111206038,Yu2024,PhysRevLett.113.023903,Wimmer2015,doi:10.1126/sciadv.aar6782,Assawaworrarit2017,PhysRevLett.108.173901,RevModPhys.88.035002}. In this Letter, we demonstrate that a truly $\mathcal{PT}$-symmetric system can break the trade-off relation between fidelity and the degree of entanglement, while ensuring the fast speed, compared to the more common passive $\mathcal{PT}$-symmetric systems. Additionally, it exhibits greater robustness against system control errors. This can lead to new insights and techniques in other quantum information processing tasks based on non-Hermitian systems, such as quantum sensing and quantum computing. 
\begin{figure}
		\centering
  \includegraphics[width=8.5cm,height=6.1cm]{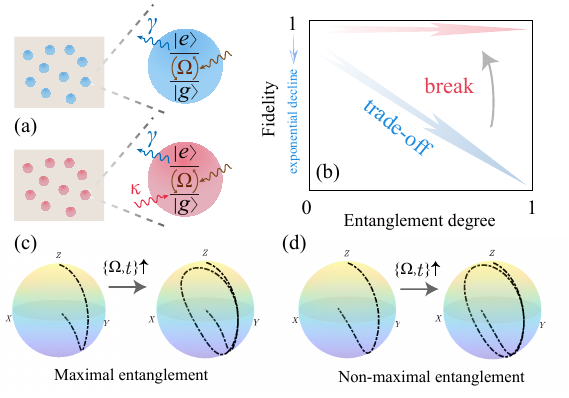}
		\caption{(a) Schematic of multi-partite all lossy EP~(top) and truly $\mathcal{PT}$-symmetric~(bottom) systems. Here $\Omega$, $\gamma$, and $\kappa$ represent the amplitude drive, loss, and gain, respectively. (b) A diagram of the trade-off between fidelity and the degree of entanglement for the all-lossy EP system, which can be broken by a truly $\mathcal{PT}$-symmetric system. Schematic of the dynamic trajectory of the reduced single qubit on the Bloch sphere to prepare maximal (c) and non-maximal (d) entanglement. For global entanglement, the trajectory lies within the Bloch sphere when the system is in specific entanglement while passing through the origin of the Bloch sphere when the system is in maximal entanglement.}
         \label{fig11}
	\end{figure}
 
\emph{General theory for many-qubit entanglement.---} For the preparation of multi-qubit entanglement as depicted in Fig.~\ref{fig11}(a), the Hamiltonian at the simplest level can be described by
\begin{eqnarray}\label{eq1}
H=\sum_{n=1}^{N}H_n+J\sum_{n\neq m=1}^N\sigma_n^{+}\sigma_m^{-},
\end{eqnarray}
where the coupling strength between two qubits is denoted by $J$, for the passive $\mathcal{PT}$-symmetric scenarios, where the loss occurs in the upper energy level $|e\rangle$, the Hamiltonian of the individual qubit is given by $H_n=-i\gamma_n/2 |e\rangle_n\langle e|+\Omega_n \sigma_n^x$, where $\gamma_n$ and $\Omega_n$ are the decay rate and drive amplitude, respectively. The $n$-th qubit operators are $\sigma_n^{+}=|e\rangle_n\langle g|, \sigma_n^{-}=|g\rangle_n\langle e|$, and $\sigma_n^x=|e\rangle_n\langle g|+|g\rangle_n\langle e|$. Its two eigenvalues are $E_{\pm}=(-i\gamma_n\pm\sqrt{-\gamma_n^2+16\Omega_n^2})/4$, featuring an EP at $\Omega_n=\gamma_n/4$. 
	\begin{figure*}
		\centering
		\includegraphics[width=\textwidth]{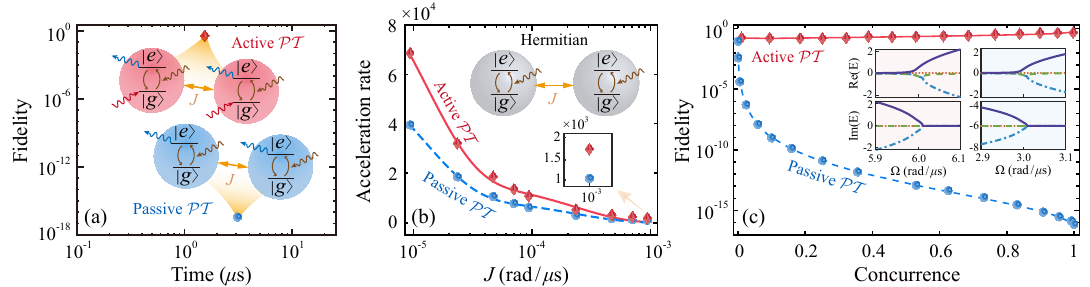}
		\caption{(a) The duration required to prepare the maximal entanglement and the corresponding fidelity for active and passive $\mathcal{PT}$-symmetric systems, where the ideal maximal entanglement state is $(-|ee\rangle+|eg\rangle+|ge\rangle+|gg\rangle)/2$ for calculating the fidelity. (b) The acceleration rate, expressed as $t_H/t$, with respect to coupling strengths. Here, $t$ is the time required to achieve the maximal entanglement for either the passive or active approach, and $t_H$ denotes the corresponding duration for the Hermitian system, where the drive amplitude $\Omega$ of the Hermitian system is set the same as for the active system. (c) The relation between the fidelity and concurrence for the active and passive $\mathcal{PT}$-symmetric systems. Inset: The real and imaginary components of the eigenvalues for both systems. The initial state is $|gg\rangle$ and $\gamma_{1(2)}=\kappa_{1(2)}=12~{\rm \mu s^{-1}}$, $J=0.001~\rm{rad/\mu s}$ for (a, c). The optimal $\Omega$ to achieve maximal or specific concurrence is obtained numerically. In (a), the optimal $\Omega$ is found as $6.23~\rm{rad/\mu s}$ and $3.152~\rm{rad/\mu s}$ for active and passive $\mathcal{PT}$-symmetric systems, respectively.
         }
         \label{fig1}
	\end{figure*}

 The system evolution state can be written as $|\psi(t)\rangle=\sum_nc_n e^{-i t \rm{Re}(E_n)}e^{t \rm{Im}(E_n)}|E_n\rangle$, where $E_n$ and $|E_n\rangle$ are eigenvalues and eigenstates of the system, respectively, and $c_n$ are the coefficients determined by the initial state. 
 Conventionally, the process of post-selection or re-normalized states is involved in evaluating the performance of non-Hermitian systems during entanglement preparation, which fails to account for a crucial factor, i.e., the success rate~\cite{PhysRevLett.124.020501,Tang2016,doi:10.1126/sciadv.adk7616}, here we use the associated fidelity to measure the performance of the system, which is defined as $F=|\langle\psi_f|\psi_i\rangle|^2$, where $|\psi_i\rangle$ is the ideal maximal entanglement and $|\psi_f\rangle$ is the practical final state without normalization.
 The evolution of the system indicates that the imaginary component, Im$(E_n)$, leads to an exponential attenuation or amplification of the probability amplitude over time; whereas the real component, Re$(E_n)$, induces oscillations in the state~\cite{Yu2021}. In the fully lossy EP system, preparing a high degree of entanglement necessitates a relatively long evolution time and the presence of negative imaginary components in the eigenvalue, significantly decreasing the probability amplitude of the evolution state, resulting in low fidelity. This underscores a critical phenomenon: the increase of entanglement is accompanied by an exponential decrease in fidelity, revealing a distinct trade-off between the fidelity and the degree of entanglement. As shown in Fig.~\ref{fig11}(b), the fidelity is negligible for the high-degree entanglement. 
 
To prepare maximal or non-maximal entanglement, numerical simulation methods are used to identify the optimal parameters \{$\Omega,t$\} for preparing the desired entanglement in the shortest possible time. To have a deeper physical insight into the optimal parameter selection and entanglement preparation, we analyze the dynamic trajectory of the reduced single qubit~(after tracing out surplus qubits)~\cite{1232957}, as illustrated in Fig.~\ref{fig11}(c, d), the maximal~(specific) entanglement can typically be prepared within a single cycle for systems with a small number of qubits. However, as the number of qubits increases, both the optimal parameters $\Omega$ and $t$ increase, necessitating multi-cycle dynamics for the reduced single qubit.

 To improve the fidelity for entanglement preparation, we introduce a gain into the system, the Hamiltonian of the system is similar to Eq.~(\ref{eq1}) but with the single-qubit Hamiltonian $H_{n}$ changed as
	\begin{equation}\label{e3}
		H'_n=-i\gamma_n/2|e\rangle\langle e|+i\kappa_n/2|g\rangle\langle g|+\Omega_n \sigma_n^x.
	\end{equation}
The trade-off relationship breaks as the gain increases and becomes entirely obsolete when the gain is balanced with the loss. This balance facilitates the achievement of maximal entanglement with high fidelity. The underlying reason is that in the truly $\mathcal{PT}$-symmetric system, the lost energy can be effectively compensated, allowing the system's properties to approximate those of a closed system. Consequently, this enables the realization of high-fidelity entanglement preparation with enhanced speed.
	
	\par
\emph{Examples for two-qubit.---} For the two-qubit system depicted in Fig.~\ref{fig1}(a), 
 we first confine our analysis to the all lossy EP system and assume $\gamma_n = \gamma$, $\Omega_n = \Omega$, for simplicity. The coupling between qubits lifts the degeneracy of the system, reducing the two-qubit system from fourth-order~($J=0$) to second-order EP. The real-to-complex spectral phase transition at the EP is critical to improving the entanglement preparation~\cite{PhysRevLett.131.100202}, and the concurrence is used to quantify the degree of entanglement~\cite{Wootters2001EntanglementOF}. Note that for calculating the concurrence, the final state is re-normalized.
	
 In comparison to Hermitian systems, requiring significant durations (inversely proportional to qubit coupling) to prepare the maximal entanglement, non-Hermitian systems can greatly accelerate this process. The acceleration rate for preparing the maximal entanglement under different coupling strengths is shown in Fig.~\ref{fig1}(b)~(blue dots), indicating \emph{a reduction of the time duration by four orders of magnitude} compared to Hermitian systems~($\gamma=0$). 
	
 Considering the performance of the fidelity during entanglement preparation, Fig.~\ref{fig1}(c) is the relation between the fidelity and concurrence for active and passive $\mathcal{PT}$-symmetric systems, which highlights a distinct trade-off between the concurrence and fidelity for the passive $\mathcal{PT}$-symmetric system. However, this trade-off relation can be effectively broken by the truly $\mathcal{PT}$-symmetric system, allowing the preparation of high degree of entanglement with high fidelity, i.e., for maximal concurrence, the fidelity has been improved by \emph{more than ten orders of magnitude}. Figure~\ref{fig1}(a) illustrates the fidelity and time required for maximal entanglement of truly $\mathcal{PT}$-symmetric and all lossy systems, respectively. The fidelity is approximately $10^{-17}$ for all lossy systems, which is negligible. However, the truly $\mathcal{PT}$-symmetric system led to a fidelity close to unity, which means a significant enhancement on the fidelity. Furthermore, it also exhibited a further acceleration effect compared to all lossy EP scenarios for the maximal entanglement preparation, as shown in Fig.~\ref{fig1}(b).		
  
 To further reveal the reasons behind the aforementioned results, we consider a gain factor $\kappa_n$ in the ground state, assuming $\kappa_n=\kappa=\gamma_n=\gamma$, Eq.~(\ref{e3}) becomes a truly $\mathcal{PT}$-symmetric Hamiltonian, the eigenvalues of which are given by $\pm1/2\sqrt{-\gamma^2+4\Omega^2}$. For the two-qubit system, as shown in the inset in Fig.~\ref{fig1}(c), when the drive amplitude is small ($\Omega < \gamma/2$), the gain is unable to compensate for the loss, the system is in non-equilibrium, one of the eigenmodes undergoes exponential decay while another amplifies, with the remaining modes unchanged with time~\cite{2019,El-Ganainy2018}. Conversely, within the $\mathcal{PT}$-symmetric phase~($\Omega > \gamma/2$), the gain can fully compensate for the loss, stabilizing the system at a real eigenfrequency, and the system behaves like a closed system~\cite{2019,El-Ganainy2018}. In contrast, the passive $\mathcal{PT}$-symmetric system, regardless of the magnitude of the driving amplitude, is incapable of counterbalancing the effect of loss, leading consistently to imaginary eigenvalues. Consequently, this underscores the potential of the truly $\mathcal{PT}$-symmetric system to significantly boost the fidelity of entanglement preparation. 

\begin{figure}
		\centering
		\includegraphics[width=8.8cm,height=6.9cm]{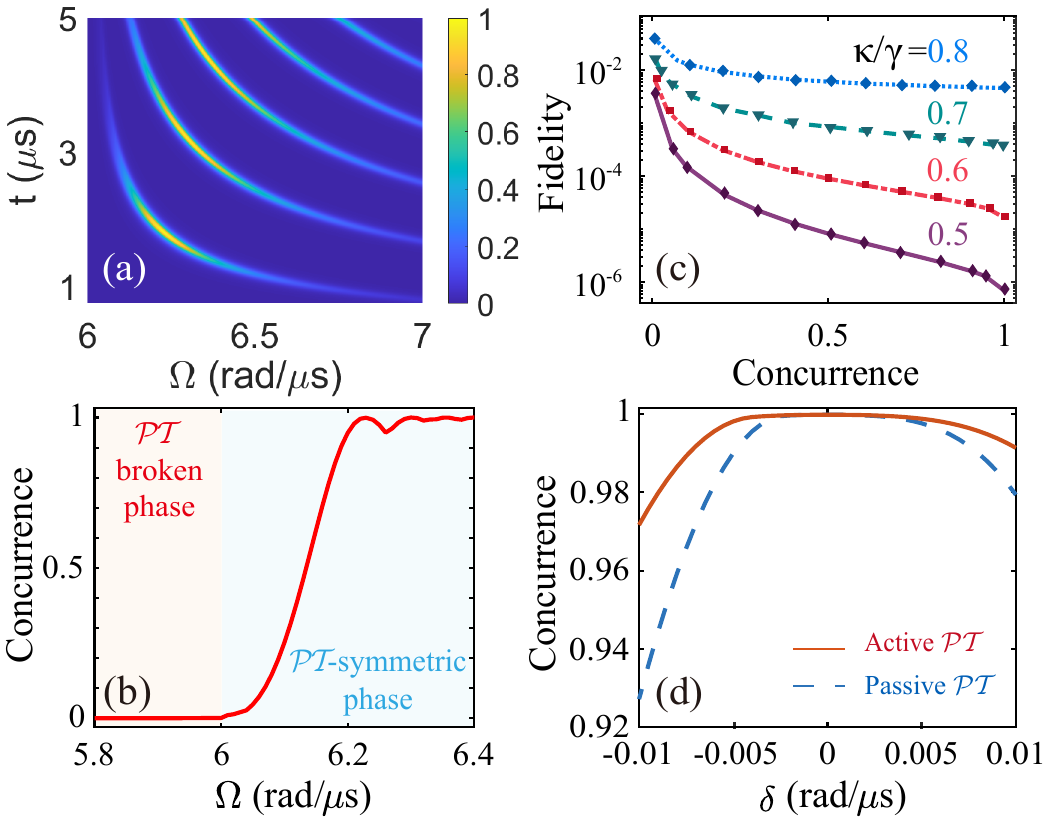}
		\caption{(a) Evolution of the concurrence with the drive amplitude $\Omega$. (b) Upper bound of the concurrence for different $\Omega$. (c) Fidelity versus concurrence for unbalanced gain-loss. (d) Concurrence versus the non-resonant driving error.}\label{fig2}
	\end{figure}

 Figure~\ref{fig2}(a) illustrates how to identify the optimal parameters \{$\Omega$, $t$\} to maximize the concurrence. Under the premise of achieving the maximal fidelity shown in Fig.~\ref{fig1}(c), the optimal parameters we choose correspond to \{$6.23~\rm{rad/\mu s}$, $1.8~\rm{\mu s}$\}. Increasing of $\Omega$, the maximal entanglement can still be prepared by prolonging the evolution time. Figure~\ref{fig2}(b) shows the upper limit of concurrence achievable under various values of $\Omega$, the entanglement always remains minimal and almost zero in the $\mathcal{PT}$-broken phase. However, upon transitioning into the $\mathcal{PT}$-symmetric phase, the entanglement gradually reaches its maximum and continues to exist with further increases in $\Omega$. The entanglement prepared within the $\mathcal{PT}$-broken region is characterized as steady-state entanglement~\cite{Supplement}.
	
	\par
To obtain more physical intuition on the improvement in fidelity, we here consider the impact of unbalanced gain-loss on the entanglement preparation. Note that for the preparation of maximal entanglement, the optimal parameters vary for different ratios of gain-loss, which is shown in~\cite{Supplement}. It is observable that as the discrepancy of gain-loss increases, the fidelity of preparing maximal entanglement decreases exponentially, as depicted in Fig.~\ref{fig2}(c). For a certain gain, it can be seen that there still exists a trade-off relation. As the discrepancy between gain and loss decreases, this relation gradually breaks down, and the truly $\mathcal{PT}$-symmetric system becomes optimal for the preparation of maximal entanglement. The fidelity is relatively high when the ratio $\kappa/\gamma$ is greater than 0.8, which indicates that the experimental requirements of our scheme can be relaxed. Furthermore, we examine the effects of off-resonant driving $\delta$ applied to the level $|g\rangle$, assuming, for simplicity, uniform error across all three qubits, the acceleration effect observed in a truly $\mathcal{PT}$-symmetric system potentially leads to reduced error accumulation, leading to enhanced robustness, as demonstrated in Fig~\ref{fig2}(d). 

To obtain repetitive analytical results, we now treat the weak interaction Hamiltonian $H=J(\sigma_1^{+}\sigma_2^{-}+\sigma_1^{-}\sigma_2^{+})$ as a perturbation, thereby calculating the concurrence. By projecting the evolution state onto $|\psi(t)\rangle=\sum_{j=1}^4\langle\Psi'_j|\psi_0\rangle e^{-it\Lambda_j}|\Psi_j\rangle$, where $\Lambda_j$ and $|\Psi_j\rangle$ represent the eigenvalues and eigenstates of the system, and $\langle\Psi'_j|$ denotes the left eigenstate within the biorthogonal basis, the concurrence can be obtained after normalizing the evolving state
\begin{equation}C=2\frac{|A|}{B},
\end{equation}which agrees well with the numerical findings~\cite{Supplement}.
 With the optimal parameters, the form of the maximal entangled state $|\psi_f\rangle$ is exactly $|\psi_i\rangle$ after normalization.
 
 \begin{figure}
		\centering
		\includegraphics[width=8.9cm,height=7.5cm]{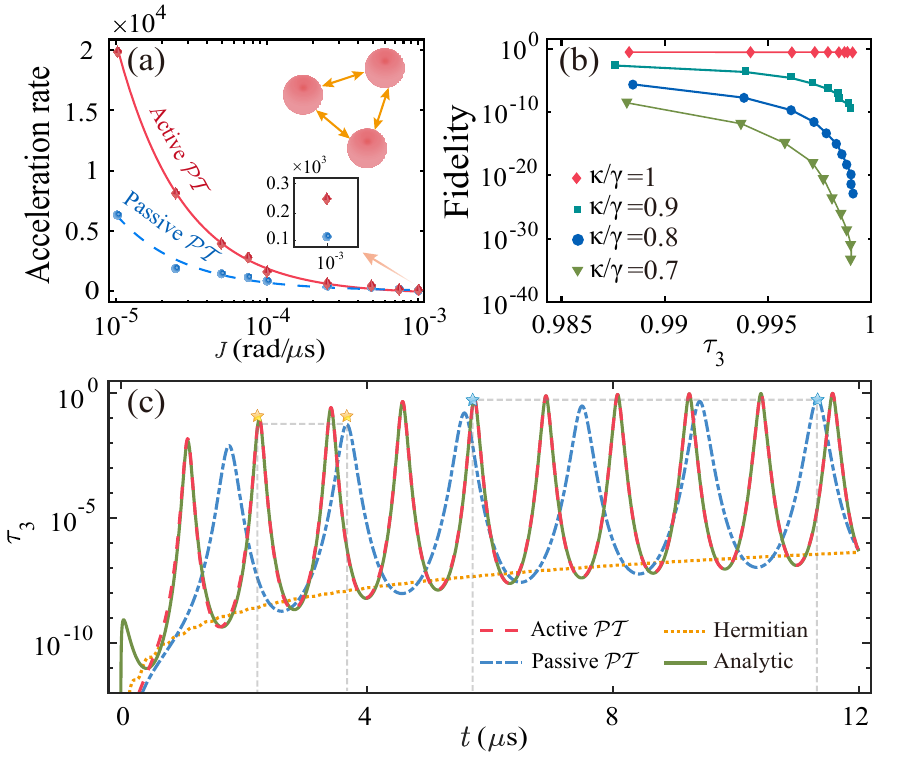}
		\caption{(a) The acceleration rate of active and passive $\mathcal{PT}$-symmetric systems. Inset: schematic diagram for preparing the triplet entanglement. (b) Fidelity of the residual tangle for a truly $\mathcal{PT}$-symmetric system~($\kappa/\gamma =1$) and unbalanced cases~($\kappa/\gamma = 0.9, 0.8, 0.7$). (c) Evolution of $\tau_3$ for active, passive $\mathcal{PT}$-symmetric, and Hermitian systems. The gray dashed line is to show, for the same degree of entanglement, the active $\mathcal{PT}$-symmetric system has an accelerated effect compared to the passive counterpart. The initial state is $|ggg\rangle$ and the other parameters are $J=0.001~\rm{rad/\mu s}$, $\kappa=\gamma=12~\rm{\mu s^{-1}}$. The optimal $\Omega$ can be found numerically as 6.58 and $3.42~\rm{rad/\mu s}$ for active and passive $\mathcal{PT}$-symmetric systems, respectively~\cite{Supplement}. The green line shows the analytic result calculated by perturbation theory for the truly $\mathcal{PT}$-symmetric Hamiltonian.
		}\label{fig3}
	\end{figure}	
 
\emph{Examples for three-qubit.---} For the generality of the above theory, we now investigate the $\mathcal{PT}$ symmetry improved triplet entanglement~[see the inset in Fig.~\ref{fig3}(a)]. Now $
	H=\sum_{n=1}^{3}H'_n+J\sum_{n\neq m=1}^3\sigma_n^{+}\sigma_m^{-}$
	for the triplet $\mathcal{PT}$-symmetric system. We now assume that the normalized evolutionary state of the system as $|\psi'\rangle=|\psi\rangle/||\psi\rangle|=a_1|ggg\rangle+a_2|gge\rangle+a_3|geg\rangle+a_4|gee\rangle+a_5|egg\rangle+a_6|ege\rangle+a_7|eeg\rangle+a_8|eee\rangle$. To quantify the genuine three-qubit entanglement, the residual tangle  $\tau_3=C_{A|BC}^2-C_{AB}^2-C_{AC}^2$ is introduced~\cite{PhysRevA.61.052306}, where $C_{AB(C)}^2$ quantifies the entanglement between qubits $A$ and $B(C)$, $C_{A|BC}^2$ quantifies the qubit $A$ and the composite qubit system $BC$,
	which can be further expressed as~\cite{PhysRevLett.97.260502}
	\begin{equation}\label{eq2}
		\tau_3=4|d_1-2d_2+4d_3|,
	\end{equation}
	where $d_1=a_1^2a_8^2+a_2^2a_7^2+a_3^2a_6^2+a_4^2a_5^2$, $d_2=a_1a_8a_4a_5+a_1a_8a_3a_6+a_1a_8a_2a_7 +a_3a_4a_5a_6+a_4a_5a_2a_7+a_2a_7a_3a_6$ and $d_3=a_1a_7a_4a_6+a_2a_8a_3a_5$. Through solving the Schr\"{o}dinger equation, we can obtain the fidelity and $\tau_3$ numerically.

A high fidelity can be realized for a high degree of entanglement preparation in the truly $\mathcal{PT}$-symmetric system, as shown in Fig.~\ref{fig3}(b) (depicted by the red diamonds). The results show that at the maximum entanglement attachment of the three-body system, the fidelity improves by \emph{several dozen orders of magnitude} as the parameters approach the gain-loss balance, which simultaneously implies that the trade-off relationship is broken. The detailed entanglement preparation process is shown in the Supplementary Materials~\cite{Supplement}. Moreover, the fidelity of achieving similar entanglement decreases exponentially as the gain-loss ratio decreases. When the ratio of gain-loss falls below 0.8, achieving the maximal entanglement becomes impossible, stressing that it is better to maintain balanced gain and loss for preparing a high degree of entanglement with high fidelity.

In addition, an enhanced acceleration effect is also observed compared to the passive $\mathcal{PT}$-symmetric system, as depicted in Fig.~\ref{fig3}(a). We stress that several Rabi-like cycles are considered to choose the optimal parameters corresponding to a the high degree of entanglement, with $\tau_3$ close to unity. We numerically analyzed the value of $\tau_3$ as a function of the drive amplitude $\Omega$ and evolution time, identifying the upper bound of entanglement for various drive amplitudes, which exhibit an oscillatory and increasing trend with the fidelity, as described in detail in \cite{Supplement}. Here we focus on preparing a high-degree of entanglement; thus presenting only the extremum points of $\tau_3$ in Fig.~\ref{fig3}(b).

For the eigenvalues $\Lambda_j$ and eigenstates $|\Psi_j\rangle$, the system state can be represented as $|\psi(t)\rangle=\sum_{j=1}^8\langle\Psi'_j|\psi_0\rangle e^{-it\Lambda_j}|\Psi_j\rangle$. Assuming that the initial state $|\psi_0\rangle$ is $|ggg\rangle$, the evolution of entanglement can be obtained as $\tau_3=|A/4\eta^{12}B^2|$~\cite{Supplement}. From Fig.~\ref{fig3}(c), we can see that the perturbation analytic result is in good agreement with the numerical simulations. It shows that the attainment of maximal entanglement undergoes several Rabi-like cycles within the three-qubit system. Notably, the truly $\mathcal{PT}$-symmetric system exhibits a distinct acceleration effect compared to all lossy and Hermitian systems. The truly $\mathcal{PT}$-symmetric system also has stronger robustness against detuning error.
	\par
\emph{Discussions.---}	As the coupling strength is reduced, the time required to achieve maximal entanglement is extended for both Hermitian and non-Hermitian dynamics. However, the acceleration factor of the truly $\mathcal{PT}$-symmetric system would be further improved, as illustrated in Figs.~\ref{fig1}(b) and~\ref{fig3}(a). Moreover, an increase in coupling strength well beyond $J=0.001~\rm{rad/\mu s}$ still allows for the attainment of maximal entanglement by appropriately tuning the parameter $\Omega$, along with observable acceleration effects~\cite{Supplement}. Also, satisfying the condition $4\Omega_1^2-\gamma_1^2=4\Omega_2^2-\gamma_2^2=4\Omega_3^2-\gamma_3^2$ preserves a high degree of entanglement and enhances the acceleration effect, thereby overcoming the challenge of maintaining identical parameters across all three qubits~\cite{Supplement}. If a method for quantifying the $n$-qubit~$(n\geq 4)$ entanglement exists, the truly $\mathcal{PT}$-symmetric improved entanglement preparation may also be valid.

\emph{Conclusions.---}  Our study shows that a truly $\mathcal{PT}$-symmetric system can be employed to accelerate the preparation of multi-qubit entanglement by \emph{several orders of magnitude} compared to the conventional Hermitian system. More importantly, the trade-off relation between the fidelity and the degree of entanglement predicted by the passive $\mathcal{PT}$-symmetric system can be broken by the truly $\mathcal{PT}$-symmetric system. The main reason is due to the fact that the energy is conserved in the $\mathcal{PT}$-symmetric phase region of the truly $\mathcal{PT}$-symmetric system. A detailed analysis was conducted on the influence of the unbalanced gain-loss on the preparation of the maximal entanglement. Our finding paves a new way for high-efficiency multi-qubit entanglement preparation and is meaningful for large-scale quantum information.
	
\begin{acknowledgements}	\emph{Acknowledgements.}-
We would like to thank Prof. \c{S}. K. {\"O}zdemir for discussions. This work was supported by the National Key R\&D Program of China (Grants No. 2022YFA1404500), Cross-disciplinary Innovative Research Group Project of Henan Province (Grant No. 232300421004), the National Natural Science Foundation of China (Grants No. 12274376, 12125406, and 12074232). and a major science and technology project of Henan Province under Grant No. 221100210400, and the Natural Science Foundation of Henan Province under Grant No. 232300421075 and 212300410085. H.J. is supported by the NSFC (11935006), the Science and Technology Innovation Program of Hunan Province (Grant No. 2020RC4047), National Key R\&D Program of China (No. 2024YFE0102400) and Hunan provincial major sci-tech program (2023ZJ1010). F.N. is supported in part by: Nippon Telegraph and Telephone Corporation (NTT) Research, the Japan Science and Technology Agency (JST) [via the CREST Quantum Frontiers program, the Quantum Leap Flagship Program (Q-LEAP), and the Moonshot R\&D Grant Number JPMJMS2061], and the Office of Naval Research (ONR) Global (via Grant No. N62909-23-1-2074).
\end{acknowledgements}	

\bibliography{REV}

\begin{thebibliography}{87}%
\makeatletter
\providecommand \@ifxundefined [1]{%
 \@ifx{#1\undefined}
}%
\providecommand \@ifnum [1]{%
 \ifnum #1\expandafter \@firstoftwo
 \else \expandafter \@secondoftwo
 \fi
}%
\providecommand \@ifx [1]{%
 \ifx #1\expandafter \@firstoftwo
 \else \expandafter \@secondoftwo
 \fi
}%
\providecommand \natexlab [1]{#1}%
\providecommand \enquote  [1]{``#1''}%
\providecommand \bibnamefont  [1]{#1}%
\providecommand \bibfnamefont [1]{#1}%
\providecommand \citenamefont [1]{#1}%
\providecommand \href@noop [0]{\@secondoftwo}%
\providecommand \href [0]{\begingroup \@sanitize@url \@href}%
\providecommand \@href[1]{\@@startlink{#1}\@@href}%
\providecommand \@@href[1]{\endgroup#1\@@endlink}%
\providecommand \@sanitize@url [0]{\catcode `\\12\catcode `\$12\catcode
  `\&12\catcode `\#12\catcode `\^12\catcode `\_12\catcode `\%12\relax}%
\providecommand \@@startlink[1]{}%
\providecommand \@@endlink[0]{}%
\providecommand \url  [0]{\begingroup\@sanitize@url \@url }%
\providecommand \@url [1]{\endgroup\@href {#1}{\urlprefix }}%
\providecommand \urlprefix  [0]{URL }%
\providecommand \Eprint [0]{\href }%
\providecommand \doibase [0]{http://dx.doi.org/}%
\providecommand \selectlanguage [0]{\@gobble}%
\providecommand \bibinfo  [0]{\@secondoftwo}%
\providecommand \bibfield  [0]{\@secondoftwo}%
\providecommand \translation [1]{[#1]}%
\providecommand \BibitemOpen [0]{}%
\providecommand \bibitemStop [0]{}%
\providecommand \bibitemNoStop [0]{.\EOS\space}%
\providecommand \EOS [0]{\spacefactor3000\relax}%
\providecommand \BibitemShut  [1]{\csname bibitem#1\endcsname}%
\let\auto@bib@innerbib\@empty
\bibitem [{\citenamefont {Horodecki}\ \emph {et~al.}(2009)\citenamefont
  {Horodecki}, \citenamefont {Horodecki}, \citenamefont {Horodecki},\ and\
  \citenamefont {Horodecki}}]{RevModPhys.81.865}%
  \BibitemOpen
  \bibfield  {author} {\bibinfo {author} {\bibfnamefont {R.}~\bibnamefont
  {Horodecki}}, \bibinfo {author} {\bibfnamefont {P.}~\bibnamefont
  {Horodecki}}, \bibinfo {author} {\bibfnamefont {M.}~\bibnamefont
  {Horodecki}}, \ and\ \bibinfo {author} {\bibfnamefont {K.}~\bibnamefont
  {Horodecki}},\ }\href {\doibase 10.1103/RevModPhys.81.865} {\bibfield
  {journal} {\bibinfo  {journal} {Rev. Mod. Phys.}\ }\textbf {\bibinfo {volume}
  {81}},\ \bibinfo {pages} {865} (\bibinfo {year} {2009})}\BibitemShut
  {NoStop}%
\bibitem [{\citenamefont {Steane}(1998)}]{Andrew}%
  \BibitemOpen
  \bibfield  {author} {\bibinfo {author} {\bibfnamefont {A.}~\bibnamefont
  {Steane}},\ }\href {\doibase 10./0034-5/61/2/002} {\bibfield  {journal}
  {\bibinfo  {journal} {Reports on Progress in Physics}\ }\textbf {\bibinfo
  {volume} {61}},\ \bibinfo {pages} {117} (\bibinfo {year} {1998})}\BibitemShut
  {NoStop}%
\bibitem [{\citenamefont {Jozsa}\ and\ \citenamefont {Linden}(2003)}]{1097}%
  \BibitemOpen
  \bibfield  {author} {\bibinfo {author} {\bibfnamefont {R.}~\bibnamefont
  {Jozsa}}\ and\ \bibinfo {author} {\bibfnamefont {N.}~\bibnamefont {Linden}},\
  }\href {\doibase 10.1098/rspa.2002.1097} {\bibfield  {journal} {\bibinfo
  {journal} {Proceedings of the Royal Society of London. Series A:
  Mathematical, Physical and Engineering Sciences}\ }\textbf {\bibinfo {volume}
  {459}},\ \bibinfo {pages} {2011} (\bibinfo {year} {2003})}\BibitemShut
  {NoStop}%
\bibitem [{\citenamefont {Walther}\ \emph {et~al.}(2005)\citenamefont
  {Walther}, \citenamefont {Resch}, \citenamefont {Rudolph}, \citenamefont
  {Schenck}, \citenamefont {Weinfurter}, \citenamefont {Vedral}, \citenamefont
  {Aspelmeyer},\ and\ \citenamefont {Zeilinger}}]{Walther2005}%
  \BibitemOpen
  \bibfield  {author} {\bibinfo {author} {\bibfnamefont {P.}~\bibnamefont
  {Walther}}, \bibinfo {author} {\bibfnamefont {K.~J.}\ \bibnamefont {Resch}},
  \bibinfo {author} {\bibfnamefont {T.}~\bibnamefont {Rudolph}}, \bibinfo
  {author} {\bibfnamefont {E.}~\bibnamefont {Schenck}}, \bibinfo {author}
  {\bibfnamefont {H.}~\bibnamefont {Weinfurter}}, \bibinfo {author}
  {\bibfnamefont {V.}~\bibnamefont {Vedral}}, \bibinfo {author} {\bibfnamefont
  {M.}~\bibnamefont {Aspelmeyer}}, \ and\ \bibinfo {author} {\bibfnamefont
  {A.}~\bibnamefont {Zeilinger}},\ }\href {\doibase 10.1038/nature03347}
  {\bibfield  {journal} {\bibinfo  {journal} {Nature}\ }\textbf {\bibinfo
  {volume} {434}},\ \bibinfo {pages} {169} (\bibinfo {year}
  {2005})}\BibitemShut {NoStop}%
\bibitem [{\citenamefont {Bouwmeester}\ \emph {et~al.}(1997)\citenamefont
  {Bouwmeester}, \citenamefont {Pan}, \citenamefont {Mattle}, \citenamefont
  {Eibl}, \citenamefont {Weinfurter},\ and\ \citenamefont
  {Zeilinger}}]{Bouwmeester1997}%
  \BibitemOpen
  \bibfield  {author} {\bibinfo {author} {\bibfnamefont {D.}~\bibnamefont
  {Bouwmeester}}, \bibinfo {author} {\bibfnamefont {J.-W.}\ \bibnamefont
  {Pan}}, \bibinfo {author} {\bibfnamefont {K.}~\bibnamefont {Mattle}},
  \bibinfo {author} {\bibfnamefont {M.}~\bibnamefont {Eibl}}, \bibinfo {author}
  {\bibfnamefont {H.}~\bibnamefont {Weinfurter}}, \ and\ \bibinfo {author}
  {\bibfnamefont {A.}~\bibnamefont {Zeilinger}},\ }\href {\doibase
  10.1038/37539} {\bibfield  {journal} {\bibinfo  {journal} {Nature}\ }\textbf
  {\bibinfo {volume} {390}},\ \bibinfo {pages} {575} (\bibinfo {year}
  {1997})}\BibitemShut {NoStop}%
\bibitem [{\citenamefont {Ursin}\ \emph {et~al.}(2007)\citenamefont {Ursin},
  \citenamefont {Tiefenbacher}, \citenamefont {Schmitt-Manderbach},
  \citenamefont {Weier}, \citenamefont {Scheidl}, \citenamefont {Lindenthal},
  \citenamefont {Blauensteiner}, \citenamefont {Jennewein}, \citenamefont
  {Perdigues}, \citenamefont {Trojek}, \citenamefont {{\"O}mer}, \citenamefont
  {F{\"u}rst}, \citenamefont {Meyenburg}, \citenamefont {Rarity}, \citenamefont
  {Sodnik}, \citenamefont {Barbieri}, \citenamefont {Weinfurter},\ and\
  \citenamefont {Zeilinger}}]{Ursin2007}%
  \BibitemOpen
  \bibfield  {author} {\bibinfo {author} {\bibfnamefont {R.}~\bibnamefont
  {Ursin}}, \bibinfo {author} {\bibfnamefont {F.}~\bibnamefont {Tiefenbacher}},
  \bibinfo {author} {\bibfnamefont {T.}~\bibnamefont {Schmitt-Manderbach}},
  \bibinfo {author} {\bibfnamefont {H.}~\bibnamefont {Weier}}, \bibinfo
  {author} {\bibfnamefont {T.}~\bibnamefont {Scheidl}}, \bibinfo {author}
  {\bibfnamefont {M.}~\bibnamefont {Lindenthal}}, \bibinfo {author}
  {\bibfnamefont {B.}~\bibnamefont {Blauensteiner}}, \bibinfo {author}
  {\bibfnamefont {T.}~\bibnamefont {Jennewein}}, \bibinfo {author}
  {\bibfnamefont {J.}~\bibnamefont {Perdigues}}, \bibinfo {author}
  {\bibfnamefont {P.}~\bibnamefont {Trojek}}, \bibinfo {author} {\bibfnamefont
  {B.}~\bibnamefont {{\"O}mer}}, \bibinfo {author} {\bibfnamefont
  {M.}~\bibnamefont {F{\"u}rst}}, \bibinfo {author} {\bibfnamefont
  {M.}~\bibnamefont {Meyenburg}}, \bibinfo {author} {\bibfnamefont
  {J.}~\bibnamefont {Rarity}}, \bibinfo {author} {\bibfnamefont
  {Z.}~\bibnamefont {Sodnik}}, \bibinfo {author} {\bibfnamefont
  {C.}~\bibnamefont {Barbieri}}, \bibinfo {author} {\bibfnamefont
  {H.}~\bibnamefont {Weinfurter}}, \ and\ \bibinfo {author} {\bibfnamefont
  {A.}~\bibnamefont {Zeilinger}},\ }\href {\doibase 10.1038/nphys629}
  {\bibfield  {journal} {\bibinfo  {journal} {Nature Physics}\ }\textbf
  {\bibinfo {volume} {3}},\ \bibinfo {pages} {481} (\bibinfo {year}
  {2007})}\BibitemShut {NoStop}%
\bibitem [{\citenamefont {Olmschenk}\ \emph {et~al.}(2009)\citenamefont
  {Olmschenk}, \citenamefont {Matsukevich}, \citenamefont {Maunz},
  \citenamefont {Hayes}, \citenamefont {Duan},\ and\ \citenamefont
  {Monroe}}]{doi:10.1126/science.1167209}%
  \BibitemOpen
  \bibfield  {author} {\bibinfo {author} {\bibfnamefont {S.}~\bibnamefont
  {Olmschenk}}, \bibinfo {author} {\bibfnamefont {D.~N.}\ \bibnamefont
  {Matsukevich}}, \bibinfo {author} {\bibfnamefont {P.}~\bibnamefont {Maunz}},
  \bibinfo {author} {\bibfnamefont {D.}~\bibnamefont {Hayes}}, \bibinfo
  {author} {\bibfnamefont {L.-M.}\ \bibnamefont {Duan}}, \ and\ \bibinfo
  {author} {\bibfnamefont {C.}~\bibnamefont {Monroe}},\ }\href {\doibase
  10.1126/science.1167209} {\bibfield  {journal} {\bibinfo  {journal}
  {Science}\ }\textbf {\bibinfo {volume} {323}},\ \bibinfo {pages} {486}
  (\bibinfo {year} {2009})}\BibitemShut {NoStop}%
\bibitem [{\citenamefont {Yin}\ \emph {et~al.}(2012)\citenamefont {Yin},
  \citenamefont {Ren}, \citenamefont {Lu}, \citenamefont {Cao}, \citenamefont
  {Yong}, \citenamefont {Wu}, \citenamefont {Liu}, \citenamefont {Liao},
  \citenamefont {Zhou}, \citenamefont {Jiang}, \citenamefont {Cai},
  \citenamefont {Xu}, \citenamefont {Pan}, \citenamefont {Jia}, \citenamefont
  {Huang}, \citenamefont {Yin}, \citenamefont {Wang}, \citenamefont {Chen},
  \citenamefont {Peng},\ and\ \citenamefont {Pan}}]{Yin2012}%
  \BibitemOpen
  \bibfield  {author} {\bibinfo {author} {\bibfnamefont {J.}~\bibnamefont
  {Yin}}, \bibinfo {author} {\bibfnamefont {J.-G.}\ \bibnamefont {Ren}},
  \bibinfo {author} {\bibfnamefont {H.}~\bibnamefont {Lu}}, \bibinfo {author}
  {\bibfnamefont {Y.}~\bibnamefont {Cao}}, \bibinfo {author} {\bibfnamefont
  {H.-L.}\ \bibnamefont {Yong}}, \bibinfo {author} {\bibfnamefont {Y.-P.}\
  \bibnamefont {Wu}}, \bibinfo {author} {\bibfnamefont {C.}~\bibnamefont
  {Liu}}, \bibinfo {author} {\bibfnamefont {S.-K.}\ \bibnamefont {Liao}},
  \bibinfo {author} {\bibfnamefont {F.}~\bibnamefont {Zhou}}, \bibinfo {author}
  {\bibfnamefont {Y.}~\bibnamefont {Jiang}}, \bibinfo {author} {\bibfnamefont
  {X.-D.}\ \bibnamefont {Cai}}, \bibinfo {author} {\bibfnamefont
  {P.}~\bibnamefont {Xu}}, \bibinfo {author} {\bibfnamefont {G.-S.}\
  \bibnamefont {Pan}}, \bibinfo {author} {\bibfnamefont {J.-J.}\ \bibnamefont
  {Jia}}, \bibinfo {author} {\bibfnamefont {Y.-M.}\ \bibnamefont {Huang}},
  \bibinfo {author} {\bibfnamefont {H.}~\bibnamefont {Yin}}, \bibinfo {author}
  {\bibfnamefont {J.-Y.}\ \bibnamefont {Wang}}, \bibinfo {author}
  {\bibfnamefont {Y.-A.}\ \bibnamefont {Chen}}, \bibinfo {author}
  {\bibfnamefont {C.-Z.}\ \bibnamefont {Peng}}, \ and\ \bibinfo {author}
  {\bibfnamefont {J.-W.}\ \bibnamefont {Pan}},\ }\href {\doibase
  10.1038/nature11332} {\bibfield  {journal} {\bibinfo  {journal} {Nature}\
  }\textbf {\bibinfo {volume} {488}},\ \bibinfo {pages} {185} (\bibinfo {year}
  {2012})}\BibitemShut {NoStop}%
\bibitem [{\citenamefont {Ma}\ \emph {et~al.}(2012)\citenamefont {Ma},
  \citenamefont {Herbst}, \citenamefont {Scheidl}, \citenamefont {Wang},
  \citenamefont {Kropatschek}, \citenamefont {Naylor}, \citenamefont
  {Wittmann}, \citenamefont {Mech}, \citenamefont {Kofler}, \citenamefont
  {Anisimova}, \citenamefont {Makarov}, \citenamefont {Jennewein},
  \citenamefont {Ursin},\ and\ \citenamefont {Zeilinger}}]{Ma2012}%
  \BibitemOpen
  \bibfield  {author} {\bibinfo {author} {\bibfnamefont {X.-S.}\ \bibnamefont
  {Ma}}, \bibinfo {author} {\bibfnamefont {T.}~\bibnamefont {Herbst}}, \bibinfo
  {author} {\bibfnamefont {T.}~\bibnamefont {Scheidl}}, \bibinfo {author}
  {\bibfnamefont {D.}~\bibnamefont {Wang}}, \bibinfo {author} {\bibfnamefont
  {S.}~\bibnamefont {Kropatschek}}, \bibinfo {author} {\bibfnamefont
  {W.}~\bibnamefont {Naylor}}, \bibinfo {author} {\bibfnamefont
  {B.}~\bibnamefont {Wittmann}}, \bibinfo {author} {\bibfnamefont
  {A.}~\bibnamefont {Mech}}, \bibinfo {author} {\bibfnamefont {J.}~\bibnamefont
  {Kofler}}, \bibinfo {author} {\bibfnamefont {E.}~\bibnamefont {Anisimova}},
  \bibinfo {author} {\bibfnamefont {V.}~\bibnamefont {Makarov}}, \bibinfo
  {author} {\bibfnamefont {T.}~\bibnamefont {Jennewein}}, \bibinfo {author}
  {\bibfnamefont {R.}~\bibnamefont {Ursin}}, \ and\ \bibinfo {author}
  {\bibfnamefont {A.}~\bibnamefont {Zeilinger}},\ }\href {\doibase
  10.1038/nature11472} {\bibfield  {journal} {\bibinfo  {journal} {Nature}\
  }\textbf {\bibinfo {volume} {489}},\ \bibinfo {pages} {269} (\bibinfo {year}
  {2012})}\BibitemShut {NoStop}%
\bibitem [{\citenamefont {Roos}\ \emph {et~al.}(2006)\citenamefont {Roos},
  \citenamefont {Chwalla}, \citenamefont {Kim}, \citenamefont {Riebe},\ and\
  \citenamefont {Blatt}}]{Roos2006}%
  \BibitemOpen
  \bibfield  {author} {\bibinfo {author} {\bibfnamefont {C.~F.}\ \bibnamefont
  {Roos}}, \bibinfo {author} {\bibfnamefont {M.}~\bibnamefont {Chwalla}},
  \bibinfo {author} {\bibfnamefont {K.}~\bibnamefont {Kim}}, \bibinfo {author}
  {\bibfnamefont {M.}~\bibnamefont {Riebe}}, \ and\ \bibinfo {author}
  {\bibfnamefont {R.}~\bibnamefont {Blatt}},\ }\href {\doibase
  10.1038/nature05101} {\bibfield  {journal} {\bibinfo  {journal} {Nature}\
  }\textbf {\bibinfo {volume} {443}},\ \bibinfo {pages} {316} (\bibinfo {year}
  {2006})}\BibitemShut {NoStop}%
\bibitem [{\citenamefont {Riedel}\ \emph {et~al.}(2010)\citenamefont {Riedel},
  \citenamefont {B{\"o}hi}, \citenamefont {Li}, \citenamefont {H{\"a}nsch},
  \citenamefont {Sinatra},\ and\ \citenamefont {Treutlein}}]{Riedel2010}%
  \BibitemOpen
  \bibfield  {author} {\bibinfo {author} {\bibfnamefont {M.~F.}\ \bibnamefont
  {Riedel}}, \bibinfo {author} {\bibfnamefont {P.}~\bibnamefont {B{\"o}hi}},
  \bibinfo {author} {\bibfnamefont {Y.}~\bibnamefont {Li}}, \bibinfo {author}
  {\bibfnamefont {T.~W.}\ \bibnamefont {H{\"a}nsch}}, \bibinfo {author}
  {\bibfnamefont {A.}~\bibnamefont {Sinatra}}, \ and\ \bibinfo {author}
  {\bibfnamefont {P.}~\bibnamefont {Treutlein}},\ }\href {\doibase
  10.1038/nature08988} {\bibfield  {journal} {\bibinfo  {journal} {Nature}\
  }\textbf {\bibinfo {volume} {464}},\ \bibinfo {pages} {1170} (\bibinfo {year}
  {2010})}\BibitemShut {NoStop}%
\bibitem [{\citenamefont {Napolitano}\ \emph {et~al.}(2011)\citenamefont
  {Napolitano}, \citenamefont {Koschorreck}, \citenamefont {Dubost},
  \citenamefont {Behbood}, \citenamefont {Sewell},\ and\ \citenamefont
  {Mitchell}}]{Napolitano2011}%
  \BibitemOpen
  \bibfield  {author} {\bibinfo {author} {\bibfnamefont {M.}~\bibnamefont
  {Napolitano}}, \bibinfo {author} {\bibfnamefont {M.}~\bibnamefont
  {Koschorreck}}, \bibinfo {author} {\bibfnamefont {B.}~\bibnamefont {Dubost}},
  \bibinfo {author} {\bibfnamefont {N.}~\bibnamefont {Behbood}}, \bibinfo
  {author} {\bibfnamefont {R.~J.}\ \bibnamefont {Sewell}}, \ and\ \bibinfo
  {author} {\bibfnamefont {M.~W.}\ \bibnamefont {Mitchell}},\ }\href {\doibase
  10.1038/nature09778} {\bibfield  {journal} {\bibinfo  {journal} {Nature}\
  }\textbf {\bibinfo {volume} {471}},\ \bibinfo {pages} {486} (\bibinfo {year}
  {2011})}\BibitemShut {NoStop}%
\bibitem [{\citenamefont {Pezz\`e}\ \emph {et~al.}(2018)\citenamefont
  {Pezz\`e}, \citenamefont {Smerzi}, \citenamefont {Oberthaler}, \citenamefont
  {Schmied},\ and\ \citenamefont {Treutlein}}]{RevModPhys}%
  \BibitemOpen
  \bibfield  {author} {\bibinfo {author} {\bibfnamefont {L.}~\bibnamefont
  {Pezz\`e}}, \bibinfo {author} {\bibfnamefont {A.}~\bibnamefont {Smerzi}},
  \bibinfo {author} {\bibfnamefont {M.~K.}\ \bibnamefont {Oberthaler}},
  \bibinfo {author} {\bibfnamefont {R.}~\bibnamefont {Schmied}}, \ and\
  \bibinfo {author} {\bibfnamefont {P.}~\bibnamefont {Treutlein}},\ }\href
  {\doibase 10.1103/RevModPhys.90.035005} {\bibfield  {journal} {\bibinfo
  {journal} {Rev. Mod. Phys.}\ }\textbf {\bibinfo {volume} {90}},\ \bibinfo
  {pages} {035005} (\bibinfo {year} {2018})}\BibitemShut {NoStop}%
\bibitem [{\citenamefont {Gardiner}\ and\ \citenamefont
  {Zoller}(2004)}]{Gardiner2004QuantumNA}%
  \BibitemOpen
  \bibfield  {author} {\bibinfo {author} {\bibfnamefont {C.~W.}\ \bibnamefont
  {Gardiner}}\ and\ \bibinfo {author} {\bibfnamefont {P.}~\bibnamefont
  {Zoller}}\ }(\bibinfo  {publisher} {Springer, Berlin},\ \bibinfo {year}
  {2004})\BibitemShut {NoStop}%
\bibitem [{\citenamefont {Schlosshauer}(2019)}]{SCHLOSSHAUER20191}%
  \BibitemOpen
  \bibfield  {author} {\bibinfo {author} {\bibfnamefont {M.}~\bibnamefont
  {Schlosshauer}},\ }\href {\doibase
  https://doi.org/10.1016/j.physrep.2019.10.001} {\bibfield  {journal}
  {\bibinfo  {journal} {Physics Reports}\ }\textbf {\bibinfo {volume} {831}},\
  \bibinfo {pages} {1} (\bibinfo {year} {2019})}\BibitemShut {NoStop}%
\bibitem [{\citenamefont {Zurek}\ and\ \citenamefont
  {Paz}(1994)}]{PhysRevLett.72.2508}%
  \BibitemOpen
  \bibfield  {author} {\bibinfo {author} {\bibfnamefont {W.~H.}\ \bibnamefont
  {Zurek}}\ and\ \bibinfo {author} {\bibfnamefont {J.~P.}\ \bibnamefont
  {Paz}},\ }\href {\doibase 10.1103/PhysRevLett.72.2508} {\bibfield  {journal}
  {\bibinfo  {journal} {Phys. Rev. Lett.}\ }\textbf {\bibinfo {volume} {72}},\
  \bibinfo {pages} {2508} (\bibinfo {year} {1994})}\BibitemShut {NoStop}%
\bibitem [{\citenamefont {Zurek}(2003)}]{RevModPhys.75.715}%
  \BibitemOpen
  \bibfield  {author} {\bibinfo {author} {\bibfnamefont {W.~H.}\ \bibnamefont
  {Zurek}},\ }\href {\doibase 10.1103/RevModPhys.75.715} {\bibfield  {journal}
  {\bibinfo  {journal} {Rev. Mod. Phys.}\ }\textbf {\bibinfo {volume} {75}},\
  \bibinfo {pages} {715} (\bibinfo {year} {2003})}\BibitemShut {NoStop}%
\bibitem [{\citenamefont {Bender}\ \emph {et~al.}(2007)\citenamefont {Bender},
  \citenamefont {Brody}, \citenamefont {Jones},\ and\ \citenamefont
  {Meister}}]{PhysRevLett.98.040403}%
  \BibitemOpen
  \bibfield  {author} {\bibinfo {author} {\bibfnamefont {C.~M.}\ \bibnamefont
  {Bender}}, \bibinfo {author} {\bibfnamefont {D.~C.}\ \bibnamefont {Brody}},
  \bibinfo {author} {\bibfnamefont {H.~F.}\ \bibnamefont {Jones}}, \ and\
  \bibinfo {author} {\bibfnamefont {B.~K.}\ \bibnamefont {Meister}},\ }\href
  {\doibase 10.1103/PhysRevLett.98.040403} {\bibfield  {journal} {\bibinfo
  {journal} {Phys. Rev. Lett.}\ }\textbf {\bibinfo {volume} {98}},\ \bibinfo
  {pages} {040403} (\bibinfo {year} {2007})}\BibitemShut {NoStop}%
\bibitem [{\citenamefont {G\"unther}\ and\ \citenamefont
  {Samsonov}(2008)}]{PhysRevLett.101.230404}%
  \BibitemOpen
  \bibfield  {author} {\bibinfo {author} {\bibfnamefont {U.}~\bibnamefont
  {G\"unther}}\ and\ \bibinfo {author} {\bibfnamefont {B.~F.}\ \bibnamefont
  {Samsonov}},\ }\href {\doibase 10.1103/PhysRevLett.101.230404} {\bibfield
  {journal} {\bibinfo  {journal} {Phys. Rev. Lett.}\ }\textbf {\bibinfo
  {volume} {101}},\ \bibinfo {pages} {230404} (\bibinfo {year}
  {2008})}\BibitemShut {NoStop}%
\bibitem [{\citenamefont {Li}\ \emph {et~al.}(2023)\citenamefont {Li},
  \citenamefont {Chen}, \citenamefont {Abbasi}, \citenamefont {Murch},\ and\
  \citenamefont {Whaley}}]{PhysRevLett.131.100202}%
  \BibitemOpen
  \bibfield  {author} {\bibinfo {author} {\bibfnamefont {Z.-Z.}\ \bibnamefont
  {Li}}, \bibinfo {author} {\bibfnamefont {W.}~\bibnamefont {Chen}}, \bibinfo
  {author} {\bibfnamefont {M.}~\bibnamefont {Abbasi}}, \bibinfo {author}
  {\bibfnamefont {K.~W.}\ \bibnamefont {Murch}}, \ and\ \bibinfo {author}
  {\bibfnamefont {K.~B.}\ \bibnamefont {Whaley}},\ }\href {\doibase
  10.1103/PhysRevLett.131.100202} {\bibfield  {journal} {\bibinfo  {journal}
  {Phys. Rev. Lett.}\ }\textbf {\bibinfo {volume} {131}},\ \bibinfo {pages}
  {100202} (\bibinfo {year} {2023})}\BibitemShut {NoStop}%
\bibitem [{\citenamefont {Tang}\ \emph {et~al.}(2024)\citenamefont {Tang},
  \citenamefont {Chen}, \citenamefont {Tang},\ and\ \citenamefont
  {Zhang}}]{Tang2024Topologically}%
  \BibitemOpen
  \bibfield  {author} {\bibinfo {author} {\bibfnamefont {Z.}~\bibnamefont
  {Tang}}, \bibinfo {author} {\bibfnamefont {T.}~\bibnamefont {Chen}}, \bibinfo
  {author} {\bibfnamefont {X.}~\bibnamefont {Tang}}, \ and\ \bibinfo {author}
  {\bibfnamefont {X.}~\bibnamefont {Zhang}},\ }\href {\doibase
  10.1038/s41377-024-01514-1} {\bibfield  {journal} {\bibinfo  {journal}
  {Light: Science {\&} Applications}\ }\textbf {\bibinfo {volume} {13}},\
  \bibinfo {pages} {167} (\bibinfo {year} {2024})}\BibitemShut {NoStop}%
\bibitem [{\citenamefont {Wu}\ \emph {et~al.}(2019)\citenamefont {Wu},
  \citenamefont {Liu}, \citenamefont {Geng}, \citenamefont {Song},
  \citenamefont {Ye}, \citenamefont {Duan}, \citenamefont {Rong},\ and\
  \citenamefont {Du}}]{doi:10.1126/science.aaw8205}%
  \BibitemOpen
  \bibfield  {author} {\bibinfo {author} {\bibfnamefont {Y.}~\bibnamefont
  {Wu}}, \bibinfo {author} {\bibfnamefont {W.}~\bibnamefont {Liu}}, \bibinfo
  {author} {\bibfnamefont {J.}~\bibnamefont {Geng}}, \bibinfo {author}
  {\bibfnamefont {X.}~\bibnamefont {Song}}, \bibinfo {author} {\bibfnamefont
  {X.}~\bibnamefont {Ye}}, \bibinfo {author} {\bibfnamefont {C.-K.}\
  \bibnamefont {Duan}}, \bibinfo {author} {\bibfnamefont {X.}~\bibnamefont
  {Rong}}, \ and\ \bibinfo {author} {\bibfnamefont {J.}~\bibnamefont {Du}},\
  }\href {\doibase 10.1126/science.aaw8205} {\bibfield  {journal} {\bibinfo
  {journal} {Science}\ }\textbf {\bibinfo {volume} {364}},\ \bibinfo {pages}
  {878} (\bibinfo {year} {2019})}\BibitemShut {NoStop}%
\bibitem [{\citenamefont {Dogra}\ \emph {et~al.}(2021)\citenamefont {Dogra},
  \citenamefont {Melnikov},\ and\ \citenamefont {Paraoanu}}]{Dogra2021}%
  \BibitemOpen
  \bibfield  {author} {\bibinfo {author} {\bibfnamefont {S.}~\bibnamefont
  {Dogra}}, \bibinfo {author} {\bibfnamefont {A.~A.}\ \bibnamefont {Melnikov}},
  \ and\ \bibinfo {author} {\bibfnamefont {G.~S.}\ \bibnamefont {Paraoanu}},\
  }\href {\doibase 10.1038/s42005-021-00534-2} {\bibfield  {journal} {\bibinfo
  {journal} {Communications Physics}\ }\textbf {\bibinfo {volume} {4}},\
  \bibinfo {pages} {26} (\bibinfo {year} {2021})}\BibitemShut {NoStop}%
\bibitem [{\citenamefont {Chu}\ \emph {et~al.}(2020)\citenamefont {Chu},
  \citenamefont {Liu}, \citenamefont {Liu},\ and\ \citenamefont
  {Cai}}]{PhysRevLett.124.020501}%
  \BibitemOpen
  \bibfield  {author} {\bibinfo {author} {\bibfnamefont {Y.}~\bibnamefont
  {Chu}}, \bibinfo {author} {\bibfnamefont {Y.}~\bibnamefont {Liu}}, \bibinfo
  {author} {\bibfnamefont {H.}~\bibnamefont {Liu}}, \ and\ \bibinfo {author}
  {\bibfnamefont {J.}~\bibnamefont {Cai}},\ }\href {\doibase
  10.1103/PhysRevLett.124.020501} {\bibfield  {journal} {\bibinfo  {journal}
  {Phys. Rev. Lett.}\ }\textbf {\bibinfo {volume} {124}},\ \bibinfo {pages}
  {020501} (\bibinfo {year} {2020})}\BibitemShut {NoStop}%
\bibitem [{\citenamefont {Tang}\ \emph {et~al.}(2016)\citenamefont {Tang},
  \citenamefont {Wang}, \citenamefont {Yu}, \citenamefont {He}, \citenamefont
  {Xu}, \citenamefont {Liu}, \citenamefont {Chen}, \citenamefont {Sun},
  \citenamefont {Sun}, \citenamefont {Han}, \citenamefont {Li},\ and\
  \citenamefont {Guo}}]{Tang2016}%
  \BibitemOpen
  \bibfield  {author} {\bibinfo {author} {\bibfnamefont {J.-S.}\ \bibnamefont
  {Tang}}, \bibinfo {author} {\bibfnamefont {Y.-T.}\ \bibnamefont {Wang}},
  \bibinfo {author} {\bibfnamefont {S.}~\bibnamefont {Yu}}, \bibinfo {author}
  {\bibfnamefont {D.-Y.}\ \bibnamefont {He}}, \bibinfo {author} {\bibfnamefont
  {J.-S.}\ \bibnamefont {Xu}}, \bibinfo {author} {\bibfnamefont {B.-H.}\
  \bibnamefont {Liu}}, \bibinfo {author} {\bibfnamefont {G.}~\bibnamefont
  {Chen}}, \bibinfo {author} {\bibfnamefont {Y.-N.}\ \bibnamefont {Sun}},
  \bibinfo {author} {\bibfnamefont {K.}~\bibnamefont {Sun}}, \bibinfo {author}
  {\bibfnamefont {Y.-J.}\ \bibnamefont {Han}}, \bibinfo {author} {\bibfnamefont
  {C.-F.}\ \bibnamefont {Li}}, \ and\ \bibinfo {author} {\bibfnamefont {G.-C.}\
  \bibnamefont {Guo}},\ }\href {\doibase 10.1038/nphoton.2016.144} {\bibfield
  {journal} {\bibinfo  {journal} {Nature Photonics}\ }\textbf {\bibinfo
  {volume} {10}},\ \bibinfo {pages} {642} (\bibinfo {year} {2016})}\BibitemShut
  {NoStop}%
\bibitem [{\citenamefont {Yu}\ \emph {et~al.}(2024{\natexlab{a}})\citenamefont
  {Yu}, \citenamefont {Zhao}, \citenamefont {Li}, \citenamefont {Hu},
  \citenamefont {Duan}, \citenamefont {Yuan},\ and\ \citenamefont
  {Zhang}}]{doi:10.1126/sciadv.adk7616}%
  \BibitemOpen
  \bibfield  {author} {\bibinfo {author} {\bibfnamefont {X.}~\bibnamefont
  {Yu}}, \bibinfo {author} {\bibfnamefont {X.}~\bibnamefont {Zhao}}, \bibinfo
  {author} {\bibfnamefont {L.}~\bibnamefont {Li}}, \bibinfo {author}
  {\bibfnamefont {X.-M.}\ \bibnamefont {Hu}}, \bibinfo {author} {\bibfnamefont
  {X.}~\bibnamefont {Duan}}, \bibinfo {author} {\bibfnamefont {H.}~\bibnamefont
  {Yuan}}, \ and\ \bibinfo {author} {\bibfnamefont {C.}~\bibnamefont {Zhang}},\
  }\href {\doibase 10.1126/sciadv.adk7616} {\bibfield  {journal} {\bibinfo
  {journal} {Science Advances}\ }\textbf {\bibinfo {volume} {10}},\ \bibinfo
  {pages} {eadk7616} (\bibinfo {year} {2024}{\natexlab{a}})}\BibitemShut
  {NoStop}%
\bibitem [{\citenamefont {Feng}\ \emph {et~al.}(2014)\citenamefont {Feng},
  \citenamefont {Wong}, \citenamefont {Ma}, \citenamefont {Wang},\ and\
  \citenamefont {Zhang}}]{doi:10.1126/science.1258479}%
  \BibitemOpen
  \bibfield  {author} {\bibinfo {author} {\bibfnamefont {L.}~\bibnamefont
  {Feng}}, \bibinfo {author} {\bibfnamefont {Z.~J.}\ \bibnamefont {Wong}},
  \bibinfo {author} {\bibfnamefont {R.-M.}\ \bibnamefont {Ma}}, \bibinfo
  {author} {\bibfnamefont {Y.}~\bibnamefont {Wang}}, \ and\ \bibinfo {author}
  {\bibfnamefont {X.}~\bibnamefont {Zhang}},\ }\href {\doibase
  10.1126/science.1258479} {\bibfield  {journal} {\bibinfo  {journal}
  {Science}\ }\textbf {\bibinfo {volume} {346}},\ \bibinfo {pages} {972}
  (\bibinfo {year} {2014})}\BibitemShut {NoStop}%
\bibitem [{\citenamefont {Gao}\ \emph {et~al.}(2015)\citenamefont {Gao},
  \citenamefont {Estrecho}, \citenamefont {Bliokh}, \citenamefont {Liew},
  \citenamefont {Fraser}, \citenamefont {Brodbeck}, \citenamefont {Kamp},
  \citenamefont {Schneider}, \citenamefont {H{\"o}fling}, \citenamefont
  {Yamamoto}, \citenamefont {Nori}, \citenamefont {Kivshar}, \citenamefont
  {Truscott}, \citenamefont {Dall},\ and\ \citenamefont
  {Ostrovskaya}}]{Gao2015}%
  \BibitemOpen
  \bibfield  {author} {\bibinfo {author} {\bibfnamefont {T.}~\bibnamefont
  {Gao}}, \bibinfo {author} {\bibfnamefont {E.}~\bibnamefont {Estrecho}},
  \bibinfo {author} {\bibfnamefont {K.~Y.}\ \bibnamefont {Bliokh}}, \bibinfo
  {author} {\bibfnamefont {T.~C.~H.}\ \bibnamefont {Liew}}, \bibinfo {author}
  {\bibfnamefont {M.~D.}\ \bibnamefont {Fraser}}, \bibinfo {author}
  {\bibfnamefont {S.}~\bibnamefont {Brodbeck}}, \bibinfo {author}
  {\bibfnamefont {M.}~\bibnamefont {Kamp}}, \bibinfo {author} {\bibfnamefont
  {C.}~\bibnamefont {Schneider}}, \bibinfo {author} {\bibfnamefont
  {S.}~\bibnamefont {H{\"o}fling}}, \bibinfo {author} {\bibfnamefont
  {Y.}~\bibnamefont {Yamamoto}}, \bibinfo {author} {\bibfnamefont
  {F.}~\bibnamefont {Nori}}, \bibinfo {author} {\bibfnamefont {Y.~S.}\
  \bibnamefont {Kivshar}}, \bibinfo {author} {\bibfnamefont {A.~G.}\
  \bibnamefont {Truscott}}, \bibinfo {author} {\bibfnamefont {R.~G.}\
  \bibnamefont {Dall}}, \ and\ \bibinfo {author} {\bibfnamefont {E.~A.}\
  \bibnamefont {Ostrovskaya}},\ }\href {\doibase 10.1038/nature15522}
  {\bibfield  {journal} {\bibinfo  {journal} {Nature}\ }\textbf {\bibinfo
  {volume} {526}},\ \bibinfo {pages} {554} (\bibinfo {year}
  {2015})}\BibitemShut {NoStop}%
\bibitem [{\citenamefont {Peng}\ \emph {et~al.}(2014)\citenamefont {Peng},
  \citenamefont {{\"O}zdemir}, \citenamefont {Lei}, \citenamefont {Monifi},
  \citenamefont {Gianfreda}, \citenamefont {Long}, \citenamefont {Fan},
  \citenamefont {Nori}, \citenamefont {Bender},\ and\ \citenamefont
  {Yang}}]{Peng2014}%
  \BibitemOpen
  \bibfield  {author} {\bibinfo {author} {\bibfnamefont {B.}~\bibnamefont
  {Peng}}, \bibinfo {author} {\bibfnamefont {{\c{S}}.~K.}\ \bibnamefont
  {{\"O}zdemir}}, \bibinfo {author} {\bibfnamefont {F.}~\bibnamefont {Lei}},
  \bibinfo {author} {\bibfnamefont {F.}~\bibnamefont {Monifi}}, \bibinfo
  {author} {\bibfnamefont {M.}~\bibnamefont {Gianfreda}}, \bibinfo {author}
  {\bibfnamefont {G.~L.}\ \bibnamefont {Long}}, \bibinfo {author}
  {\bibfnamefont {S.}~\bibnamefont {Fan}}, \bibinfo {author} {\bibfnamefont
  {F.}~\bibnamefont {Nori}}, \bibinfo {author} {\bibfnamefont {C.~M.}\
  \bibnamefont {Bender}}, \ and\ \bibinfo {author} {\bibfnamefont
  {L.}~\bibnamefont {Yang}},\ }\href {\doibase 10.1038/nphys2927} {\bibfield
  {journal} {\bibinfo  {journal} {Nature Physics}\ }\textbf {\bibinfo {volume}
  {10}},\ \bibinfo {pages} {394} (\bibinfo {year} {2014})}\BibitemShut
  {NoStop}%
\bibitem [{\citenamefont {Jahromi}\ \emph {et~al.}(2017)\citenamefont
  {Jahromi}, \citenamefont {Hassan}, \citenamefont {Christodoulides},\ and\
  \citenamefont {Abouraddy}}]{Jahromi2017}%
  \BibitemOpen
  \bibfield  {author} {\bibinfo {author} {\bibfnamefont {A.~K.}\ \bibnamefont
  {Jahromi}}, \bibinfo {author} {\bibfnamefont {A.~U.}\ \bibnamefont {Hassan}},
  \bibinfo {author} {\bibfnamefont {D.~N.}\ \bibnamefont {Christodoulides}}, \
  and\ \bibinfo {author} {\bibfnamefont {A.~F.}\ \bibnamefont {Abouraddy}},\
  }\href {\doibase 10.1038/s41467-017-00958-x} {\bibfield  {journal} {\bibinfo
  {journal} {Nature Communications}\ }\textbf {\bibinfo {volume} {8}},\
  \bibinfo {pages} {1359} (\bibinfo {year} {2017})}\BibitemShut {NoStop}%
\bibitem [{\citenamefont {Chang}\ \emph {et~al.}(2014)\citenamefont {Chang},
  \citenamefont {Jiang}, \citenamefont {Hua}, \citenamefont {Yang},
  \citenamefont {Wen}, \citenamefont {Jiang}, \citenamefont {Li}, \citenamefont
  {Wang},\ and\ \citenamefont {Xiao}}]{Chang2014}%
  \BibitemOpen
  \bibfield  {author} {\bibinfo {author} {\bibfnamefont {L.}~\bibnamefont
  {Chang}}, \bibinfo {author} {\bibfnamefont {X.}~\bibnamefont {Jiang}},
  \bibinfo {author} {\bibfnamefont {S.}~\bibnamefont {Hua}}, \bibinfo {author}
  {\bibfnamefont {C.}~\bibnamefont {Yang}}, \bibinfo {author} {\bibfnamefont
  {J.}~\bibnamefont {Wen}}, \bibinfo {author} {\bibfnamefont {L.}~\bibnamefont
  {Jiang}}, \bibinfo {author} {\bibfnamefont {G.}~\bibnamefont {Li}}, \bibinfo
  {author} {\bibfnamefont {G.}~\bibnamefont {Wang}}, \ and\ \bibinfo {author}
  {\bibfnamefont {M.}~\bibnamefont {Xiao}},\ }\href {\doibase
  10.1038/nphoton.2014.133} {\bibfield  {journal} {\bibinfo  {journal} {Nature
  Photonics}\ }\textbf {\bibinfo {volume} {8}},\ \bibinfo {pages} {524}
  (\bibinfo {year} {2014})}\BibitemShut {NoStop}%
\bibitem [{\citenamefont {Cao}\ \emph {et~al.}(2022)\citenamefont {Cao},
  \citenamefont {Wang}, \citenamefont {Chen}, \citenamefont {Hu}, \citenamefont
  {Wang}, \citenamefont {Yang},\ and\ \citenamefont {Zhang}}]{Cao2022}%
  \BibitemOpen
  \bibfield  {author} {\bibinfo {author} {\bibfnamefont {W.}~\bibnamefont
  {Cao}}, \bibinfo {author} {\bibfnamefont {C.}~\bibnamefont {Wang}}, \bibinfo
  {author} {\bibfnamefont {W.}~\bibnamefont {Chen}}, \bibinfo {author}
  {\bibfnamefont {S.}~\bibnamefont {Hu}}, \bibinfo {author} {\bibfnamefont
  {H.}~\bibnamefont {Wang}}, \bibinfo {author} {\bibfnamefont {L.}~\bibnamefont
  {Yang}}, \ and\ \bibinfo {author} {\bibfnamefont {X.}~\bibnamefont {Zhang}},\
  }\href {\doibase 10.1038/s41565-021-01038-4} {\bibfield  {journal} {\bibinfo
  {journal} {Nature Nanotechnology}\ }\textbf {\bibinfo {volume} {17}},\
  \bibinfo {pages} {262} (\bibinfo {year} {2022})}\BibitemShut {NoStop}%
\bibitem [{\citenamefont {Zhang}\ \emph {et~al.}(2020)\citenamefont {Zhang},
  \citenamefont {Li}, \citenamefont {Wang}, \citenamefont {Feng}, \citenamefont
  {Guan},\ and\ \citenamefont {Yao}}]{Zhang2020}%
  \BibitemOpen
  \bibfield  {author} {\bibinfo {author} {\bibfnamefont {J.}~\bibnamefont
  {Zhang}}, \bibinfo {author} {\bibfnamefont {L.}~\bibnamefont {Li}}, \bibinfo
  {author} {\bibfnamefont {G.}~\bibnamefont {Wang}}, \bibinfo {author}
  {\bibfnamefont {X.}~\bibnamefont {Feng}}, \bibinfo {author} {\bibfnamefont
  {B.-O.}\ \bibnamefont {Guan}}, \ and\ \bibinfo {author} {\bibfnamefont
  {J.}~\bibnamefont {Yao}},\ }\href {\doibase 10.1038/s41467-020-16705-8}
  {\bibfield  {journal} {\bibinfo  {journal} {Nature Communications}\ }\textbf
  {\bibinfo {volume} {11}},\ \bibinfo {pages} {3217} (\bibinfo {year}
  {2020})}\BibitemShut {NoStop}%
\bibitem [{\citenamefont {El-Ganainy}\ \emph {et~al.}(2007)\citenamefont
  {El-Ganainy}, \citenamefont {Makris}, \citenamefont {Christodoulides},\ and\
  \citenamefont {Musslimani}}]{El-Ganainy:07}%
  \BibitemOpen
  \bibfield  {author} {\bibinfo {author} {\bibfnamefont {R.}~\bibnamefont
  {El-Ganainy}}, \bibinfo {author} {\bibfnamefont {K.~G.}\ \bibnamefont
  {Makris}}, \bibinfo {author} {\bibfnamefont {D.~N.}\ \bibnamefont
  {Christodoulides}}, \ and\ \bibinfo {author} {\bibfnamefont {Z.~H.}\
  \bibnamefont {Musslimani}},\ }\href {\doibase 10.1364/OL.32.002632}
  {\bibfield  {journal} {\bibinfo  {journal} {Opt. Lett.}\ }\textbf {\bibinfo
  {volume} {32}},\ \bibinfo {pages} {2632} (\bibinfo {year}
  {2007})}\BibitemShut {NoStop}%
\bibitem [{\citenamefont {R{\"u}ter}\ \emph {et~al.}(2010)\citenamefont
  {R{\"u}ter}, \citenamefont {Makris}, \citenamefont {El-Ganainy},
  \citenamefont {Christodoulides}, \citenamefont {Segev},\ and\ \citenamefont
  {Kip}}]{Ruter2010}%
  \BibitemOpen
  \bibfield  {author} {\bibinfo {author} {\bibfnamefont {C.~E.}\ \bibnamefont
  {R{\"u}ter}}, \bibinfo {author} {\bibfnamefont {K.~G.}\ \bibnamefont
  {Makris}}, \bibinfo {author} {\bibfnamefont {R.}~\bibnamefont {El-Ganainy}},
  \bibinfo {author} {\bibfnamefont {D.~N.}\ \bibnamefont {Christodoulides}},
  \bibinfo {author} {\bibfnamefont {M.}~\bibnamefont {Segev}}, \ and\ \bibinfo
  {author} {\bibfnamefont {D.}~\bibnamefont {Kip}},\ }\href {\doibase
  10.1038/nphys1515} {\bibfield  {journal} {\bibinfo  {journal} {Nature
  Physics}\ }\textbf {\bibinfo {volume} {6}},\ \bibinfo {pages} {192} (\bibinfo
  {year} {2010})}\BibitemShut {NoStop}%
\bibitem [{\citenamefont {Liu}\ \emph {et~al.}(2018)\citenamefont {Liu},
  \citenamefont {Hao}, \citenamefont {Li}, \citenamefont {Capmany},
  \citenamefont {Zhu},\ and\ \citenamefont {Li}}]{Liu2018}%
  \BibitemOpen
  \bibfield  {author} {\bibinfo {author} {\bibfnamefont {Y.}~\bibnamefont
  {Liu}}, \bibinfo {author} {\bibfnamefont {T.}~\bibnamefont {Hao}}, \bibinfo
  {author} {\bibfnamefont {W.}~\bibnamefont {Li}}, \bibinfo {author}
  {\bibfnamefont {J.}~\bibnamefont {Capmany}}, \bibinfo {author} {\bibfnamefont
  {N.}~\bibnamefont {Zhu}}, \ and\ \bibinfo {author} {\bibfnamefont
  {M.}~\bibnamefont {Li}},\ }\href {\doibase 10.1038/s41377-018-0035-8}
  {\bibfield  {journal} {\bibinfo  {journal} {Light: Science {\&}
  Applications}\ }\textbf {\bibinfo {volume} {7}},\ \bibinfo {pages} {38}
  (\bibinfo {year} {2018})}\BibitemShut {NoStop}%
\bibitem [{\citenamefont {{\"O}zdemir}\ \emph {et~al.}(2019)\citenamefont
  {{\"O}zdemir}, \citenamefont {Rotter}, \citenamefont {Nori},\ and\
  \citenamefont {Yang}}]{2019}%
  \BibitemOpen
  \bibfield  {author} {\bibinfo {author} {\bibfnamefont {{\c{S}}.~K.}\
  \bibnamefont {{\"O}zdemir}}, \bibinfo {author} {\bibfnamefont
  {S.}~\bibnamefont {Rotter}}, \bibinfo {author} {\bibfnamefont
  {F.}~\bibnamefont {Nori}}, \ and\ \bibinfo {author} {\bibfnamefont
  {L.}~\bibnamefont {Yang}},\ }\href {\doibase 10.1038/s41563-019-0304-9}
  {\bibfield  {journal} {\bibinfo  {journal} {Nature Materials}\ }\textbf
  {\bibinfo {volume} {18}},\ \bibinfo {pages} {783} (\bibinfo {year}
  {2019})}\BibitemShut {NoStop}%
\bibitem [{\citenamefont {Miri}\ and\ \citenamefont
  {Alù}(2019)}]{doi:10.1126/science.aar7709}%
  \BibitemOpen
  \bibfield  {author} {\bibinfo {author} {\bibfnamefont {M.-A.}\ \bibnamefont
  {Miri}}\ and\ \bibinfo {author} {\bibfnamefont {A.}~\bibnamefont {Alù}},\
  }\href {\doibase 10.1126/science.aar7709} {\bibfield  {journal} {\bibinfo
  {journal} {Science}\ }\textbf {\bibinfo {volume} {363}},\ \bibinfo {pages}
  {eaar7709} (\bibinfo {year} {2019})}\BibitemShut {NoStop}%
\bibitem [{\citenamefont {Feng}\ \emph {et~al.}(2017)\citenamefont {Feng},
  \citenamefont {El-Ganainy},\ and\ \citenamefont {Ge}}]{Feng2017}%
  \BibitemOpen
  \bibfield  {author} {\bibinfo {author} {\bibfnamefont {L.}~\bibnamefont
  {Feng}}, \bibinfo {author} {\bibfnamefont {R.}~\bibnamefont {El-Ganainy}}, \
  and\ \bibinfo {author} {\bibfnamefont {L.}~\bibnamefont {Ge}},\ }\href
  {\doibase 10.1038/s41566-017-0031-1} {\bibfield  {journal} {\bibinfo
  {journal} {Nature Photonics}\ }\textbf {\bibinfo {volume} {11}},\ \bibinfo
  {pages} {752} (\bibinfo {year} {2017})}\BibitemShut {NoStop}%
\bibitem [{\citenamefont {Xiao}\ \emph {et~al.}(2021)\citenamefont {Xiao},
  \citenamefont {Deng}, \citenamefont {Wang}, \citenamefont {Wang},
  \citenamefont {Yi},\ and\ \citenamefont {Xue}}]{PhysRevLett.126.230402}%
  \BibitemOpen
  \bibfield  {author} {\bibinfo {author} {\bibfnamefont {L.}~\bibnamefont
  {Xiao}}, \bibinfo {author} {\bibfnamefont {T.}~\bibnamefont {Deng}}, \bibinfo
  {author} {\bibfnamefont {K.}~\bibnamefont {Wang}}, \bibinfo {author}
  {\bibfnamefont {Z.}~\bibnamefont {Wang}}, \bibinfo {author} {\bibfnamefont
  {W.}~\bibnamefont {Yi}}, \ and\ \bibinfo {author} {\bibfnamefont
  {P.}~\bibnamefont {Xue}},\ }\href {\doibase 10.1103/PhysRevLett.126.230402}
  {\bibfield  {journal} {\bibinfo  {journal} {Phys. Rev. Lett.}\ }\textbf
  {\bibinfo {volume} {126}},\ \bibinfo {pages} {230402} (\bibinfo {year}
  {2021})}\BibitemShut {NoStop}%
\bibitem [{\citenamefont {Zhao}\ and\ \citenamefont
  {Feng}(2018)}]{10.1093/nsr/nwy011}%
  \BibitemOpen
  \bibfield  {author} {\bibinfo {author} {\bibfnamefont {H.}~\bibnamefont
  {Zhao}}\ and\ \bibinfo {author} {\bibfnamefont {L.}~\bibnamefont {Feng}},\
  }\href {\doibase 10.1093/nsr/nwy011} {\bibfield  {journal} {\bibinfo
  {journal} {National Science Review}\ }\textbf {\bibinfo {volume} {5}},\
  \bibinfo {pages} {183} (\bibinfo {year} {2018})}\BibitemShut {NoStop}%
\bibitem [{\citenamefont {Li}\ \emph {et~al.}(2020)\citenamefont {Li},
  \citenamefont {Cao}, \citenamefont {Zhi}, \citenamefont {Zhang},
  \citenamefont {Zou}, \citenamefont {Feng}, \citenamefont {Guan},\ and\
  \citenamefont {Yao}}]{Li2020}%
  \BibitemOpen
  \bibfield  {author} {\bibinfo {author} {\bibfnamefont {L.}~\bibnamefont
  {Li}}, \bibinfo {author} {\bibfnamefont {Y.}~\bibnamefont {Cao}}, \bibinfo
  {author} {\bibfnamefont {Y.}~\bibnamefont {Zhi}}, \bibinfo {author}
  {\bibfnamefont {J.}~\bibnamefont {Zhang}}, \bibinfo {author} {\bibfnamefont
  {Y.}~\bibnamefont {Zou}}, \bibinfo {author} {\bibfnamefont {X.}~\bibnamefont
  {Feng}}, \bibinfo {author} {\bibfnamefont {B.-O.}\ \bibnamefont {Guan}}, \
  and\ \bibinfo {author} {\bibfnamefont {J.}~\bibnamefont {Yao}},\ }\href
  {\doibase 10.1038/s41377-020-00407-3} {\bibfield  {journal} {\bibinfo
  {journal} {Light: Science {\&} Applications}\ }\textbf {\bibinfo {volume}
  {9}},\ \bibinfo {pages} {169} (\bibinfo {year} {2020})}\BibitemShut {NoStop}%
\bibitem [{\citenamefont {Kremer}\ \emph {et~al.}(2019)\citenamefont {Kremer},
  \citenamefont {Biesenthal}, \citenamefont {Maczewsky}, \citenamefont
  {Heinrich}, \citenamefont {Thomale},\ and\ \citenamefont
  {Szameit}}]{kremer2019}%
  \BibitemOpen
  \bibfield  {author} {\bibinfo {author} {\bibfnamefont {M.}~\bibnamefont
  {Kremer}}, \bibinfo {author} {\bibfnamefont {T.}~\bibnamefont {Biesenthal}},
  \bibinfo {author} {\bibfnamefont {L.~J.}\ \bibnamefont {Maczewsky}}, \bibinfo
  {author} {\bibfnamefont {M.}~\bibnamefont {Heinrich}}, \bibinfo {author}
  {\bibfnamefont {R.}~\bibnamefont {Thomale}}, \ and\ \bibinfo {author}
  {\bibfnamefont {A.}~\bibnamefont {Szameit}},\ }\href {\doibase
  10.1038/s41467-018-08104-x} {\bibfield  {journal} {\bibinfo  {journal}
  {Nature Communications}\ }\textbf {\bibinfo {volume} {10}},\ \bibinfo {pages}
  {435} (\bibinfo {year} {2019})}\BibitemShut {NoStop}%
\bibitem [{\citenamefont {Zhu}\ \emph {et~al.}(2014)\citenamefont {Zhu},
  \citenamefont {Ramezani}, \citenamefont {Shi}, \citenamefont {Zhu},\ and\
  \citenamefont {Zhang}}]{PhysRevX.4.031042}%
  \BibitemOpen
  \bibfield  {author} {\bibinfo {author} {\bibfnamefont {X.}~\bibnamefont
  {Zhu}}, \bibinfo {author} {\bibfnamefont {H.}~\bibnamefont {Ramezani}},
  \bibinfo {author} {\bibfnamefont {C.}~\bibnamefont {Shi}}, \bibinfo {author}
  {\bibfnamefont {J.}~\bibnamefont {Zhu}}, \ and\ \bibinfo {author}
  {\bibfnamefont {X.}~\bibnamefont {Zhang}},\ }\href {\doibase
  10.1103/PhysRevX.4.031042} {\bibfield  {journal} {\bibinfo  {journal} {Phys.
  Rev. X}\ }\textbf {\bibinfo {volume} {4}},\ \bibinfo {pages} {031042}
  (\bibinfo {year} {2014})}\BibitemShut {NoStop}%
\bibitem [{\citenamefont {Shi}\ \emph {et~al.}(2016)\citenamefont {Shi},
  \citenamefont {Dubois}, \citenamefont {Chen}, \citenamefont {Cheng},
  \citenamefont {Ramezani}, \citenamefont {Wang},\ and\ \citenamefont
  {Zhang}}]{Shi2016}%
  \BibitemOpen
  \bibfield  {author} {\bibinfo {author} {\bibfnamefont {C.}~\bibnamefont
  {Shi}}, \bibinfo {author} {\bibfnamefont {M.}~\bibnamefont {Dubois}},
  \bibinfo {author} {\bibfnamefont {Y.}~\bibnamefont {Chen}}, \bibinfo {author}
  {\bibfnamefont {L.}~\bibnamefont {Cheng}}, \bibinfo {author} {\bibfnamefont
  {H.}~\bibnamefont {Ramezani}}, \bibinfo {author} {\bibfnamefont
  {Y.}~\bibnamefont {Wang}}, \ and\ \bibinfo {author} {\bibfnamefont
  {X.}~\bibnamefont {Zhang}},\ }\href {\doibase 10.1038/ncomms11110} {\bibfield
   {journal} {\bibinfo  {journal} {Nature Communications}\ }\textbf {\bibinfo
  {volume} {7}},\ \bibinfo {pages} {11110} (\bibinfo {year}
  {2016})}\BibitemShut {NoStop}%
\bibitem [{\citenamefont {Regensburger}\ \emph {et~al.}(2012)\citenamefont
  {Regensburger}, \citenamefont {Bersch}, \citenamefont {Miri}, \citenamefont
  {Onishchukov}, \citenamefont {Christodoulides},\ and\ \citenamefont
  {Peschel}}]{Regensburger2012}%
  \BibitemOpen
  \bibfield  {author} {\bibinfo {author} {\bibfnamefont {A.}~\bibnamefont
  {Regensburger}}, \bibinfo {author} {\bibfnamefont {C.}~\bibnamefont
  {Bersch}}, \bibinfo {author} {\bibfnamefont {M.-A.}\ \bibnamefont {Miri}},
  \bibinfo {author} {\bibfnamefont {G.}~\bibnamefont {Onishchukov}}, \bibinfo
  {author} {\bibfnamefont {D.~N.}\ \bibnamefont {Christodoulides}}, \ and\
  \bibinfo {author} {\bibfnamefont {U.}~\bibnamefont {Peschel}},\ }\href
  {\doibase 10.1038/nature11298} {\bibfield  {journal} {\bibinfo  {journal}
  {Nature}\ }\textbf {\bibinfo {volume} {488}},\ \bibinfo {pages} {167}
  (\bibinfo {year} {2012})}\BibitemShut {NoStop}%
\bibitem [{\citenamefont {Zhang}\ \emph {et~al.}(2016)\citenamefont {Zhang},
  \citenamefont {Zhang}, \citenamefont {Sheng}, \citenamefont {Yang},
  \citenamefont {Miri}, \citenamefont {Christodoulides}, \citenamefont {He},
  \citenamefont {Zhang},\ and\ \citenamefont {Xiao}}]{PhysRevLett.117.123601}%
  \BibitemOpen
  \bibfield  {author} {\bibinfo {author} {\bibfnamefont {Z.}~\bibnamefont
  {Zhang}}, \bibinfo {author} {\bibfnamefont {Y.}~\bibnamefont {Zhang}},
  \bibinfo {author} {\bibfnamefont {J.}~\bibnamefont {Sheng}}, \bibinfo
  {author} {\bibfnamefont {L.}~\bibnamefont {Yang}}, \bibinfo {author}
  {\bibfnamefont {M.-A.}\ \bibnamefont {Miri}}, \bibinfo {author}
  {\bibfnamefont {D.~N.}\ \bibnamefont {Christodoulides}}, \bibinfo {author}
  {\bibfnamefont {B.}~\bibnamefont {He}}, \bibinfo {author} {\bibfnamefont
  {Y.}~\bibnamefont {Zhang}}, \ and\ \bibinfo {author} {\bibfnamefont
  {M.}~\bibnamefont {Xiao}},\ }\href {\doibase 10.1103/PhysRevLett.117.123601}
  {\bibfield  {journal} {\bibinfo  {journal} {Phys. Rev. Lett.}\ }\textbf
  {\bibinfo {volume} {117}},\ \bibinfo {pages} {123601} (\bibinfo {year}
  {2016})}\BibitemShut {NoStop}%
\bibitem [{\citenamefont {Makris}\ \emph {et~al.}(2010)\citenamefont {Makris},
  \citenamefont {El-Ganainy}, \citenamefont {Christodoulides},\ and\
  \citenamefont {Musslimani}}]{PhysRevA.81.063807}%
  \BibitemOpen
  \bibfield  {author} {\bibinfo {author} {\bibfnamefont {K.~G.}\ \bibnamefont
  {Makris}}, \bibinfo {author} {\bibfnamefont {R.}~\bibnamefont {El-Ganainy}},
  \bibinfo {author} {\bibfnamefont {D.~N.}\ \bibnamefont {Christodoulides}}, \
  and\ \bibinfo {author} {\bibfnamefont {Z.~H.}\ \bibnamefont {Musslimani}},\
  }\href {\doibase 10.1103/PhysRevA.81.063807} {\bibfield  {journal} {\bibinfo
  {journal} {Phys. Rev. A}\ }\textbf {\bibinfo {volume} {81}},\ \bibinfo
  {pages} {063807} (\bibinfo {year} {2010})}\BibitemShut {NoStop}%
\bibitem [{\citenamefont {Bender}\ and\ \citenamefont
  {Boettcher}(1998)}]{PhysRevLett.80.5243}%
  \BibitemOpen
  \bibfield  {author} {\bibinfo {author} {\bibfnamefont {C.~M.}\ \bibnamefont
  {Bender}}\ and\ \bibinfo {author} {\bibfnamefont {S.}~\bibnamefont
  {Boettcher}},\ }\href {\doibase 10.1103/PhysRevLett.80.5243} {\bibfield
  {journal} {\bibinfo  {journal} {Phys. Rev. Lett.}\ }\textbf {\bibinfo
  {volume} {80}},\ \bibinfo {pages} {5243} (\bibinfo {year}
  {1998})}\BibitemShut {NoStop}%
\bibitem [{\citenamefont {Bender}\ \emph {et~al.}(1999)\citenamefont {Bender},
  \citenamefont {Boettcher},\ and\ \citenamefont
  {Meisinger}}]{10.1063/1.532860}%
  \BibitemOpen
  \bibfield  {author} {\bibinfo {author} {\bibfnamefont {C.~M.}\ \bibnamefont
  {Bender}}, \bibinfo {author} {\bibfnamefont {S.}~\bibnamefont {Boettcher}}, \
  and\ \bibinfo {author} {\bibfnamefont {P.~N.}\ \bibnamefont {Meisinger}},\
  }\href {\doibase 10.1063/1.532860} {\bibfield  {journal} {\bibinfo  {journal}
  {Journal of Mathematical Physics}\ }\textbf {\bibinfo {volume} {40}},\
  \bibinfo {pages} {2201} (\bibinfo {year} {1999})}\BibitemShut {NoStop}%
\bibitem [{\citenamefont {Bender}(2007)}]{Bender_2007}%
  \BibitemOpen
  \bibfield  {author} {\bibinfo {author} {\bibfnamefont {C.~M.}\ \bibnamefont
  {Bender}},\ }\href {\doibase 10.1088/0034-4885/70/6/R03} {\bibfield
  {journal} {\bibinfo  {journal} {Reports on Progress in Physics}\ }\textbf
  {\bibinfo {volume} {70}},\ \bibinfo {pages} {947} (\bibinfo {year}
  {2007})}\BibitemShut {NoStop}%
\bibitem [{\citenamefont {El-Ganainy}\ \emph
  {et~al.}(2018{\natexlab{a}})\citenamefont {El-Ganainy}, \citenamefont
  {Makris}, \citenamefont {Khajavikhan}, \citenamefont {Musslimani},
  \citenamefont {Rotter},\ and\ \citenamefont
  {Christodoulides}}]{ElGanainy2018}%
  \BibitemOpen
  \bibfield  {author} {\bibinfo {author} {\bibfnamefont {R.}~\bibnamefont
  {El-Ganainy}}, \bibinfo {author} {\bibfnamefont {K.~G.}\ \bibnamefont
  {Makris}}, \bibinfo {author} {\bibfnamefont {M.}~\bibnamefont {Khajavikhan}},
  \bibinfo {author} {\bibfnamefont {Z.~H.}\ \bibnamefont {Musslimani}},
  \bibinfo {author} {\bibfnamefont {S.}~\bibnamefont {Rotter}}, \ and\ \bibinfo
  {author} {\bibfnamefont {D.~N.}\ \bibnamefont {Christodoulides}},\ }\href
  {\doibase 10.1038/nphys4323} {\bibfield  {journal} {\bibinfo  {journal}
  {Nature Physics}\ }\textbf {\bibinfo {volume} {14}},\ \bibinfo {pages} {11}
  (\bibinfo {year} {2018}{\natexlab{a}})}\BibitemShut {NoStop}%
\bibitem [{\citenamefont {Mostafazadeh}(2003)}]{Mostafazadeh_2003}%
  \BibitemOpen
  \bibfield  {author} {\bibinfo {author} {\bibfnamefont {A.}~\bibnamefont
  {Mostafazadeh}},\ }\href {\doibase 10.1088/0305-4470/36/25/312} {\bibfield
  {journal} {\bibinfo  {journal} {Journal of Physics A: Mathematical and
  General}\ }\textbf {\bibinfo {volume} {36}},\ \bibinfo {pages} {7081}
  (\bibinfo {year} {2003})}\BibitemShut {NoStop}%
\bibitem [{\citenamefont {Mostafazadeh}(2002)}]{10.1063/1.1418246}%
  \BibitemOpen
  \bibfield  {author} {\bibinfo {author} {\bibfnamefont {A.}~\bibnamefont
  {Mostafazadeh}},\ }\href {\doibase 10.1063/1.1418246} {\bibfield  {journal}
  {\bibinfo  {journal} {Journal of Mathematical Physics}\ }\textbf {\bibinfo
  {volume} {43}},\ \bibinfo {pages} {205} (\bibinfo {year} {2002})}\BibitemShut
  {NoStop}%
\bibitem [{\citenamefont {Yuto~Ashida}\ and\ \citenamefont
  {Ueda}(2020)}]{1876991}%
  \BibitemOpen
  \bibfield  {author} {\bibinfo {author} {\bibfnamefont {Z.~G.}\ \bibnamefont
  {Yuto~Ashida}}\ and\ \bibinfo {author} {\bibfnamefont {M.}~\bibnamefont
  {Ueda}},\ }\href {\doibase 10.1080/00018732.2021.1876991} {\bibfield
  {journal} {\bibinfo  {journal} {Advances in Physics}\ }\textbf {\bibinfo
  {volume} {69}},\ \bibinfo {pages} {249} (\bibinfo {year} {2020})}\BibitemShut
  {NoStop}%
\bibitem [{\citenamefont {Makris}\ \emph {et~al.}(2008)\citenamefont {Makris},
  \citenamefont {El-Ganainy}, \citenamefont {Christodoulides},\ and\
  \citenamefont {Musslimani}}]{PhysRevLett.100.103904}%
  \BibitemOpen
  \bibfield  {author} {\bibinfo {author} {\bibfnamefont {K.~G.}\ \bibnamefont
  {Makris}}, \bibinfo {author} {\bibfnamefont {R.}~\bibnamefont {El-Ganainy}},
  \bibinfo {author} {\bibfnamefont {D.~N.}\ \bibnamefont {Christodoulides}}, \
  and\ \bibinfo {author} {\bibfnamefont {Z.~H.}\ \bibnamefont {Musslimani}},\
  }\href {\doibase 10.1103/PhysRevLett.100.103904} {\bibfield  {journal}
  {\bibinfo  {journal} {Phys. Rev. Lett.}\ }\textbf {\bibinfo {volume} {100}},\
  \bibinfo {pages} {103904} (\bibinfo {year} {2008})}\BibitemShut {NoStop}%
\bibitem [{\citenamefont {Ashida}\ \emph {et~al.}(2017)\citenamefont {Ashida},
  \citenamefont {Furukawa},\ and\ \citenamefont {Ueda}}]{Ashida2017}%
  \BibitemOpen
  \bibfield  {author} {\bibinfo {author} {\bibfnamefont {Y.}~\bibnamefont
  {Ashida}}, \bibinfo {author} {\bibfnamefont {S.}~\bibnamefont {Furukawa}}, \
  and\ \bibinfo {author} {\bibfnamefont {M.}~\bibnamefont {Ueda}},\ }\href
  {\doibase 10.1038/ncomms15791} {\bibfield  {journal} {\bibinfo  {journal}
  {Nature Communications}\ }\textbf {\bibinfo {volume} {8}},\ \bibinfo {pages}
  {15791} (\bibinfo {year} {2017})}\BibitemShut {NoStop}%
\bibitem [{\citenamefont {Weimann}\ \emph {et~al.}(2017)\citenamefont
  {Weimann}, \citenamefont {Kremer}, \citenamefont {Plotnik}, \citenamefont
  {Lumer}, \citenamefont {Nolte}, \citenamefont {Makris}, \citenamefont
  {Segev}, \citenamefont {Rechtsman},\ and\ \citenamefont
  {Szameit}}]{Weimann2017}%
  \BibitemOpen
  \bibfield  {author} {\bibinfo {author} {\bibfnamefont {S.}~\bibnamefont
  {Weimann}}, \bibinfo {author} {\bibfnamefont {M.}~\bibnamefont {Kremer}},
  \bibinfo {author} {\bibfnamefont {Y.}~\bibnamefont {Plotnik}}, \bibinfo
  {author} {\bibfnamefont {Y.}~\bibnamefont {Lumer}}, \bibinfo {author}
  {\bibfnamefont {S.}~\bibnamefont {Nolte}}, \bibinfo {author} {\bibfnamefont
  {K.~G.}\ \bibnamefont {Makris}}, \bibinfo {author} {\bibfnamefont
  {M.}~\bibnamefont {Segev}}, \bibinfo {author} {\bibfnamefont {M.~C.}\
  \bibnamefont {Rechtsman}}, \ and\ \bibinfo {author} {\bibfnamefont
  {A.}~\bibnamefont {Szameit}},\ }\href {\doibase 10.1038/nmat4811} {\bibfield
  {journal} {\bibinfo  {journal} {Nature Materials}\ }\textbf {\bibinfo
  {volume} {16}},\ \bibinfo {pages} {433} (\bibinfo {year} {2017})}\BibitemShut
  {NoStop}%
\bibitem [{\citenamefont {Ge}\ and\ \citenamefont
  {Stone}(2014)}]{PhysRevX.4.031011}%
  \BibitemOpen
  \bibfield  {author} {\bibinfo {author} {\bibfnamefont {L.}~\bibnamefont
  {Ge}}\ and\ \bibinfo {author} {\bibfnamefont {A.~D.}\ \bibnamefont {Stone}},\
  }\href {\doibase 10.1103/PhysRevX.4.031011} {\bibfield  {journal} {\bibinfo
  {journal} {Phys. Rev. X}\ }\textbf {\bibinfo {volume} {4}},\ \bibinfo {pages}
  {031011} (\bibinfo {year} {2014})}\BibitemShut {NoStop}%
\bibitem [{\citenamefont {Kawabata}\ \emph {et~al.}(2017)\citenamefont
  {Kawabata}, \citenamefont {Ashida},\ and\ \citenamefont
  {Ueda}}]{PhysRevLett.119.190401}%
  \BibitemOpen
  \bibfield  {author} {\bibinfo {author} {\bibfnamefont {K.}~\bibnamefont
  {Kawabata}}, \bibinfo {author} {\bibfnamefont {Y.}~\bibnamefont {Ashida}}, \
  and\ \bibinfo {author} {\bibfnamefont {M.}~\bibnamefont {Ueda}},\ }\href
  {\doibase 10.1103/PhysRevLett.119.190401} {\bibfield  {journal} {\bibinfo
  {journal} {Phys. Rev. Lett.}\ }\textbf {\bibinfo {volume} {119}},\ \bibinfo
  {pages} {190401} (\bibinfo {year} {2017})}\BibitemShut {NoStop}%
\bibitem [{\citenamefont {Ge}\ and\ \citenamefont {El-Ganainy}(2016)}]{Ge2016}%
  \BibitemOpen
  \bibfield  {author} {\bibinfo {author} {\bibfnamefont {L.}~\bibnamefont
  {Ge}}\ and\ \bibinfo {author} {\bibfnamefont {R.}~\bibnamefont
  {El-Ganainy}},\ }\href {\doibase 10.1038/srep24889} {\bibfield  {journal}
  {\bibinfo  {journal} {Scientific Reports}\ }\textbf {\bibinfo {volume} {6}},\
  \bibinfo {pages} {24889} (\bibinfo {year} {2016})}\BibitemShut {NoStop}%
\bibitem [{\citenamefont {Ge}\ \emph {et~al.}(2012)\citenamefont {Ge},
  \citenamefont {Chong},\ and\ \citenamefont {Stone}}]{PhysRevA.85.023802}%
  \BibitemOpen
  \bibfield  {author} {\bibinfo {author} {\bibfnamefont {L.}~\bibnamefont
  {Ge}}, \bibinfo {author} {\bibfnamefont {Y.~D.}\ \bibnamefont {Chong}}, \
  and\ \bibinfo {author} {\bibfnamefont {A.~D.}\ \bibnamefont {Stone}},\ }\href
  {\doibase 10.1103/PhysRevA.85.023802} {\bibfield  {journal} {\bibinfo
  {journal} {Phys. Rev. A}\ }\textbf {\bibinfo {volume} {85}},\ \bibinfo
  {pages} {023802} (\bibinfo {year} {2012})}\BibitemShut {NoStop}%
\bibitem [{\citenamefont {Fleury}\ \emph {et~al.}(2015)\citenamefont {Fleury},
  \citenamefont {Sounas},\ and\ \citenamefont {Al{\`u}}}]{Fleury2015}%
  \BibitemOpen
  \bibfield  {author} {\bibinfo {author} {\bibfnamefont {R.}~\bibnamefont
  {Fleury}}, \bibinfo {author} {\bibfnamefont {D.}~\bibnamefont {Sounas}}, \
  and\ \bibinfo {author} {\bibfnamefont {A.}~\bibnamefont {Al{\`u}}},\ }\href
  {\doibase 10.1038/ncomms6905} {\bibfield  {journal} {\bibinfo  {journal}
  {Nature Communications}\ }\textbf {\bibinfo {volume} {6}},\ \bibinfo {pages}
  {5905} (\bibinfo {year} {2015})}\BibitemShut {NoStop}%
\bibitem [{\citenamefont {Hodaei}\ \emph {et~al.}(2017)\citenamefont {Hodaei},
  \citenamefont {Hassan}, \citenamefont {Wittek}, \citenamefont
  {Garcia-Gracia}, \citenamefont {El-Ganainy}, \citenamefont
  {Christodoulides},\ and\ \citenamefont {Khajavikhan}}]{Hodaei2017}%
  \BibitemOpen
  \bibfield  {author} {\bibinfo {author} {\bibfnamefont {H.}~\bibnamefont
  {Hodaei}}, \bibinfo {author} {\bibfnamefont {A.~U.}\ \bibnamefont {Hassan}},
  \bibinfo {author} {\bibfnamefont {S.}~\bibnamefont {Wittek}}, \bibinfo
  {author} {\bibfnamefont {H.}~\bibnamefont {Garcia-Gracia}}, \bibinfo {author}
  {\bibfnamefont {R.}~\bibnamefont {El-Ganainy}}, \bibinfo {author}
  {\bibfnamefont {D.~N.}\ \bibnamefont {Christodoulides}}, \ and\ \bibinfo
  {author} {\bibfnamefont {M.}~\bibnamefont {Khajavikhan}},\ }\href {\doibase
  10.1038/nature23280} {\bibfield  {journal} {\bibinfo  {journal} {Nature}\
  }\textbf {\bibinfo {volume} {548}},\ \bibinfo {pages} {187} (\bibinfo {year}
  {2017})}\BibitemShut {NoStop}%
\bibitem [{\citenamefont {Liu}\ \emph {et~al.}(2016)\citenamefont {Liu},
  \citenamefont {Zhang}, \citenamefont {Kaya~{\"O}zdemir}, \citenamefont
  {Peng}, \citenamefont {Jing}, \citenamefont {L\"u}, \citenamefont {Li},
  \citenamefont {Yang}, \citenamefont {Nori},\ and\ \citenamefont
  {Liu}}]{110802}%
  \BibitemOpen
  \bibfield  {author} {\bibinfo {author} {\bibfnamefont {Z.-P.}\ \bibnamefont
  {Liu}}, \bibinfo {author} {\bibfnamefont {J.}~\bibnamefont {Zhang}}, \bibinfo
  {author} {\bibfnamefont {{\c{S}}.}~\bibnamefont {Kaya~{\"O}zdemir}}, \bibinfo
  {author} {\bibfnamefont {B.}~\bibnamefont {Peng}}, \bibinfo {author}
  {\bibfnamefont {H.}~\bibnamefont {Jing}}, \bibinfo {author} {\bibfnamefont
  {X.-Y.}\ \bibnamefont {L\"u}}, \bibinfo {author} {\bibfnamefont {C.-W.}\
  \bibnamefont {Li}}, \bibinfo {author} {\bibfnamefont {L.}~\bibnamefont
  {Yang}}, \bibinfo {author} {\bibfnamefont {F.}~\bibnamefont {Nori}}, \ and\
  \bibinfo {author} {\bibfnamefont {Y.-x.}\ \bibnamefont {Liu}},\ }\href
  {\doibase 10.1103/PhysRevLett.117.110802} {\bibfield  {journal} {\bibinfo
  {journal} {Phys. Rev. Lett.}\ }\textbf {\bibinfo {volume} {117}},\ \bibinfo
  {pages} {110802} (\bibinfo {year} {2016})}\BibitemShut {NoStop}%
\bibitem [{\citenamefont {Chen}\ \emph {et~al.}(2017)\citenamefont {Chen},
  \citenamefont {Kaya~{\"O}zdemir}, \citenamefont {Zhao}, \citenamefont
  {Wiersig},\ and\ \citenamefont {Yang}}]{Chen2017}%
  \BibitemOpen
  \bibfield  {author} {\bibinfo {author} {\bibfnamefont {W.}~\bibnamefont
  {Chen}}, \bibinfo {author} {\bibfnamefont {{\c{S}}.}~\bibnamefont
  {Kaya~{\"O}zdemir}}, \bibinfo {author} {\bibfnamefont {G.}~\bibnamefont
  {Zhao}}, \bibinfo {author} {\bibfnamefont {J.}~\bibnamefont {Wiersig}}, \
  and\ \bibinfo {author} {\bibfnamefont {L.}~\bibnamefont {Yang}},\ }\href
  {\doibase 10.1038/nature23281} {\bibfield  {journal} {\bibinfo  {journal}
  {Nature}\ }\textbf {\bibinfo {volume} {548}},\ \bibinfo {pages} {192}
  (\bibinfo {year} {2017})}\BibitemShut {NoStop}%
\bibitem [{\citenamefont {Yu}\ \emph {et~al.}(2020)\citenamefont {Yu},
  \citenamefont {Meng}, \citenamefont {Tang}, \citenamefont {Xu}, \citenamefont
  {Wang}, \citenamefont {Yin}, \citenamefont {Ke}, \citenamefont {Liu},
  \citenamefont {Li}, \citenamefont {Yang}, \citenamefont {Chen}, \citenamefont
  {Han}, \citenamefont {Li},\ and\ \citenamefont
  {Guo}}]{PhysRevLett.125.240506}%
  \BibitemOpen
  \bibfield  {author} {\bibinfo {author} {\bibfnamefont {S.}~\bibnamefont
  {Yu}}, \bibinfo {author} {\bibfnamefont {Y.}~\bibnamefont {Meng}}, \bibinfo
  {author} {\bibfnamefont {J.-S.}\ \bibnamefont {Tang}}, \bibinfo {author}
  {\bibfnamefont {X.-Y.}\ \bibnamefont {Xu}}, \bibinfo {author} {\bibfnamefont
  {Y.-T.}\ \bibnamefont {Wang}}, \bibinfo {author} {\bibfnamefont
  {P.}~\bibnamefont {Yin}}, \bibinfo {author} {\bibfnamefont {Z.-J.}\
  \bibnamefont {Ke}}, \bibinfo {author} {\bibfnamefont {W.}~\bibnamefont
  {Liu}}, \bibinfo {author} {\bibfnamefont {Z.-P.}\ \bibnamefont {Li}},
  \bibinfo {author} {\bibfnamefont {Y.-Z.}\ \bibnamefont {Yang}}, \bibinfo
  {author} {\bibfnamefont {G.}~\bibnamefont {Chen}}, \bibinfo {author}
  {\bibfnamefont {Y.-J.}\ \bibnamefont {Han}}, \bibinfo {author} {\bibfnamefont
  {C.-F.}\ \bibnamefont {Li}}, \ and\ \bibinfo {author} {\bibfnamefont {G.-C.}\
  \bibnamefont {Guo}},\ }\href {\doibase 10.1103/PhysRevLett.125.240506}
  {\bibfield  {journal} {\bibinfo  {journal} {Phys. Rev. Lett.}\ }\textbf
  {\bibinfo {volume} {125}},\ \bibinfo {pages} {240506} (\bibinfo {year}
  {2020})}\BibitemShut {NoStop}%
\bibitem [{\citenamefont {Jing}\ \emph {et~al.}(2015)\citenamefont {Jing},
  \citenamefont {{\"O}zdemir}, \citenamefont {Geng}, \citenamefont {Zhang},
  \citenamefont {L{\"u}}, \citenamefont {Peng}, \citenamefont {Yang},\ and\
  \citenamefont {Nori}}]{Jing2015}%
  \BibitemOpen
  \bibfield  {author} {\bibinfo {author} {\bibfnamefont {H.}~\bibnamefont
  {Jing}}, \bibinfo {author} {\bibfnamefont {{\c{S}}.~K.}\ \bibnamefont
  {{\"O}zdemir}}, \bibinfo {author} {\bibfnamefont {Z.}~\bibnamefont {Geng}},
  \bibinfo {author} {\bibfnamefont {J.}~\bibnamefont {Zhang}}, \bibinfo
  {author} {\bibfnamefont {X.-Y.}\ \bibnamefont {L{\"u}}}, \bibinfo {author}
  {\bibfnamefont {B.}~\bibnamefont {Peng}}, \bibinfo {author} {\bibfnamefont
  {L.}~\bibnamefont {Yang}}, \ and\ \bibinfo {author} {\bibfnamefont
  {F.}~\bibnamefont {Nori}},\ }\href {\doibase 10.1038/srep09663} {\bibfield
  {journal} {\bibinfo  {journal} {Scientific Reports}\ }\textbf {\bibinfo
  {volume} {5}},\ \bibinfo {pages} {9663} (\bibinfo {year} {2015})}\BibitemShut
  {NoStop}%
\bibitem [{\citenamefont {Hodaei}\ \emph {et~al.}(2014)\citenamefont {Hodaei},
  \citenamefont {Miri}, \citenamefont {Heinrich}, \citenamefont
  {Christodoulides},\ and\ \citenamefont
  {Khajavikhan}}]{doi:10.1126/science.1258480}%
  \BibitemOpen
  \bibfield  {author} {\bibinfo {author} {\bibfnamefont {H.}~\bibnamefont
  {Hodaei}}, \bibinfo {author} {\bibfnamefont {M.-A.}\ \bibnamefont {Miri}},
  \bibinfo {author} {\bibfnamefont {M.}~\bibnamefont {Heinrich}}, \bibinfo
  {author} {\bibfnamefont {D.~N.}\ \bibnamefont {Christodoulides}}, \ and\
  \bibinfo {author} {\bibfnamefont {M.}~\bibnamefont {Khajavikhan}},\ }\href
  {\doibase 10.1126/science.1258480} {\bibfield  {journal} {\bibinfo  {journal}
  {Science}\ }\textbf {\bibinfo {volume} {346}},\ \bibinfo {pages} {975}
  (\bibinfo {year} {2014})}\BibitemShut {NoStop}%
\bibitem [{\citenamefont {Liu}\ \emph {et~al.}(2017)\citenamefont {Liu},
  \citenamefont {Li}, \citenamefont {Guzzon}, \citenamefont {Norberg},
  \citenamefont {Parker}, \citenamefont {Lu}, \citenamefont {Coldren},\ and\
  \citenamefont {Yao}}]{lllLiu2017}%
  \BibitemOpen
  \bibfield  {author} {\bibinfo {author} {\bibfnamefont {W.}~\bibnamefont
  {Liu}}, \bibinfo {author} {\bibfnamefont {M.}~\bibnamefont {Li}}, \bibinfo
  {author} {\bibfnamefont {R.~S.}\ \bibnamefont {Guzzon}}, \bibinfo {author}
  {\bibfnamefont {E.~J.}\ \bibnamefont {Norberg}}, \bibinfo {author}
  {\bibfnamefont {J.~S.}\ \bibnamefont {Parker}}, \bibinfo {author}
  {\bibfnamefont {M.}~\bibnamefont {Lu}}, \bibinfo {author} {\bibfnamefont
  {L.~A.}\ \bibnamefont {Coldren}}, \ and\ \bibinfo {author} {\bibfnamefont
  {J.}~\bibnamefont {Yao}},\ }\href {\doibase 10.1038/ncomms15389} {\bibfield
  {journal} {\bibinfo  {journal} {Nature Communications}\ }\textbf {\bibinfo
  {volume} {8}},\ \bibinfo {pages} {15389} (\bibinfo {year}
  {2017})}\BibitemShut {NoStop}%
\bibitem [{\citenamefont {Jing}\ \emph {et~al.}(2014)\citenamefont {Jing},
  \citenamefont {\"Ozdemir}, \citenamefont {L\"u}, \citenamefont {Zhang},
  \citenamefont {Yang},\ and\ \citenamefont {Nori}}]{PhysRevLett.113.053604}%
  \BibitemOpen
  \bibfield  {author} {\bibinfo {author} {\bibfnamefont {H.}~\bibnamefont
  {Jing}}, \bibinfo {author} {\bibfnamefont {S.~K.}\ \bibnamefont {\"Ozdemir}},
  \bibinfo {author} {\bibfnamefont {X.-Y.}\ \bibnamefont {L\"u}}, \bibinfo
  {author} {\bibfnamefont {J.}~\bibnamefont {Zhang}}, \bibinfo {author}
  {\bibfnamefont {L.}~\bibnamefont {Yang}}, \ and\ \bibinfo {author}
  {\bibfnamefont {F.}~\bibnamefont {Nori}},\ }\href {\doibase
  10.1103/PhysRevLett.113.053604} {\bibfield  {journal} {\bibinfo  {journal}
  {Phys. Rev. Lett.}\ }\textbf {\bibinfo {volume} {113}},\ \bibinfo {pages}
  {053604} (\bibinfo {year} {2014})}\BibitemShut {NoStop}%
\bibitem [{\citenamefont {Zhang}\ \emph {et~al.}(2018)\citenamefont {Zhang},
  \citenamefont {Peng}, \citenamefont {{\"O}zdemir}, \citenamefont {Pichler},
  \citenamefont {Krimer}, \citenamefont {Zhao}, \citenamefont {Nori},
  \citenamefont {Liu}, \citenamefont {Rotter},\ and\ \citenamefont
  {Yang}}]{Zhang2018}%
  \BibitemOpen
  \bibfield  {author} {\bibinfo {author} {\bibfnamefont {J.}~\bibnamefont
  {Zhang}}, \bibinfo {author} {\bibfnamefont {B.}~\bibnamefont {Peng}},
  \bibinfo {author} {\bibfnamefont {{\c{S}}.~K.}\ \bibnamefont {{\"O}zdemir}},
  \bibinfo {author} {\bibfnamefont {K.}~\bibnamefont {Pichler}}, \bibinfo
  {author} {\bibfnamefont {D.~O.}\ \bibnamefont {Krimer}}, \bibinfo {author}
  {\bibfnamefont {G.}~\bibnamefont {Zhao}}, \bibinfo {author} {\bibfnamefont
  {F.}~\bibnamefont {Nori}}, \bibinfo {author} {\bibfnamefont {Y.-x.}\
  \bibnamefont {Liu}}, \bibinfo {author} {\bibfnamefont {S.}~\bibnamefont
  {Rotter}}, \ and\ \bibinfo {author} {\bibfnamefont {L.}~\bibnamefont
  {Yang}},\ }\href {\doibase 10.1038/s41566-018-0213-5} {\bibfield  {journal}
  {\bibinfo  {journal} {Nature Photonics}\ }\textbf {\bibinfo {volume} {12}},\
  \bibinfo {pages} {479} (\bibinfo {year} {2018})}\BibitemShut {NoStop}%
\bibitem [{\citenamefont {Feng}\ \emph {et~al.}(2011)\citenamefont {Feng},
  \citenamefont {Ayache}, \citenamefont {Huang}, \citenamefont {Xu},
  \citenamefont {Lu}, \citenamefont {Chen}, \citenamefont {Fainman},\ and\
  \citenamefont {Scherer}}]{11111206038}%
  \BibitemOpen
  \bibfield  {author} {\bibinfo {author} {\bibfnamefont {L.}~\bibnamefont
  {Feng}}, \bibinfo {author} {\bibfnamefont {M.}~\bibnamefont {Ayache}},
  \bibinfo {author} {\bibfnamefont {J.}~\bibnamefont {Huang}}, \bibinfo
  {author} {\bibfnamefont {Y.-L.}\ \bibnamefont {Xu}}, \bibinfo {author}
  {\bibfnamefont {M.-H.}\ \bibnamefont {Lu}}, \bibinfo {author} {\bibfnamefont
  {Y.-F.}\ \bibnamefont {Chen}}, \bibinfo {author} {\bibfnamefont
  {Y.}~\bibnamefont {Fainman}}, \ and\ \bibinfo {author} {\bibfnamefont
  {A.}~\bibnamefont {Scherer}},\ }\href {\doibase 10.1126/science.1206038}
  {\bibfield  {journal} {\bibinfo  {journal} {Science}\ }\textbf {\bibinfo
  {volume} {333}},\ \bibinfo {pages} {729} (\bibinfo {year}
  {2011})}\BibitemShut {NoStop}%
\bibitem [{\citenamefont {Yu}\ \emph {et~al.}(2024{\natexlab{b}})\citenamefont
  {Yu}, \citenamefont {Xue}, \citenamefont {Guo}, \citenamefont {Chan},
  \citenamefont {Terh}, \citenamefont {Soci}, \citenamefont {Zhang},\ and\
  \citenamefont {Chong}}]{Yu2024}%
  \BibitemOpen
  \bibfield  {author} {\bibinfo {author} {\bibfnamefont {L.}~\bibnamefont
  {Yu}}, \bibinfo {author} {\bibfnamefont {H.}~\bibnamefont {Xue}}, \bibinfo
  {author} {\bibfnamefont {R.}~\bibnamefont {Guo}}, \bibinfo {author}
  {\bibfnamefont {E.~A.}\ \bibnamefont {Chan}}, \bibinfo {author}
  {\bibfnamefont {Y.~Y.}\ \bibnamefont {Terh}}, \bibinfo {author}
  {\bibfnamefont {C.}~\bibnamefont {Soci}}, \bibinfo {author} {\bibfnamefont
  {B.}~\bibnamefont {Zhang}}, \ and\ \bibinfo {author} {\bibfnamefont {Y.~D.}\
  \bibnamefont {Chong}},\ }\href@noop {} {\bibfield  {journal} {\bibinfo
  {journal} {Nature}\ }\textbf {\bibinfo {volume} {632}},\ \bibinfo {pages}
  {63} (\bibinfo {year} {2024}{\natexlab{b}})}\BibitemShut {NoStop}%
\bibitem [{\citenamefont {Fleury}\ \emph {et~al.}(2014)\citenamefont {Fleury},
  \citenamefont {Sounas},\ and\ \citenamefont
  {Al\`u}}]{PhysRevLett.113.023903}%
  \BibitemOpen
  \bibfield  {author} {\bibinfo {author} {\bibfnamefont {R.}~\bibnamefont
  {Fleury}}, \bibinfo {author} {\bibfnamefont {D.~L.}\ \bibnamefont {Sounas}},
  \ and\ \bibinfo {author} {\bibfnamefont {A.}~\bibnamefont {Al\`u}},\ }\href
  {\doibase 10.1103/PhysRevLett.113.023903} {\bibfield  {journal} {\bibinfo
  {journal} {Phys. Rev. Lett.}\ }\textbf {\bibinfo {volume} {113}},\ \bibinfo
  {pages} {023903} (\bibinfo {year} {2014})}\BibitemShut {NoStop}%
\bibitem [{\citenamefont {Wimmer}\ \emph {et~al.}(2015)\citenamefont {Wimmer},
  \citenamefont {Regensburger}, \citenamefont {Miri}, \citenamefont {Bersch},
  \citenamefont {Christodoulides},\ and\ \citenamefont {Peschel}}]{Wimmer2015}%
  \BibitemOpen
  \bibfield  {author} {\bibinfo {author} {\bibfnamefont {M.}~\bibnamefont
  {Wimmer}}, \bibinfo {author} {\bibfnamefont {A.}~\bibnamefont
  {Regensburger}}, \bibinfo {author} {\bibfnamefont {M.-A.}\ \bibnamefont
  {Miri}}, \bibinfo {author} {\bibfnamefont {C.}~\bibnamefont {Bersch}},
  \bibinfo {author} {\bibfnamefont {D.~N.}\ \bibnamefont {Christodoulides}}, \
  and\ \bibinfo {author} {\bibfnamefont {U.}~\bibnamefont {Peschel}},\ }\href
  {\doibase 10.1038/ncomms8782} {\bibfield  {journal} {\bibinfo  {journal}
  {Nature Communications}\ }\textbf {\bibinfo {volume} {6}},\ \bibinfo {pages}
  {7782} (\bibinfo {year} {2015})}\BibitemShut {NoStop}%
\bibitem [{\citenamefont {Zhang}\ and\ \citenamefont
  {Yao}(2018)}]{doi:10.1126/sciadv.aar6782}%
  \BibitemOpen
  \bibfield  {author} {\bibinfo {author} {\bibfnamefont {J.}~\bibnamefont
  {Zhang}}\ and\ \bibinfo {author} {\bibfnamefont {J.}~\bibnamefont {Yao}},\
  }\href {\doibase 10.1126/sciadv.aar6782} {\bibfield  {journal} {\bibinfo
  {journal} {Science Advances}\ }\textbf {\bibinfo {volume} {4}},\ \bibinfo
  {pages} {eaar6782} (\bibinfo {year} {2018})}\BibitemShut {NoStop}%
\bibitem [{\citenamefont {Assawaworrarit}\ \emph {et~al.}(2017)\citenamefont
  {Assawaworrarit}, \citenamefont {Yu},\ and\ \citenamefont
  {Fan}}]{Assawaworrarit2017}%
  \BibitemOpen
  \bibfield  {author} {\bibinfo {author} {\bibfnamefont {S.}~\bibnamefont
  {Assawaworrarit}}, \bibinfo {author} {\bibfnamefont {X.}~\bibnamefont {Yu}},
  \ and\ \bibinfo {author} {\bibfnamefont {S.}~\bibnamefont {Fan}},\ }\href
  {\doibase 10.1038/nature22404} {\bibfield  {journal} {\bibinfo  {journal}
  {Nature}\ }\textbf {\bibinfo {volume} {546}},\ \bibinfo {pages} {387}
  (\bibinfo {year} {2017})}\BibitemShut {NoStop}%
\bibitem [{\citenamefont {Liertzer}\ \emph {et~al.}(2012)\citenamefont
  {Liertzer}, \citenamefont {Ge}, \citenamefont {Cerjan}, \citenamefont
  {Stone}, \citenamefont {T\"ureci},\ and\ \citenamefont
  {Rotter}}]{PhysRevLett.108.173901}%
  \BibitemOpen
  \bibfield  {author} {\bibinfo {author} {\bibfnamefont {M.}~\bibnamefont
  {Liertzer}}, \bibinfo {author} {\bibfnamefont {L.}~\bibnamefont {Ge}},
  \bibinfo {author} {\bibfnamefont {A.}~\bibnamefont {Cerjan}}, \bibinfo
  {author} {\bibfnamefont {A.~D.}\ \bibnamefont {Stone}}, \bibinfo {author}
  {\bibfnamefont {H.~E.}\ \bibnamefont {T\"ureci}}, \ and\ \bibinfo {author}
  {\bibfnamefont {S.}~\bibnamefont {Rotter}},\ }\href {\doibase
  10.1103/PhysRevLett.108.173901} {\bibfield  {journal} {\bibinfo  {journal}
  {Phys. Rev. Lett.}\ }\textbf {\bibinfo {volume} {108}},\ \bibinfo {pages}
  {173901} (\bibinfo {year} {2012})}\BibitemShut {NoStop}%
\bibitem [{\citenamefont {Konotop}\ \emph {et~al.}(2016)\citenamefont
  {Konotop}, \citenamefont {Yang},\ and\ \citenamefont
  {Zezyulin}}]{RevModPhys.88.035002}%
  \BibitemOpen
  \bibfield  {author} {\bibinfo {author} {\bibfnamefont {V.~V.}\ \bibnamefont
  {Konotop}}, \bibinfo {author} {\bibfnamefont {J.}~\bibnamefont {Yang}}, \
  and\ \bibinfo {author} {\bibfnamefont {D.~A.}\ \bibnamefont {Zezyulin}},\
  }\href {\doibase 10.1103/RevModPhys.88.035002} {\bibfield  {journal}
  {\bibinfo  {journal} {Rev. Mod. Phys.}\ }\textbf {\bibinfo {volume} {88}},\
  \bibinfo {pages} {035002} (\bibinfo {year} {2016})}\BibitemShut {NoStop}%
\bibitem [{\citenamefont {Yu}\ and\ \citenamefont {Vollmer}(2021)}]{Yu2021}%
  \BibitemOpen
  \bibfield  {author} {\bibinfo {author} {\bibfnamefont {D.}~\bibnamefont
  {Yu}}\ and\ \bibinfo {author} {\bibfnamefont {F.}~\bibnamefont {Vollmer}},\
  }\href {\doibase 10.1038/s42005-021-00575-7} {\bibfield  {journal} {\bibinfo
  {journal} {Communications Physics}\ }\textbf {\bibinfo {volume} {4}},\
  \bibinfo {pages} {77} (\bibinfo {year} {2021})}\BibitemShut {NoStop}%
\bibitem [{\citenamefont {Walter}\ \emph {et~al.}(2013)\citenamefont {Walter},
  \citenamefont {Doran}, \citenamefont {Gross},\ and\ \citenamefont
  {Christandl}}]{1232957}%
  \BibitemOpen
  \bibfield  {author} {\bibinfo {author} {\bibfnamefont {M.}~\bibnamefont
  {Walter}}, \bibinfo {author} {\bibfnamefont {B.}~\bibnamefont {Doran}},
  \bibinfo {author} {\bibfnamefont {D.}~\bibnamefont {Gross}}, \ and\ \bibinfo
  {author} {\bibfnamefont {M.}~\bibnamefont {Christandl}},\ }\href {\doibase
  10.1126/science.1232957} {\bibfield  {journal} {\bibinfo  {journal}
  {Science}\ }\textbf {\bibinfo {volume} {340}},\ \bibinfo {pages} {1205}
  (\bibinfo {year} {2013})}\BibitemShut {NoStop}%
\bibitem [{\citenamefont {Wootters}(2001)}]{Wootters2001EntanglementOF}%
  \BibitemOpen
  \bibfield  {author} {\bibinfo {author} {\bibfnamefont {W.~K.}\ \bibnamefont
  {Wootters}},\ }\href {https://api.semanticscholar.org/CorpusID:4688193}
  {\bibfield  {journal} {\bibinfo  {journal} {Quantum Inf. Comput.}\ }\textbf
  {\bibinfo {volume} {1}},\ \bibinfo {pages} {27} (\bibinfo {year}
  {2001})}\BibitemShut {NoStop}%
\bibitem [{\citenamefont {El-Ganainy}\ \emph
  {et~al.}(2018{\natexlab{b}})\citenamefont {El-Ganainy}, \citenamefont
  {Makris}, \citenamefont {Khajavikhan}, \citenamefont {Musslimani},
  \citenamefont {Rotter},\ and\ \citenamefont
  {Christodoulides}}]{El-Ganainy2018}%
  \BibitemOpen
  \bibfield  {author} {\bibinfo {author} {\bibfnamefont {R.}~\bibnamefont
  {El-Ganainy}}, \bibinfo {author} {\bibfnamefont {K.~G.}\ \bibnamefont
  {Makris}}, \bibinfo {author} {\bibfnamefont {M.}~\bibnamefont {Khajavikhan}},
  \bibinfo {author} {\bibfnamefont {Z.~H.}\ \bibnamefont {Musslimani}},
  \bibinfo {author} {\bibfnamefont {S.}~\bibnamefont {Rotter}}, \ and\ \bibinfo
  {author} {\bibfnamefont {D.~N.}\ \bibnamefont {Christodoulides}},\ }\href
  {\doibase 10.1038/nphys4323} {\bibfield  {journal} {\bibinfo  {journal}
  {Nature Physics}\ }\textbf {\bibinfo {volume} {14}},\ \bibinfo {pages} {11}
  (\bibinfo {year} {2018}{\natexlab{b}})}\BibitemShut {NoStop}%
\bibitem [{Sup()}]{Supplement}%
  \BibitemOpen
  \href@noop {} {\bibinfo  {journal} {See Supplemental Material for detailed
  discussion}\ }\BibitemShut {NoStop}%
\bibitem [{\citenamefont {Coffman}\ \emph {et~al.}(2000)\citenamefont
  {Coffman}, \citenamefont {Kundu},\ and\ \citenamefont
  {Wootters}}]{PhysRevA.61.052306}%
  \BibitemOpen
\bibfield  {journal} {  }\bibfield  {author} {\bibinfo {author} {\bibfnamefont
  {V.}~\bibnamefont {Coffman}}, \bibinfo {author} {\bibfnamefont
  {J.}~\bibnamefont {Kundu}}, \ and\ \bibinfo {author} {\bibfnamefont {W.~K.}\
  \bibnamefont {Wootters}},\ }\href {\doibase 10.1103/PhysRevA.61.052306}
  {\bibfield  {journal} {\bibinfo  {journal} {Phys. Rev. A}\ }\textbf {\bibinfo
  {volume} {61}},\ \bibinfo {pages} {052306} (\bibinfo {year}
  {2000})}\BibitemShut {NoStop}%
\bibitem [{\citenamefont {Lohmayer}\ \emph {et~al.}(2006)\citenamefont
  {Lohmayer}, \citenamefont {Osterloh}, \citenamefont {Siewert},\ and\
  \citenamefont {Uhlmann}}]{PhysRevLett.97.260502}%
  \BibitemOpen
  \bibfield  {author} {\bibinfo {author} {\bibfnamefont {R.}~\bibnamefont
  {Lohmayer}}, \bibinfo {author} {\bibfnamefont {A.}~\bibnamefont {Osterloh}},
  \bibinfo {author} {\bibfnamefont {J.}~\bibnamefont {Siewert}}, \ and\
  \bibinfo {author} {\bibfnamefont {A.}~\bibnamefont {Uhlmann}},\ }\href
  {\doibase 10.1103/PhysRevLett.97.260502} {\bibfield  {journal} {\bibinfo
  {journal} {Phys. Rev. Lett.}\ }\textbf {\bibinfo {volume} {97}},\ \bibinfo
  {pages} {260502} (\bibinfo {year} {2006})}\BibitemShut {NoStop}%
\end{thebibliography}%
\begin{widetext}
\clearpage
\title{Supplementary material }

	\section{I. Illustration of the two-qubit case}	\subsection{A. Eigenvalues and eigenstates of the truly $\mathcal{PT}$ symmetric system}
\par For the non-Hermitian Hamiltonian, it is essential to construct biorthogonal basis vectors, we first illustrate the construction of biorthogonal basis vectors for the single qubit in Hibert space and its dual space. The Hamiltonian of the single-qubit in active $\mathcal{PT}$ symmetric system is
		\begin{eqnarray}\label{eq2}
			H_n=-i\gamma_n/2|e\rangle\langle e|+i\kappa_n/2|g\rangle\langle g|+\Omega_n \sigma_n^x,
		\end{eqnarray}
		with $\kappa_n=\gamma_n=\gamma$, the eigenvalues and eigenstates of the system are
		\begin{eqnarray}\label{eq2}
			E_{\pm}=\pm1/2\sqrt{-\gamma^2+4\Omega^2},\notag
			\\ |\psi_{\pm}\rangle=\frac{1}{2\sqrt{2}\Omega}
			\begin{pmatrix}
				i\gamma\pm\sqrt{-\gamma^2+4\Omega^2}\\2\Omega
			\end{pmatrix}.
		\end{eqnarray}
		\par To construct the biorthogonal basis, the Hamiltonian in the dual space of $H_n$ needs to consider
		\begin{eqnarray}\label{eq3}
			H'_n=i\gamma_n/2|e\rangle\langle e|-i\kappa_n/2|g\rangle\langle g|+\Omega_n \sigma_n^x,
		\end{eqnarray}
		with the eigenvalues and eigenstates
		\begin{eqnarray}\label{eq4}
			E'_{\pm}=\pm1/2\sqrt{-\gamma^2+4\Omega^2},\notag
			\\ |\psi'_{\pm}\rangle=\frac{1}{2\sqrt{2}\Omega}
			\begin{pmatrix}
				-i\gamma\pm\sqrt{-\gamma^2+4\Omega^2}\\2\Omega
			\end{pmatrix}.
		\end{eqnarray}
		\par It is easy to see that $|\psi_{\pm}\rangle$ does not satisfy the orthogonality due to the non-Hermitian nature of the Hamiltonian, i.e., $\langle\psi_+|\psi_-\rangle\neq0$, but with the dual Hamiltonian $H'_n$, there have $\langle\psi'_+|\psi_-\rangle=\langle\psi'_-|\psi_+\rangle=0$, and the normalized biorthogonal eigenstates can be constructed $|\overline{\psi}_{\pm}\rangle=|\psi_{\pm}\rangle/\sqrt{\langle\psi'_{\pm}|\psi_{\pm}\rangle}$.
	
		\par For the two-qubit non-Hermitian Hamiltonian without the coupling between qubits $H_0=\sum_{n=1}^2-i\gamma_n/2|e\rangle\langle e|+i\kappa_n/2|g\rangle\langle g|+\Omega_n \sigma_n^x$, suppose $\kappa_n=\gamma_n=\gamma$ and $\eta=\sqrt{-\gamma^2+4\Omega^2}$, the first two eigenvalues and the corresponding eigenstates are	\begin{eqnarray}\label{eq5}
			E_{++}=\eta&,& E_{--}=-\eta,\notag
			\\ |\psi_{++}\rangle=|\psi_{+}\rangle|\psi_{+}\rangle=
			\frac{1}{8\Omega^2}
			\begin{pmatrix}
				(i\gamma+\eta)^2\\2\Omega(i\gamma+\eta)\\2\Omega(i\gamma+\eta)\\4\Omega^2
			\end{pmatrix}&,&
			|\psi_{--}\rangle=|\psi_{-}\rangle|\psi_{-}\rangle=
			\frac{1}{8\Omega^2}
			\begin{pmatrix}
				(i\gamma-\eta)^2\\2\Omega(i\gamma-\eta)\\2\Omega(i\gamma-\eta)\\4\Omega^2
			\end{pmatrix}.
		\end{eqnarray}
		In the dual space, the eigenvalues and eigenstates of the Hamiltonian $H'_0=\sum_{n=1}^2i\gamma_n/2|e\rangle\langle e|-i\kappa_n/2|g\rangle\langle g|+\Omega_n \sigma_n^x$ are
		\begin{eqnarray}\label{eq6}
			E'_{++}=\eta&,& E'_{--}=-\eta,\notag
			\\ |\psi'_{++}\rangle=|\psi'_{+}\rangle|\psi'_{+}\rangle=
			\frac{1}{8\Omega^2}
			\begin{pmatrix}
				(-i\gamma+\eta)^2\\2\Omega(-i\gamma+\eta)\\2\Omega(-i\gamma+\eta)\\4\Omega^2
			\end{pmatrix}&,&
			|\psi'_{--}\rangle=|\psi'_{-}\rangle|\psi'_{-}\rangle=
			\frac{1}{8\Omega^2}
			\begin{pmatrix}
				(-i\gamma-\eta)^2\\2\Omega(-i\gamma-\eta)\\2\Omega(-i\gamma-\eta)\\4\Omega^2
			\end{pmatrix}.
		\end{eqnarray}
		
		The corresponding normalized biorthogonal eigenstates can be constructed as $|\overline{\psi}_{++}\rangle=|\psi_{++}\rangle/\sqrt{\langle\psi'_{++}|\psi_{++}\rangle}$ and $|\overline{\psi}_{--}\rangle=|\psi_{--}\rangle/\sqrt{\langle\psi'_{--}|\psi_{--}\rangle}$.
		\par The other two eigenvalues and eigenstates are given by
		\begin{eqnarray}\label{eq7}
			E_{+-}=E_{-+}&=&0,\notag
			\\ |\psi_{+-}\rangle=|\psi_{+}\rangle|\psi_{-}\rangle=
			\frac{1}{4\Omega}
			\begin{pmatrix}
				-2\Omega\\i\gamma+\eta\\i\gamma-\eta\\2\Omega
			\end{pmatrix}&,&
			|\psi_{-+}\rangle=|\psi_{-}\rangle|\psi_{+}\rangle=
			\frac{1}{4\Omega}
			\begin{pmatrix}
				-2\Omega\\i\gamma-\eta\\i\gamma+\eta\\2\Omega
			\end{pmatrix},
		\end{eqnarray}
		Similarly, in the dual space, the corresponding eigenvalues and eigenstates are
		\begin{eqnarray}\label{eq7}
			E'_{+-}=E'_{-+}&=&0,\notag
			\\ |\psi'_{+-}\rangle=|\psi'_{+}\rangle|\psi'_{-}\rangle=
			\frac{1}{4\Omega}
			\begin{pmatrix}
				-2\Omega\\-i\gamma+\eta\\-i\gamma-\eta\\2\Omega
			\end{pmatrix}&,&
			|\psi'_{-+}\rangle=|\psi'_{-}\rangle|\psi'_{+}\rangle=
			\frac{1}{4\Omega}
			\begin{pmatrix}
				-2\Omega\\-i\gamma-\eta\\-i\gamma+\eta\\2\Omega
			\end{pmatrix},
		\end{eqnarray}
		the normalized biorthogonal basis are $|\overline{\psi}_{+-}\rangle=|\psi_{+-}\rangle/\sqrt{\langle\psi'_{+-}|\psi_{+-}\rangle}$ and $|\overline{\psi}_{-+}\rangle=|\psi_{-+}\rangle/\sqrt{\langle\psi'_{-+}|\psi_{-+}\rangle}$.
		\par At this point, we have constructed a complete set of biorthogonal basis of the Hamiltonian $H_0$, including $\{|\overline{\psi}_{++}\rangle,|\overline{\psi}_{--}\rangle,|\overline{\psi}_{+-}\rangle,|\overline{\psi}_{-+}\rangle\}$ and corresponding $\{|\overline{\psi'}_{++}\rangle,|\overline{\psi'}_{--}\rangle,|\overline{\psi'}_{+-}\rangle,|\overline{\psi'}_{-+}\rangle\}$ in the dual space. It can be verified that these eigenvectors are not only normalized but also orthogonal to each other.
		\par Consider the interaction Hamiltonian $H_I=J(\sigma_1^{+}\sigma_2^{-}+\sigma_1^{-}\sigma_2^{+})$, which we treat as the perturbation due to the weak coupling between qubits. Next, we use the degenerate and non-degenerate perturbation theory to solve the system's eigenvalues and eigenstates[1].
		\par  First consider the perturbation theory in the degenerate subspace, the perturbation matrix can be written as
		\begin{eqnarray}\label{eq13}
			H_{i}=
			\begin{pmatrix}
				\langle\overline{\psi'}_{+-}|H_I|\overline{\psi}_{+-}\rangle&\langle\overline{\psi'}_{+-}|H_I|\overline{\psi}_{-+}\rangle\\\langle\overline{\psi'}_{-+}|H_I|\overline{\psi}_{+-}\rangle&\langle\overline{\psi'}_{-+}|H_I|\overline{\psi}_{-+}\rangle
			\end{pmatrix}
			=
			\begin{pmatrix}
				\frac{2J\Omega^2}{\gamma^2-4\Omega^2}&\frac{J(\gamma^2-2\Omega^2)}{\gamma^2-4\Omega^2}\\\frac{J(\gamma^2-2\Omega^2)}{\gamma^2-4\Omega^2}&\frac{2J\Omega^2}{\gamma^2-4\Omega^2}
			\end{pmatrix}.
		\end{eqnarray}
		The eigenvalues of this Hermitian matrix are $\lambda_1=-J$ and $\lambda_2=\frac{J\gamma^2}{\gamma^2-4\Omega^2}$, the corresponding eigenstates are $|\lambda_1\rangle=(-|\overline{\psi}_{+-}\rangle+|\overline{\psi}_{-+}\rangle)/\sqrt{2}$, $|\lambda_2\rangle=(|\overline{\psi}_{+-}\rangle+|\overline{\psi}_{-+}\rangle/\sqrt{2}$, which constructed a new set of basis vectors. In the dual space, the eigenstates are $|\lambda_1'\rangle=(-|\overline{\psi'}_{+-}\rangle+|\overline{\psi'}_{-+}\rangle)/\sqrt{2}$, $|\lambda_2'\rangle=(|\overline{\psi'}_{+-}\rangle+|\overline{\psi'}_{-+}\rangle/\sqrt{2}$.
		\par The first-order corrections for the eigenvalues are
		\begin{eqnarray}\label{eq14}
			\Lambda_1&=&E_{+-}+\lambda_1=-J,\notag\\
			\Lambda_2&=&E_{-+}+\lambda_2=\frac{J\gamma^2}{\gamma^2-4\Omega^2},
		\end{eqnarray}
		and the perturbed eigenstates are given by
		\begin{eqnarray}\label{eq15}
			|\Psi_1\rangle&=&|\lambda_1\rangle+\frac{\langle\overline{\psi'}_{++}|H_I|\lambda_1\rangle}{E_{+-}-E_{++}}|\overline{\psi}_{++}\rangle+\frac{\langle\overline{\psi'}_{--}|H_I|\lambda_1\rangle}{E_{+-}-E_{--}}|\overline{\psi}_{--}\rangle\notag,
		\end{eqnarray}
		\begin{eqnarray}\label{eq15}
			|\Psi_2\rangle&=&|\lambda_2\rangle+\frac{\langle\overline{\psi'}_{++}|H_I|\lambda_2\rangle}{E_{-+}-E_{++}}|\overline{\psi}_{++}\rangle+\frac{\langle\overline{\psi'}_{--}|H_I|\lambda_2\rangle}{E_{-+}-E_{--}}|\overline{\psi}_{--}\rangle,
		\end{eqnarray}
		\begin{eqnarray}\label{eq15}
			|\Psi'_1\rangle&=&|\lambda'_1\rangle+\frac{\langle\overline{\psi}_{++}|H_I|\lambda'_1\rangle}{E_{+-}-E_{++}}|\overline{\psi'}_{++}\rangle+\frac{\langle\overline{\psi}_{--}|H_I|\lambda'_1\rangle}{E_{+-}-E_{--}}|\overline{\psi'}_{--}\rangle\notag,
		\end{eqnarray}
		\begin{eqnarray}\label{eq15}
			|\Psi_2'\rangle&=&|\lambda'_2\rangle+\frac{\langle\overline{\psi}_{++}|H_I|\lambda'_2\rangle}{E_{-+}-E_{++}}|\overline{\psi'}_{++}\rangle+\frac{\langle\overline{\psi}_{--}|H_I|\lambda'_2\rangle}{E_{-+}-E_{--}}|\overline{\psi'}_{--}\rangle.
		\end{eqnarray}
		\par For non-degenerate subspace, the perturbation eigenvalue is obtained by using the first-order non-degenerate perturbation approximation
		\begin{eqnarray}\label{eq23}
			\Lambda_3=E_{++}+\langle\overline{\psi'}_{++}|H_I|\psi'_{++}\rangle=\eta-\frac{2J\Omega^2}{\eta^2},\notag
		\end{eqnarray}
		\begin{eqnarray}\label{eq24}
			\Lambda_4=E_{--}+\langle\overline{\psi'}_{--}|H_I|\psi'_{--}\rangle=-\eta-\frac{2J\Omega^2}{\eta^2},
		\end{eqnarray}
		the corresponding eigenstates are
		\begin{eqnarray}\label{eq25}
			|\Psi_3\rangle&\approx&|\overline{\psi}_{++}\rangle+\frac{\langle\overline{\psi'}_{--}|H_I|\overline{\psi}_{++}\rangle}{E_{++}-E_{--}}|\overline{\psi}_{--}\rangle,\notag
		\end{eqnarray}
		\begin{eqnarray}\label{eq25}
			|\Psi_4\rangle&\approx&|\overline{\psi}_{--}\rangle+\frac{\langle\overline{\psi'}_{++}|H_I|\overline{\psi}_{--}\rangle}{E_{--}-E_{++}}|\overline{\psi}_{++}\rangle.
		\end{eqnarray}
		\begin{eqnarray}\label{eq25}
			|\Psi'_3\rangle&\approx&|\overline{\psi'}_{++}\rangle+\frac{\langle\overline{\psi}_{--}|H_I|\overline{\psi'}_{++}\rangle}{E_{++}-E_{--}}|\overline{\psi}_{--}\rangle,\notag
		\end{eqnarray}
		\begin{eqnarray}\label{eq25}
			|\Psi'_4\rangle&\approx&|\overline{\psi'}_{--}\rangle+\frac{\langle\overline{\psi}_{++}|H_I|\overline{\psi'}_{--}\rangle}{E_{--}-E_{++}}|\overline{\psi}_{++}\rangle.
		\end{eqnarray}
		\par At present, we have used perturbation theory to give the first-order corrected eigenvalues and eigenstates $\{|\Psi_1\rangle,|\Psi_2\rangle,|\Psi_3\rangle,|\Psi_4\rangle\}$ and $\{|\Psi'_1\rangle,|\Psi'_2\rangle,|\Psi'_3\rangle,|\Psi'_4\rangle\}$, these eigenvalues are not only normalized but also orthogonal to each other, thus can form a complete set of basis.
		\subsection{B. Analytic solutions of the concurrence for the truly $\mathcal{PT}$ symmetric system}
		\par Assuming that the initial state of the system is $|\psi_0\rangle=|gg\rangle$, the system evolution state can be written as
		\begin{eqnarray}\label{eq28}
			|\psi(t)\rangle&=&U(t)|\psi_0\rangle\notag\\&=&U(t)(|\Psi_1\rangle\langle\Psi'_1|+|\Psi_2\rangle\langle\Psi'_2|+|\Psi_3\rangle\langle\Psi'_3|\notag+|\Psi_4\rangle\langle\Psi'_4|)|\psi_0\rangle\notag\\	&=&\langle\Psi'_1|\psi_0\rangle U(t)|\Psi_1\rangle+\langle\Psi'_2|\psi_0\rangle U(t)|\Psi_2\rangle+\langle\Psi'_3|\psi_0\rangle U(t)|\Psi_3\rangle+\langle\Psi'_4|\psi_0\rangle U(t)|\Psi_4\rangle\notag\\
			&=&\langle\Psi'_1|\psi_0\rangle e^{-it\Lambda_1}|\Psi_1\rangle+\langle\Psi'_2|\psi_0\rangle e^{-it\Lambda_2}|\Psi_2\rangle+\langle\Psi'_3|\psi_0\rangle e^{-it\Lambda_3}|\Psi_3\rangle+\langle\Psi'_4|\psi_0\rangle e^{-it\Lambda_4}|\Psi_4\rangle.
		\end{eqnarray}
		\par After substituting in the corresponding eigenvalues and eigenstates, the final state is $|\psi_f\rangle=a|gg\rangle+b(|ge\rangle+|eg\rangle)+c|ee\rangle$, where
		\begin{eqnarray}
			a&=&\frac{\Omega^2}{\eta^8}\left[2\eta^2e^{\frac{iJ\gamma^2t}{\eta^2}}(\eta^2+iJ\gamma)^2-\frac{\Omega^2e^{-it(\eta+\frac{2J\Omega^2}{\eta^2})}(Js+2\eta^3)^2}{2s}+\frac{\Omega^2e^{it(\eta-\frac{2J\Omega^2}{\eta^2})}(Jz+2\eta^3)^2}{2z}\right],\notag\\
		b&=&\frac{\Omega}{4\eta^8}\left[4\gamma\eta^4e^{\frac{iJt\gamma^2}{\eta^2}}(J\gamma-i\gamma^2)-\frac{2\Omega^2(2\eta^3+Js)}{s}e^{-it(\frac{2J\Omega^2}{\eta^2}+\eta)}(p+\eta^4-i\gamma J\Omega^2)+\frac{2\Omega^2}{z}(2\eta^3\right.\notag\\
			&&\left.
+Jz)e^{it(-\frac{2J\Omega^2}{\eta^2}+\eta)}(p-\eta^4+i\gamma J\Omega^2)\right],\notag\\
			c&=&\frac{-\Omega^2}{2\eta^8}\left[4\eta^2e^{\frac{iJt\gamma^2}{\eta^2}}(\eta^4+\gamma^2 J^2)+2e^{-it(\frac{2J\Omega^2}{\eta^2}+\eta)}(q+l)+2e^{it(-\frac{2J\Omega^2}{\eta^2}+\eta)}(q-l)\right],
		\end{eqnarray}
		with $s=\gamma^2-2\Omega^2+i\gamma\eta$, $z=-\gamma^2+2\Omega^2+i\gamma\eta$, $p=\eta(i\gamma^3-4i\gamma\Omega^2+J\Omega^2)$, $q=48\gamma^2\Omega^4-J^2\Omega^4-64\Omega^6+\gamma^6-12\gamma^4\Omega^2$, $l=8J\eta\Omega^4-6J\gamma^2\eta\Omega^2+J\gamma^4\eta$.
		\par The concurrence can be obtained by
		
		\begin{eqnarray}
			C=\frac{2|ac-b^2|}{|a|^2+2|b|^2+|d|^2}=2\frac{|A|}{B},
		\end{eqnarray}
		where
		\begin{eqnarray}
A&=&\frac{\Omega^2}{2} e^{-\frac{2it(\eta^3+2J\Omega^2)}{\eta^2}}[-2e^{\frac{2it(-J\eta^2+\eta^3+6J\Omega^2)}{\eta^2}}\eta^{12}+2e^{2it\eta}(\eta^6-J^2\Omega^4)(\eta^6-J^2\Omega^4+iJ\eta^4\gamma)+e^{4it\eta}J\eta^2(-2\eta^7\Omega^2\notag\\
   &&+4J\eta^4\Omega^4+J^2\eta\Omega^4(\eta^2-2\Omega^2)+iJ^2\eta^2\Omega^4\gamma+\eta^9-i\eta^8\gamma)+J(2\eta^9\Omega^2+4J\eta^6\Omega^4+2J^2\eta^3\Omega^6-J^2\eta^4\Omega^4(\eta\notag\\
   &&-i\gamma)-\eta^{11}-i\eta^{10}\gamma)],\notag
   \end{eqnarray}
\begin{eqnarray}
B&=&\frac{1}{\eta^{2}}[\eta^{12}(3\gamma^2+\eta^2)\Omega^2+4\eta^{10}(J^2+3\eta^2)\Omega^4+6J^2\eta^6(\gamma^2-3\eta^2)\Omega^6+24J^2\eta^6\Omega^8+J^4\Omega^{10}(\gamma^2+\eta^2)+4J^4\Omega^{12}\notag\\
   &&+(\eta^4+\gamma^2\Omega^2-5\eta^2\Omega^2+4\Omega^4)(\eta^6-J^2\Omega^4)^2\cos{(2t\eta)}+2\eta^6\Omega^2\cos{(t\frac{J\eta^2-\eta^3-6J\Omega^2}{\eta^2})}(\eta^2-4\Omega^2-\gamma^2)(\eta^3\notag\\
   &&+J\Omega^2)^2+2\eta^6\Omega^2\cos{(t\frac{J\eta^2+\eta^3-6J\Omega^2}{\eta^2})}(\eta^2-4\Omega^2-\gamma^2)(\eta^3-J\Omega^2)^2+4\gamma\eta^7\Omega^2\sin{(t\frac{J\eta^2+\eta^3-6J\Omega^2}{\eta^2})}\notag\\
   &&(\eta^6-J^2\Omega^4)-4\gamma\eta^7\Omega^2(\eta^6-J^2\Omega^4)\sin{(t\frac{J\eta^2-\eta^3-6J\Omega^2}{\eta^2})}+\gamma\eta(\eta^2-4\Omega^2)(\eta^{12}-J^4\Omega^8)\sin{(2t\eta)}].
		\end{eqnarray}
 
\begin{figure*}
			\centering
			\includegraphics[width=12cm,height=5.8cm]{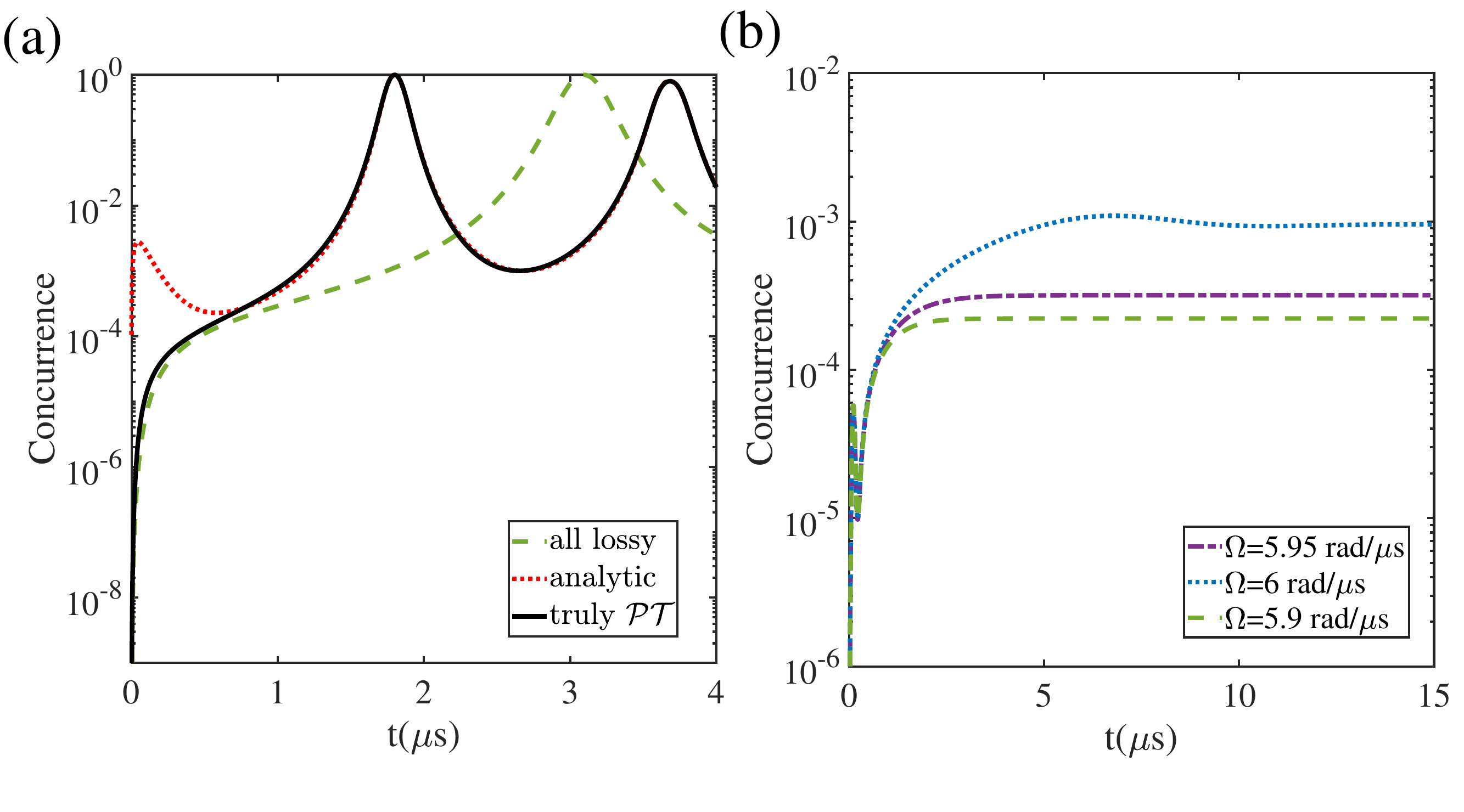}
			\caption{(a) The evolution of concurrence for truly $\mathcal{PT}$ symmetric and all lossy systems with the analytic result calculated by the perturbation theory for the truly $\mathcal{PT}$ symmetric Hamiltonian, where $\Omega=6.23~\rm{rad/\mu s}$ and $\Omega=3.152~\rm{rad/\mu s}$ for truly $\mathcal{PT}$ symmetric and all lossy systems, respectively. (b) The evolution of concurrence in the $\mathcal{PT}$-broken phase for the truly $\mathcal{PT}$ symmetric system.}\label{figs1}
		\end{figure*}
   \begin{figure*}
  \centering
			\includegraphics[width=13.5cm,height=6cm]{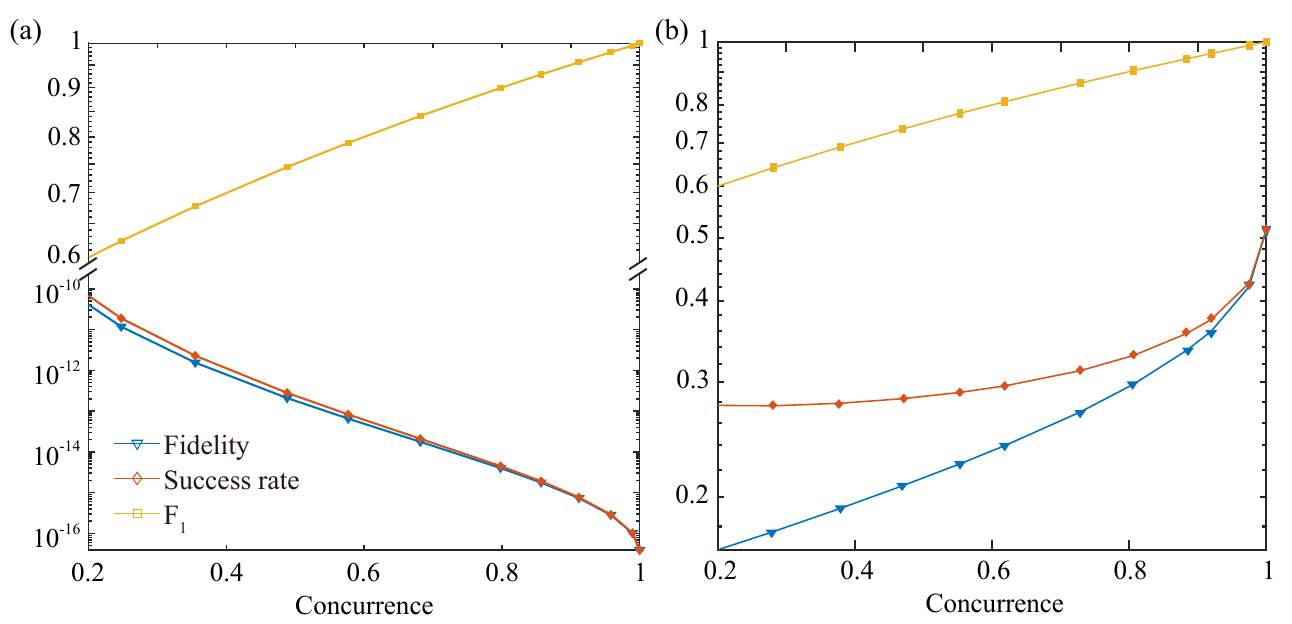}
			\caption{Fidelity, Success rate, and $F_1$ as the function of concurrence for passive (a) and active (b) $\mathcal{PT}$ symmetric systems.}\label{figs1n}
  \end{figure*}
	\par The analytical result is shown in Fig.~\ref{figs1}(a), which agrees well with the numerical simulation result except for the region close to $t=0$, the truly $\mathcal{PT}$ symmetric system has an advantage in the speed of preparing the maximal entanglement, and it's easy to verify that $C=0$ when $J=0$, there is no entanglement exists. The form of the maximal entangled state before normalization is $|\psi_f\rangle=0.3536e^{i\pi/4}|ee\rangle+0.3536e^{i5\pi/4}(|ge\rangle+|eg\rangle+|gg\rangle)$, and is $(-|ee\rangle+|ge\rangle+|eg\rangle+|gg\rangle)/2$ after normalize, which correspond to the Bell state. In addition, we give the evolution of the concurrence in Fig.~\ref{figs1}(b) in the $\mathcal{PT}$-broken phase for the truly $\mathcal{PT}$ symmetric Hamiltonian, the steady-state entanglement can be prepared, but the degree of entanglement is really low.
 \subsection{C. Fidelity and success rate for non-Hermitian systems and the trade-off relation}
 In non-Hermitian systems, the renormalization process has a significant impact on the practical success rate[2], which we define as the square of the norm of the normalization coefficient, and is often overlooked but should be considered. Fig.~\ref{figs1n} illustrates the relationship between the success rate, fidelity, and concurrence, together with the fidelity of the normalized state ($F_1=\langle\psi'_f|\psi_i\rangle$, where $|\psi'_f\rangle$ is the normalized state of $|\psi_f\rangle$, as discussed in the main text). Despite the relatively high value of $F_1$, the practical fidelity is notably low and strongly correlated with the success rate. In this study, we employ fidelity as a metric to assess system performance. In the case of passive $\mathcal{PT}$ symmetric systems, the fidelity and success rate are low for a high-degree of entanglement preparation, exhibiting a trade-off relation between the degree of entanglement and fidelity. In contrast, this trade-off relation is mitigated in active $\mathcal{PT}$ symmetric systems, where both the success rate and fidelity experience enhancements by several orders of magnitude.
 \subsection{D. Optimal parameters for the unbalanced gain-loss cases}
  Fig.~\ref{figs33} shows the optimal parameter combinations \{$\Omega,t$\} under different ratios of gain and loss for the maximal entanglement preparation, it is evident that the duration correspondingly extends as the discrepancy between gain and loss amplifies, with the Im($E_n$) decrease, thus decline the fidelity as the negative imaginary eigenvalues accumulate over time.		
	\subsection{E. Acceleration effect under strong coupling}
	Figs.~\ref{figs22} (a) and (b) shows the evolution of the concurrence for $J=0.01~\rm{rad/\mu s}$ and $J=0.1~\rm{rad/\mu s}$ under different driven amplitude $\Omega$ for the active $\mathcal{PT}$ symmetric system, where the coupling strength is much greater than $J=0.001~\rm{rad/\mu s}$, but the maximal entanglement can still be prepared when the properly driven amplitude is selected, and there is also an obvious acceleration effect compared to the conventional Hermitian system.
		\begin{figure*}
			\centering
			\includegraphics[width=6.5cm,height=5cm]{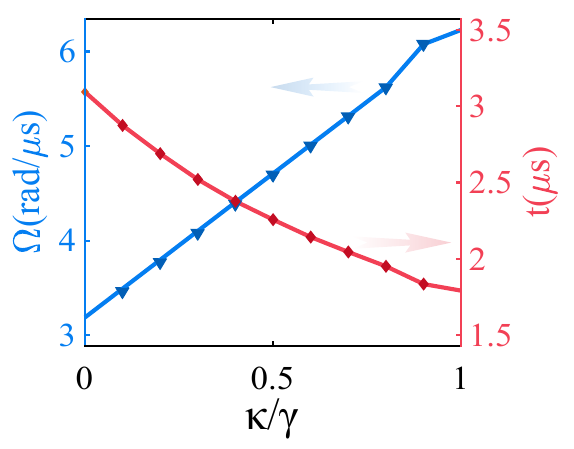}
			\caption{The optimal parameters for preparing the maximal entanglement in different ratios of gain-loss.}\label{figs33}
		\end{figure*}
 
		\begin{figure*}
			\centering
			\includegraphics[width=12cm,height=6.2cm]{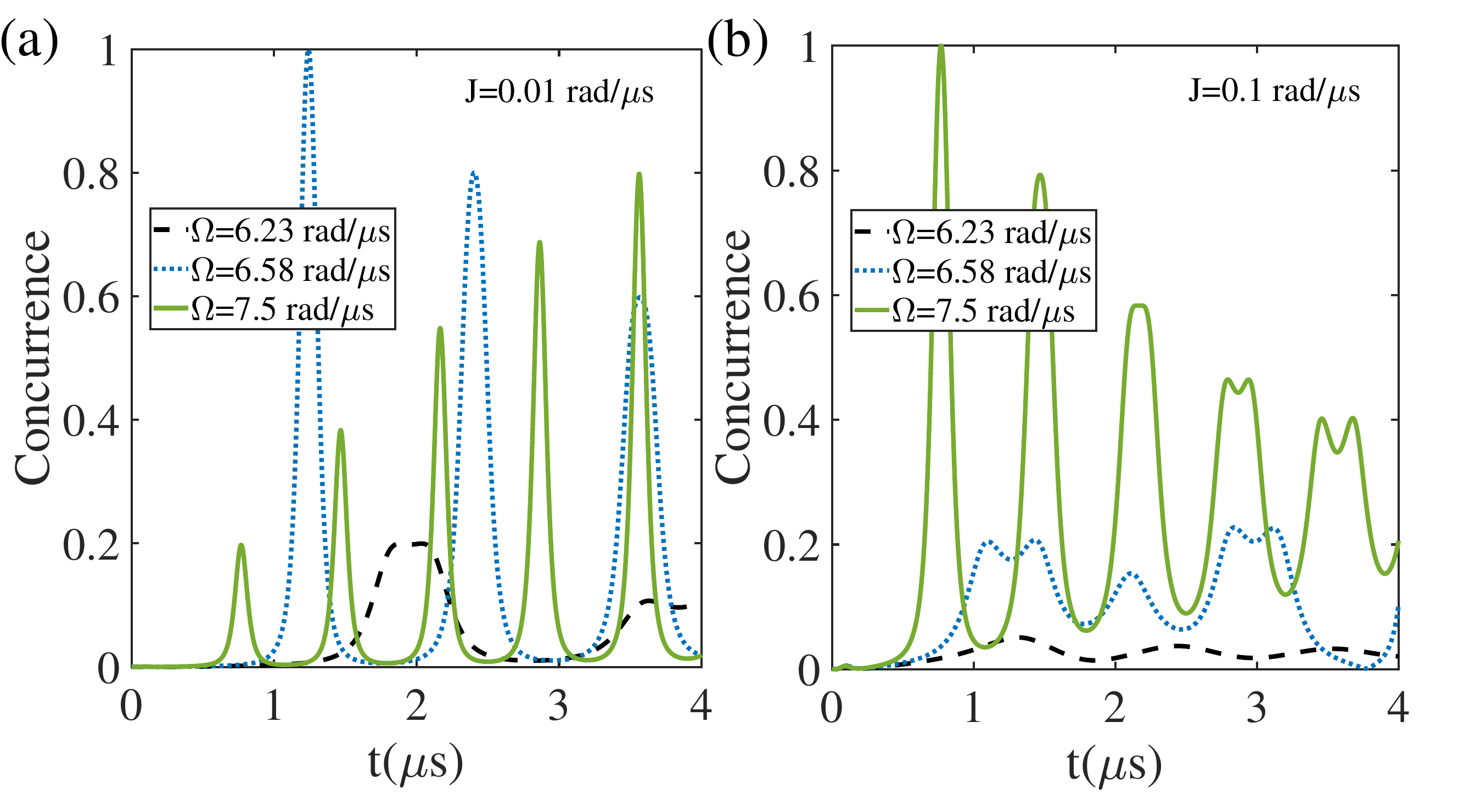}
			\caption{The evolution of the concurrence for the coupling strength $J=0.01~\rm{rad/\mu s}$ and $J=0.1~\rm{rad/\mu s}$ under different driven amplitude.}\label{figs22}
		\end{figure*}

		\section{II. Illustration of the three-qubit case}
		
		\subsection{A. Eigenvalues and eigenstates of the truly $\mathcal{PT}$ symmetric system}
		\par Next, we demonstrate the triplet non-Hermitian Hamiltonian without the coupling between qubits $H_0=\sum_{n=1}^3i\kappa_n/2\\|g\rangle\langle g|-i\gamma_n/2|e\rangle\langle e|+\Omega_n \sigma_n^x$ with $\kappa_n=\gamma_n=\gamma$. The first two eigenvalues are $E_{+++}=\frac{3\eta}{2}$, $E_{---}=-\frac{3\eta}{2}$ and the corresponding eigenstates$ |\psi_{+++}\rangle=|\psi_{+}\rangle|\psi_{+}\rangle|\psi_{+}\rangle$, $ |\psi_{---}\rangle=|\psi_{-}\rangle|\psi_{-}\rangle|\psi_{-}\rangle$.
		In the dual space, the eigenvalues and eigenstates of the Hamiltonian $H'=\sum_{n=1}^3i\gamma_n/2|e\rangle\langle e|-i\kappa_n/2|g\rangle\langle g|+\Omega_n \sigma_n^x$ are $E_{+++}'=\frac{3\eta}{2}$, $E_{---}'=-\frac{3\eta}{2}$ and $ |\psi_{+++}'\rangle=|\psi_{+}'\rangle|\psi_{+}'\rangle|\psi_{+}'\rangle$, $ |\psi_{---}'\rangle=|\psi_{-}'\rangle|\psi_{-}'\rangle|\psi_{-}'\rangle$.
		With the dual space, the corresponding normalized biorthogonal eigenstates can be constructed as $|\overline{\psi}_{+++}\rangle=|\psi_{+++}\rangle/
\sqrt{\langle\psi'_{+++}|\psi_{+++}\rangle}$ and $|\overline{\psi}_{---}\rangle=|\psi_{---}\rangle/\sqrt{\langle\psi'_{---}|\psi_{---}\rangle}$.
		\par The other three eigenvalues and eigenstates are given by $E_{+--}=E_{-+-}=E_{--+}=-\frac{\eta}{2}$ and $|\psi_{+--}\rangle=|\psi_{+}\rangle|\psi_{-}\rangle|\psi_{-}\rangle$, $|\psi_{-+-}\rangle=|\psi_{-}\rangle|\psi_{+}\rangle|\psi_{-}\rangle$, $|\psi_{--+}\rangle=|\psi_{-}\rangle|\psi_{-}\rangle|\psi_{+}\rangle$.
		In the dual space, the corresponding eigenvalues and eigenstates are given by $E'_{+--}=E'_{-+-}=E'_{--+}=-\frac{\eta}{2}$, $|\psi'_{+--}\rangle=|\psi'_{+}\rangle|\psi'_{-}\rangle|\psi'_{-}\rangle$,
			$|\psi'_{-+-}\rangle=|\psi'_{-}\rangle|\psi'_{+}\rangle|\psi'_{-}\rangle$ and $|\psi'_{--+}\rangle=|\psi'_{-}\rangle|\psi'_{-}\rangle|\psi'_{+}\rangle$.
   Similarly, the normalized biorthogoal basis are $|\overline{\psi}_{+--}\rangle=|\psi_{+--}\rangle/\sqrt{\langle\psi'_{+--}|\psi_{+--}\rangle}$, $|\overline{\psi}_{-+-}\rangle=|\psi_{-+-}\rangle/\sqrt{\langle\psi'_{-+-}|\psi_{-+-}\rangle}$ and $|\overline{\psi}_{--+}\rangle=|\psi_{--+}\rangle/\sqrt{\langle\psi'_{--+}|\psi_{--+}\rangle}$.
		\par The last three eigenvalues and eigenstates are $E_{-++}=E_{+-+}=E_{++-}=\frac{\eta}{2}$,$|\psi_{-++}\rangle=|\psi_{-}\rangle|\psi_{+}\rangle|\psi_{+}\rangle$, and $|\psi_{+-+}\rangle=|\psi_{+}\rangle|\psi_{-}\rangle|\psi_{+}\rangle$, $|\psi_{--+}\rangle=|\psi_{-}\rangle|\psi_{-}\rangle|\psi_{+}\rangle$,
		in the dual space, the eigenvalues and eigenstates are $E'_{-++}=E'_{+-+}=E'_{++-}=\frac{\eta}{2}$, $|\psi'_{-++}\rangle=|\psi'_{-}\rangle|\psi'_{+}\rangle|\psi'_{+}\rangle$, $
			|\psi'_{+-+}\rangle=|\psi'_{+}\rangle|\psi'_{-}\rangle|\psi'_{+}\rangle$,	$|\psi'_{--+}\rangle=|\psi'_{-}\rangle|\psi'_{-}\rangle|\psi'_{+}\rangle$,
		the normalized biorthogoal basis are $|\overline{\psi}_{-++}\rangle=|\psi_{-++}\rangle/\sqrt{\langle\psi'_{-++}|\psi_{-++}\rangle}$, $|\overline{\psi}_{+-+}\rangle=|\psi_{+-+}\rangle/\sqrt{\langle\psi'_{+-+}|\psi_{+-+}\rangle}$ and $|\overline{\psi}_{++-}\rangle=|\psi_{++-}\rangle/\sqrt{\langle\psi'_{++-}|\psi_{++-}\rangle}$.
		\par We have constructed a complete set of orthogonal basis of the Hamiltonian $H_0$, including normalized eigenvectors $\{|\overline{\psi}_{+++}\rangle,|\overline{\psi}_{---}\rangle,|\overline{\psi}_{+--}\rangle,|\overline{\psi}_{-+-}\rangle,|\overline{\psi}_{--+}\rangle,|\overline{\psi}_{-++}\rangle,|\overline{\psi}_{+-+}\rangle,|\overline{\psi}_{++-}\rangle\}$ accompanied with in the dual space $\{|\overline{\psi'}_{+++}\rangle,|\overline{\psi'}_{---}\rangle,|\overline{\psi'}_{+--}\rangle,|\overline{\psi'}_{-+-}\rangle,|\overline{\psi'}_{--+}\rangle,|\overline{\psi'}_{-++}\rangle,|\overline{\psi'}_{+-+}\rangle,|\overline{\psi'}_{++-}\rangle\}$. It is easily verified that these eigenvectors satisfy not only the normalization,
		\begin{eqnarray}\label{eq11}
			&&\langle\overline{\psi'}_{+++}|\overline{\psi}_{+++}\rangle=\langle\overline{\psi'}_{---}|\overline{\psi}_{---}\rangle=\langle\overline{\psi'}_{+--}|\overline{\psi}_{+--}\rangle=\langle\overline{\psi'}_{-+-}|\overline{\psi}_{-+-}\rangle\cr\cr&&=\langle\overline{\psi'}_{+--}|\overline{\psi}_{+--}\rangle=\langle\overline{\psi'}_{-++}|\overline{\psi}_{-++}\rangle=\langle\overline{\psi'}_{+-+}|\overline{\psi}_{+-+}\rangle=\langle\overline{\psi'}_{++-}|\overline{\psi}_{++-}\rangle=1,
		\end{eqnarray}
		but also orthogonality,
		\begin{eqnarray}\label{eq12}
			&&\langle\overline{\psi'}_{+++}|\overline{\psi}_{---}\rangle=\langle\overline{\psi'}_{+++}|\overline{\psi}_{+--}\rangle=\langle\overline{\psi'}_{+++}|\overline{\psi}_{-+-}\rangle=\langle\overline{\psi'}_{+++}|\overline{\psi}_{--+}\rangle\cr\cr&&=\langle\overline{\psi'}_{+++}|\overline{\psi}_{-++}\rangle=\langle\overline{\psi'}_{+++}|\overline{\psi}_{+-+}\rangle=\langle\overline{\psi'}_{+++}|\overline{\psi}_{++-}\rangle=0,\cr\cr&&\langle\overline{\psi'}_{---}|\overline{\psi}_{+++}\rangle=\langle\overline{\psi'}_{---}|\overline{\psi}_{+--}\rangle=\langle\overline{\psi'}_{---}|\overline{\psi}_{-+-}\rangle=\langle\overline{\psi'}_{---}|\overline{\psi}_{--+}\rangle\cr\cr&&=\langle\overline{\psi'}_{---}|\overline{\psi}_{-++}\rangle=\langle\overline{\psi'}_{---}|\overline{\psi}_{+-+}\rangle=\langle\overline{\psi'}_{---}|\overline{\psi}_{++-}\rangle=0,		
		\end{eqnarray}
		The remaining internal products of states are not shown here.

				\par Since the coupling between qubits is very weak, we treat the interaction Hamiltonian $H_I=J\sum_{n\neq m=1}^3\sigma_n^{+}\sigma_m^{-}$ as a perturbation. Next, we consider using degenerate and non-degenerate perturbation theory to solve the eigenvalues and eigenstates of the system.
		\par  First consider perturbation theory in the degenerate case, for the first three degenerate states, the perturbation matrix can be written as
		\begin{eqnarray}\label{eq13}
			H_{i}^1&=&
			\begin{pmatrix}
				\langle\overline{\psi'}_{+--}|H_I|\overline{\psi}_{+--}\rangle&\langle\overline{\psi'}_{+--}|H_I|\overline{\psi}_{-+-}\rangle&\langle\overline{\psi'}_{+--}|H_I|\overline{\psi}_{--+}\rangle\\\langle\overline{\psi'}_{-+-}|H_I|\overline{\psi}_{+--}\rangle&\langle\overline{\psi'}_{-+-}|H_I|\overline{\psi}_{-+-}\rangle&\langle\overline{\psi'}_{-+-}|H_I|\overline{\psi}_{--+}\rangle\\\langle\overline{\psi'}_{--+}|H_I|\overline{\psi}_{+--}\rangle&\langle\overline{\psi'}_{--+}|H_I|\overline{\psi}_{-+-}\rangle&\langle\overline{\psi'}_{--+}|H_I|\overline{\psi}_{--+}\rangle
			\end{pmatrix}\notag\\
			&=&
			\begin{pmatrix}
				\frac{2J\Omega^2}{\gamma^2-4\Omega^2}&\frac{J(\gamma^2-2\Omega^2)}{\gamma^2-4\Omega^2}&\frac{J(\gamma^2-2\Omega^2)}{\gamma^2-4\Omega^2}\\\frac{J(\gamma^2-2\Omega^2)}{\gamma^2-4\Omega^2}&\frac{2J\Omega^2}{\gamma^2-4\Omega^2}&\frac{J(\gamma^2-2\Omega^2)}{\gamma^2-4\Omega^2}\\\frac{J(\gamma^2-2\Omega^2)}{\gamma^2-4\Omega^2}&\frac{J(\gamma^2-2\Omega^2)}{\gamma^2-4\Omega^2}&\frac{2J\Omega^2}{\gamma^2-4\Omega^2}
			\end{pmatrix}.
		\end{eqnarray}
		The eigenvalues of this Hermitian matrix are $\lambda_1=\lambda_2=-{J}$ and $\lambda_3=\frac{2J(\gamma^2-\Omega^2)}{\gamma^2-4\Omega^2}$ and the corresponding eigenstates are $|\lambda_1\rangle=(-|\overline{\psi}_{+--}\rangle+|\overline{\psi}_{--+}\rangle)/\sqrt{2}$, $|\lambda_2\rangle=(-|\overline{\psi}_{+--}\rangle+2|\overline{\psi}_{-+-}\rangle-|\overline{\psi}_{--+}\rangle)/\sqrt{6}$ and $|\lambda_3\rangle=(|\overline{\psi}_{+--}\rangle+|\overline{\psi}_{-+-}\rangle+|\overline{\psi}_{--+}\rangle)/\sqrt{3}$, which constructed a new set of basis vectors, and we have used Schmidt orthogonalization here[3]. In the dual space, $|\lambda_1'\rangle=(-|\overline{\psi'}_{+--}\rangle+|\overline{\psi'}_{--+}\rangle)/\sqrt{2}$, $|\lambda_2'\rangle=(-|\overline{\psi'}_{+--}\rangle+2|\overline{\psi'}_{-+-}\rangle-|\overline{\psi'}_{--+}\rangle)/\sqrt{6}$ and $|\lambda_3'\rangle=(|\overline{\psi'}_{+--}\rangle+|\overline{\psi'}_{-+-}\rangle+|\overline{\psi'}_{--+}\rangle)/\sqrt{3}$.
		\par Under the first-order non-degenerate perturbation theory, the perturbation eigenvalues are
		\begin{eqnarray}\label{eq14}
			\Lambda_1&=&E_{+--}+\lambda_1=-\frac{\eta}{2}-J,\notag\\
			\Lambda_2&=&E_{-+-}+\lambda_2=-\frac{\eta}{2}-J,\notag\\
			\Lambda_3&=&E_{--+}+\lambda_3=-\frac{\eta}{2}+\frac{2J(\gamma^2-\Omega^2)}{\gamma^2-4\Omega^2}.
		\end{eqnarray}
		After the first-order correction of the eigenstate, the degenerate eigenstate becomes
		\begin{eqnarray}\label{eq15}
			|\Psi_1\rangle&=&|\lambda_1\rangle+\frac{\langle\overline{\psi'}_{+++}|H_I|\lambda_1\rangle}{E_{+--}-E_{+++}}|\overline{\psi}_{+++}\rangle+\frac{\langle\overline{\psi'}_{---}|H_I|\lambda_1\rangle}{E_{+--}-E_{---}}|\overline{\psi}_{---}\rangle+\frac{\langle\overline{\psi'}_{-++}|H_I|\lambda_1\rangle}{E_{+--}-E_{-++}}|\overline{\psi}_{-++}\rangle\notag\\
			&&+\frac{\langle\overline{\psi'}_{+-+}|H_I|\lambda_1\rangle}{E_{+--}-E_{+-+}}|\overline{\psi}_{+-+}\rangle+\frac{\langle\overline{\psi'}_{++-}|H_I|\lambda_1\rangle}{E_{+--}-E_{++-}}|\overline{\psi}_{++-}\rangle\approx|\lambda_1\rangle,\notag
		\end{eqnarray}
		\begin{eqnarray}\label{eq15}
			|\Psi'_1\rangle&=&|\lambda_1'\rangle+\frac{\langle\overline{\psi}_{+++}|H_I|\lambda_1'\rangle}{E_{+--}-E_{+++}}|\overline{\psi'}_{+++}\rangle+\frac{\langle\overline{\psi}_{---}|H_I|\lambda_1'\rangle}{E_{+--}-E_{---}}|\overline{\psi'}_{---}\rangle+\frac{\langle\overline{\psi}_{-++}|H_I|\lambda_1'\rangle}{E_{+--}-E_{-++}}|\overline{\psi'}_{-++}\rangle\notag\\
			&&+\frac{\langle\overline{\psi}_{+-+}|H_I|\lambda_1'\rangle}{E_{+--}-E_{+-+}}|\overline{\psi'}_{+-+}\rangle+\frac{\langle\overline{\psi}_{++-}|H_I|\lambda_1'\rangle}{E_{+--}-E_{++-}}|\overline{\psi'}_{++-}\rangle\approx|\lambda_1'\rangle,
		\end{eqnarray}
and $|\Psi_2\rangle\approx|\lambda_2\rangle, |\Psi_3\rangle\approx|\lambda_3\rangle,$	$|\Psi_2'\rangle\approx|\lambda_2'\rangle, |\Psi_3'\rangle\approx|\lambda_3'\rangle$.
		
		\par For the last three degenerate states, the perturbation matrix can be written as
		\begin{eqnarray}\label{eq18}
			H_{i}^2&=&
			\begin{pmatrix}
				\langle\overline{\psi'}_{-++}|H_I|\overline{\psi}_{-++}\rangle&\langle\overline{\psi'}_{-++}|H_I|\overline{\psi}_{+-+}\rangle&\langle\overline{\psi'}_{-++}|H_I|\overline{\psi}_{++-}\rangle\\\langle\overline{\psi'}_{+-+}|H_I|\overline{\psi}_{-++}\rangle&\langle\overline{\psi'}_{+-+}|H_I|\overline{\psi}_{+-+}\rangle&\langle\overline{\psi'}_{+-+}|H_I|\overline{\psi}_{++-}\rangle\\\langle\overline{\psi'}_{++-}|H_I|\overline{\psi}_{-++}\rangle&\langle\overline{\psi'}_{++-}|H_I|\overline{\psi}_{+-+}\rangle&\langle\overline{\psi'}_{++-}|H_I|\overline{\psi}_{++-}\rangle
			\end{pmatrix}\notag\\
			&=&
			\begin{pmatrix}
				\frac{2J\Omega^2}{\gamma^2-4\Omega^2}&\frac{J(\gamma^2-2\Omega^2)}{\gamma^2-4\Omega^2}&\frac{J(\gamma^2-2\Omega^2)}{\gamma^2-4\Omega^2}\\\frac{J(\gamma^2-2\Omega^2)}{\gamma^2-4\Omega^2}&\frac{2J\Omega^2}{\gamma^2-4\Omega^2}&\frac{J(\gamma^2-2\Omega^2)}{\gamma^2-4\Omega^2}\\\frac{J(\gamma^2-2\Omega^2)}{\gamma^2-4\Omega^2}&\frac{J(\gamma^2-2\Omega^2)}{\gamma^2-4\Omega^2}&\frac{2J\Omega^2}{\gamma^2-4\Omega^2}
			\end{pmatrix}.
		\end{eqnarray}
		The eigenvalues of this Hermitian matrix are $\lambda_4=\lambda_5=-{J}$ and $\lambda_6=\frac{2J(\gamma^2-\Omega^2)}{\gamma^2-4\Omega^2}$ and the corresponding eigenstates are $|\lambda_4\rangle=(-|\overline{\psi}_{-++}\rangle+|\overline{\psi}_{++-}\rangle)/\sqrt{2}$, $|\lambda_5\rangle=(-|\overline{\psi}_{-++}\rangle+2|\overline{\psi}_{+-+}\rangle-|\overline{\psi}_{++-}\rangle)/\sqrt{6}$ and $|\lambda_6\rangle=(|\overline{\psi}_{-++}\rangle+|\overline{\psi}_{+-+}\rangle+|\overline{\psi}_{++-}\rangle)/\sqrt{3}$, which constructed a new set of basis vectors. In the dual space, $|\lambda_4'\rangle=(-|\overline{\psi'}_{-++}\rangle+|\overline{\psi'}_{++-}\rangle)/\sqrt{2}$, $|\lambda_5'\rangle=(-|\overline{\psi'}_{-++}\rangle+2|\overline{\psi'}_{+-+}\rangle-|\overline{\psi'}_{++-}\rangle)/\sqrt{6}$ and $|\lambda_6'\rangle=(|\overline{\psi'}_{-++}\rangle+|\overline{\psi'}_{+-+}\rangle+|\overline{\psi'}_{++-}\rangle)/\sqrt{3}$.
		\par Similarly, the perturbation eigenvalue are
		\begin{eqnarray}\label{eq19}
			\Lambda_4&=&E_{-++}+\lambda_4=\frac{\eta}{2}-J,\notag\\
			\Lambda_5&=&E_{+-+}+\lambda_5=\frac{\eta}{2}-J,\notag\\
			\Lambda_6&=&E_{++-}+\lambda_6=\frac{\eta}{2}+\frac{2J(\gamma^2-\Omega^2)}{\gamma^2-4\Omega^2}.
		\end{eqnarray}
		After the first-order correction of the eigenstate, the degenerate eigenstate becomes
		\begin{eqnarray}\label{eq20}
			|\Psi_4\rangle&=&|\lambda_4\rangle+\frac{\langle\overline{\psi'}_{+++}|H_I|\lambda_4\rangle}{E_{-++}-E_{+++}}|\overline{\psi}_{+++}\rangle+\frac{\langle\overline{\psi'}_{---}|H_I|\lambda_4\rangle}{E_{-++}-E_{---}}|\overline{\psi}_{---}\rangle
			+\frac{\langle\overline{\psi'}_{+--}|H_I|\lambda_4\rangle}{E_{-++}-E_{+--}}|\overline{\psi}_{+--}\rangle\notag\\
			&&+\frac{\langle\overline{\psi'}_{-+-}|H_I|\lambda_4\rangle}{E_{-++}-E_{-+-}}|\overline{\psi}_{-+-}\rangle
			+\frac{\langle\overline{\psi'}_{--+}|H_I|\lambda_4\rangle}{E_{-++}-E_{--+}}|\overline{\psi}_{--+}\rangle\approx|\lambda_4\rangle,\notag
		\end{eqnarray}
		
		\begin{eqnarray}\label{eq20}
			|\Psi_4'\rangle&=&|\lambda_4'\rangle+\frac{\langle\overline{\psi}_{+++}|H_I|\lambda_4'\rangle}{E_{-++}-E_{+++}}|\overline{\psi'}_{+++}\rangle+\frac{\langle\overline{\psi}_{---}|H_I|\lambda_4'\rangle}{E_{-++}-E_{---}}|\overline{\psi'}_{---}\rangle
			+\frac{\langle\overline{\psi}_{+--}|H_I|\lambda_4'\rangle}{E_{-++}-E_{+--}}|\overline{\psi'}_{+--}\rangle\notag\\
			&&+\frac{\langle\overline{\psi}_{-+-}|H_I|\lambda_4'\rangle}{E_{-++}-E_{-+-}}|\overline{\psi'}_{-+-}\rangle
			+\frac{\langle\overline{\psi}_{--+}|H_I|\lambda_4'\rangle}{E_{-++}-E_{--+}}|\overline{\psi'}_{--+}\rangle\approx|\lambda_4'\rangle,
		\end{eqnarray}
		and $\Psi_5\approx|\lambda_5\rangle, \Psi_6\approx|\lambda_6\rangle$, $\Psi'_5\approx|\lambda'_5\rangle, \Psi'_6\approx|\lambda'_6\rangle$.	
		\par For non-degenerate subspaces, the perturbation eigenvalue can be obtained by using the first-order non-degenerate perturbation approximation
		\begin{eqnarray}\label{eq23}
			\Lambda_7=E_{+++}+\langle\overline{\psi'}_{+++}|H_I|\psi'_{+++}\rangle=\frac{3\eta}{2}-\frac{6J\Omega^2}{\gamma^2-4\Omega^2},\notag
		\end{eqnarray}
		\begin{eqnarray}\label{eq24}
			\Lambda_8=E_{---}+\langle\overline{\psi'}_{---}|H_I|\psi'_{---}\rangle=-\frac{3\eta}{2}-\frac{6J\Omega^2}{\gamma^2-4\Omega^2}.
		\end{eqnarray}
		The corresponding eigenstates are
		\begin{eqnarray}\label{eq25}
			|\Psi_7\rangle&=&|\overline{\psi}_{+++}\rangle+\frac{\langle\overline{\psi'}_{---}|H_I|\overline{\psi}_{+++}\rangle}{E_{+++}-E_{---}}|\overline{\psi}_{---}\rangle+\frac{\langle\overline{\psi'}_{+--}|H_I|\overline{\psi}_{+++}\rangle}{E_{+++}-E_{+--}}|\overline{\psi}_{+--}\rangle\notag\\
			&&+\frac{\langle\overline{\psi'}_{-+-}|H_I|\overline{\psi}_{+++}\rangle}{E_{+++}-E_{-+-}}|\overline{\psi}_{-+-}\rangle+\frac{\langle\overline{\psi'}_{--+}|H_I|\overline{\psi}_{+++}\rangle}{E_{+++}-E_{--+}}|\overline{\psi}_{--+}\rangle+\frac{\langle\overline{\psi'}_{-++}|H_I|\overline{\psi}_{+++}\rangle}{E_{+++}-E_{-++}}\notag\\
			&&|\overline{\psi}_{-++}\rangle+\frac{\langle\overline{\psi'}_{+-+}|H_I|\overline{\psi}_{+++}\rangle}{E_{+++}-E_{+-+}}|\overline{\psi}_{+-+}\rangle+\frac{\langle\overline{\psi'}_{++-}|H_I|\overline{\psi}_{+++}\rangle}{E_{+++}-E_{++-}}|\overline{\psi}_{++-}\rangle\notag\\
			&&\approx|\lambda_7\rangle+\frac{\langle\overline{\psi'}_{---}|H_I|\overline{\psi}_{+++}\rangle}{E_{+++}-E_{---}}|\overline{\psi}_{---}\rangle,\notag
		\end{eqnarray}
		\begin{eqnarray}\label{eq25}
			|\Psi_7'\rangle&=&|\overline{\psi'}_{+++}\rangle+\frac{\langle\overline{\psi}_{---}|H_I|\overline{\psi'}_{+++}\rangle}{E_{+++}-E_{---}}|\overline{\psi'}_{---}\rangle+\frac{\langle\overline{\psi}_{+--}|H_I|\overline{\psi'}_{+++}\rangle}{E_{+++}-E_{+--}}|\overline{\psi'}_{+--}\rangle\notag\\
			&&+\frac{\langle\overline{\psi}_{-+-}|H_I|\overline{\psi'}_{+++}\rangle}{E_{+++}-E_{-+-}}|\overline{\psi'}_{-+-}\rangle+\frac{\langle\overline{\psi}_{--+}|H_I|\overline{\psi'}_{+++}\rangle}{E_{+++}-E_{--+}}|\overline{\psi'}_{--+}\rangle+\frac{\langle\overline{\psi}_{-++}|H_I|\overline{\psi'}_{+++}\rangle}{E_{+++}-E_{-++}}\notag\\
			&&|\overline{\psi'}_{-++}\rangle+\frac{\langle\overline{\psi}_{+-+}|H_I|\overline{\psi'}_{+++}\rangle}{E_{+++}-E_{+-+}}|\overline{\psi'}_{+-+}\rangle+\frac{\langle\overline{\psi}_{++-}|H_I|\overline{\psi'}_{+++}\rangle}{E_{+++}-E_{++-}}|\overline{\psi'}_{++-}\rangle\notag\\
			&&\approx|\lambda_7'\rangle+\frac{\langle\overline{\psi}_{---}|H_I|\overline{\psi'}_{+++}\rangle}{E_{+++}-E_{---}}|\overline{\psi'}_{---}\rangle,
		\end{eqnarray}
and $|\Psi_8\rangle\approx|\lambda_8\rangle+\frac{\langle\overline{\psi'}_{+++}|H_I|\overline{\psi}_{---}\rangle}{E_{---}-E_{+++}}|\overline{\psi}_{+++}\rangle$, $|\Psi_8'\rangle\approx|\lambda_8'\rangle+\frac{\langle\overline{\psi}_{+++}|H_I|\overline{\psi'}_{---}\rangle}{E_{---}-E_{+++}}|\overline{\psi'}_{+++}\rangle$.
		
\par Due to weak coupling, the overlap between subspaces is small, and we have made further simplification by taking advantage of this feature. At present, we use perturbation theory to give the first-order corrected eigenvalues and eigenstates of the system, these eigenvalues also normalized
\begin{eqnarray}\label{eq27}
\langle\Psi'_3|\Psi_3\rangle&=&\langle\Psi'_6|\Psi_6\rangle=\langle\Psi'_7|\Psi_7\rangle=\langle\Psi'_8|\Psi_8\rangle=1+\mathcal{O}(J^2),\notag\\
			\langle\Psi'_1|\Psi_1\rangle&=&\langle\Psi'_2|\Psi_2\rangle=\langle\Psi'_4|\Psi_4\rangle=\langle\Psi'_5|\Psi_5\rangle=1.
		\end{eqnarray}
  and it is easy to verify that these base vectors are orthogonal to each other, and thus can form a complete set basis.
		
		\subsection{B. Analytic solutions of the residual tangle for the truly $\mathcal{PT}$ symmetric system}
		\par Assuming that the initial state of the system is $|\psi_0\rangle=|ggg\rangle$, the evolution state of the system can be written as
		\begin{eqnarray}\label{eq28}
			|\psi(t)\rangle&=&U(t)|\psi_0\rangle\notag\\&=&U(t)(|\Psi_1\rangle\langle\Psi'_1|+|\Psi_2\rangle\langle\Psi'_2|+|\Psi_3\rangle\langle\Psi'_3|\notag+|\Psi_4\rangle\langle\Psi'_4|+|\Psi_5\rangle\langle\Psi'_5|+|\Psi_6\rangle\langle\Psi'_6|+|\Psi_7\rangle\langle\Psi'_7|\notag+|\Psi_8\rangle\langle\Psi'_8|)|\psi_0\rangle\notag\\
			&=&\langle\Psi'_1|\psi_0\rangle U(t)|\Psi_1\rangle+\langle\Psi'_2|\psi_0\rangle U(t)|\Psi_2\rangle+\langle\Psi'_3|\psi_0\rangle U(t)|\Psi_3\rangle+\langle\Psi'_4|\psi_0\rangle U(t)|\Psi_4\rangle+\langle\Psi'_5|\psi_0\rangle U(t)|\Psi_5\rangle\notag\\&&+\langle\Psi'_6|\psi_0\rangle U(t)|\Psi_6\rangle+\langle\Psi'_7|\psi_0\rangle U(t)|\Psi_7\rangle+\langle\Psi'_8|\psi_0\rangle U(t)|\Psi_8\rangle\notag\\
			&=&\langle\Psi'_1|\psi_0\rangle e^{-it\Lambda_1}|\Psi_1\rangle+\langle\Psi'_2|\psi_0\rangle e^{-it\Lambda_2}|\Psi_2\rangle+\langle\Psi'_3|\psi_0\rangle e^{-it\Lambda_3}|\Psi_3\rangle+\langle\Psi'_4|\psi_0\rangle e^{-it\Lambda_4}|\Psi_4\rangle+\langle\Psi'_5|\psi_0\rangle e^{-it\Lambda_5}|\Psi_5\rangle\notag\\&&+\langle\Psi'_6|\psi_0\rangle e^{-it\Lambda_6}|\Psi_6\rangle+\langle\Psi'_7|\psi_0\rangle e^{-it\Lambda_7}|\Psi_7\rangle+\langle\Psi'_8|\psi_0\rangle e^{-it\Lambda_8}|\Psi_8\rangle.
		\end{eqnarray}
		\par After substituting in the corresponding eigenvalues and eigenstates, the final state is $|\psi_f\rangle=a|ggg\rangle+b(|gge\rangle+|geg\rangle+|egg\rangle)+c(|eeg\rangle+|ege\rangle+|gee\rangle)+d|eee\rangle$, where
		\begin{eqnarray}\label{eq29}
			a&=&\frac{1}{8\eta^3}\left\{12e^{\frac{it\alpha}{2\eta^2}}\Omega^2\left[-i\gamma+\eta+e^{-it\eta}(i\gamma+\eta)\right]+e^{\frac{3it(\alpha-4J\gamma^2)}{2\eta^2}}(-i\gamma+\eta)^3+e^{\frac{-3it(\alpha-4Jrr)}{2\eta^2}}(i\gamma+\eta)^3\right\},\notag\\
			b&=&\frac{e^{\frac{4itJ\beta}{\eta^2}}\Omega}{2\eta^3}\left[e^{it(\frac{-2J\beta}{\eta^2}+\frac{\eta}{2})}(-\eta^2-2\beta-i\gamma\eta)+e^{\frac{-it\alpha}{2\eta^2}}(\eta^2+2\beta-i\gamma\eta)+e^{\frac{it(4\eta^3-4pp-\alpha)}{2\eta^2}}(rr+i\gamma\eta)+\right.\notag\\
			&&\left.e^{\frac{-it(2\eta^3+4pp+\alpha)}{2\eta^2}}(-rr+i\gamma\eta)\right],\notag\\
			c&=&-\frac{\Omega^2 e^{\frac{-3it(\alpha-4Jrr)}{2\eta^2}}}{2\eta^3}\left[i\gamma(-1+e^{3it\eta}+3e^{\frac{it(2pp+\eta^3)}{\eta^2}}-3e^{\frac{2it(pp+\eta^3)}{\eta^2}})+\eta(-1-e^{3it\eta}+e^{\frac{it(2pp+\eta^3)}{\eta^2}}+e^{\frac{2it(pp+\eta^3)}{\eta^2}})\right],
			\notag\\
			d&=&\frac{\Omega^3e^{\frac{it\alpha}{2\eta^2}}}{\eta^3}\left[3-3e^{-it\eta}-e^{\frac{-it(2pp-\eta^3)}{\eta^2}}+e^{\frac{-2it(pp+\eta^3)}{\eta^2}}\right].
		\end{eqnarray}
		in which $\alpha=4J\beta+\eta^3$, $\beta=\gamma^2-\Omega^2$, $pp=J(\eta^2+2\beta)$, and $rr=\gamma^2-2\Omega^2$. After normalizing the final state, the residual tangle can be obtained
		\begin{eqnarray}\label{eq30}
			\tau_3=|\frac{A}{4\eta^{12}B^2}|,
		\end{eqnarray}
		where
		\begin{eqnarray}\label{eq31}
			\begin{split}
				A=&8xv^3\Omega ^6 e^{\frac{it ( \alpha +24  \beta  J)}{2\eta ^2}}
				+8h^3 y\Omega ^6 e^{-\frac{9 i t (\alpha -4 J rr)}{2 \eta ^2}}-3 h^2 v^2\Omega ^6e^{\frac{t (-3 i \alpha +8 i \beta  J+12 i J rr)}{\eta ^2}}+16x^2y^2\Omega ^6 e^{\frac{i \alpha  t}{\eta ^2}} \\
				&-24 xyvh\Omega ^6 e^{\frac{i t
						(-\alpha +4 \beta  J+6 J rr)}{\eta ^2}},\notag
			\end{split}
		\end{eqnarray}
		\begin{eqnarray}\label{eq32}
B&=&\frac{1}{32 \eta ^6} \left\{(\gamma^2+\eta^2)^3+96\gamma^2(\gamma^2+\eta^2)\Omega^2+48\gamma^2(13\gamma^2+5\eta^2)\Omega^4+1024\Omega^6-48\Omega^2[\gamma^4-\eta^2\Omega^2+16\Omega^4-\gamma^2(\eta^2-13\right.\notag\\
&&\left. \Omega^2)]\cos{(t\eta)}+[-\gamma^6+15\gamma^4\eta^2-15\gamma^2\eta^4+\eta^6-48\gamma^2\beta\Omega^2+48(3\gamma^2+\eta^2)\Omega^4-256\Omega^6]\cos{(3t\eta)}+2[72(\eta^2-\gamma^2)\right.\notag\\		&&\left.\Omega^4\cos{(\frac{3t\eta}{2}-\frac{2Jt\beta}{\eta^2})}+6(\gamma^4-6\gamma^2\eta^2+\eta^4)\Omega^2\cos{(\frac{5t\eta}{2}-p)}+6\Omega^2\cos{(2t\eta+q)}(5\gamma^4+\eta^4-4\eta^2\Omega^2-10\gamma^2\eta^2+\right.\notag\\		&&\left.12\gamma^2\Omega^2)+6\Omega^2(\eta^4-\gamma^4)\cos{(\frac{t\eta}{2}+p)}-6\Omega^2(6\gamma^4+4\gamma^2\eta^2-\eta^4+12\gamma^2\Omega^2+4\eta^2\Omega^2)\cos{(t\eta-q)}+24\Omega^2(\gamma^4\Omega^2-\right.\notag\\		&&\left.\eta^2\gamma^2+3\Omega^2\gamma^2-\eta^2\Omega^2)\cos{(2t\eta-q)}+48\gamma\eta\Omega^2(\gamma^2+5\Omega^2)\sin{(t\eta)}-24\Omega^2\cos{(t\eta+q)}(\gamma^4+\gamma^2\eta^2+3\gamma^2\Omega^2+\eta^2\right.\notag\\		&&\left.\Omega^2)+48\gamma\eta\Omega^2(\gamma^2+2\Omega^2)\sin{(q-\eta)}+\gamma\eta(3\gamma^4-10\gamma^2\eta^2+3\eta^4+48\gamma^2\Omega^2-48\Omega^4)\sin{(3t\eta)}+24\gamma\eta\Omega^2(\eta^2-\gamma^2)\right.\notag\\		&&\left.\sin{(\frac{5t\eta}{2}-p)}+144\gamma\eta\Omega^4\sin{(\frac{3t\eta}{2}-\frac{2Jt\beta}{\eta^2})}+12\gamma\eta\Omega^2(\gamma^2+\eta^2)\sin{(\frac{t\eta}{2}+p)}+12\gamma\eta\Omega^2\sin{(t\eta-q)}(\gamma^2+12\Omega^2+\right.\notag\\		&&\left.\eta^2)+144\gamma\eta\Omega^4\sin{(t\eta+q)}+24\gamma\eta\Omega^2(-3\gamma^2+\eta^2-4\Omega^2)\sin(2t+q)]\right\},
\end{eqnarray}
where $x=e^{-\frac{i t \left(2 pp-\eta ^3\right)}{\eta ^2}}-e^{-\frac{2 i t \left(\eta ^3+pp\right)}{\eta ^2}}+3 e^{-i \eta  t}-3$, $y=\frac{1}{8} (\eta +i \gamma )^3 e^{-\frac{3 i t (\alpha -4 J rr)}{2 \eta^2}}+\frac{1}{8} (\eta -i \gamma )^3 e^{\frac{3 i t \left(\alpha -4 \gamma ^2 J\right)}{2 \eta ^2}}+\frac{3}{2} \Omega ^2 (\eta -i \gamma ) e^{\frac{i \alpha t}{2 \eta ^2}}+\frac{3}{2} \Omega ^2 (\eta +i \gamma ) e^{\frac{t \left(i \alpha -2 i \eta ^3\right)}{2 \eta ^2}}$, $v=-\left(2 \beta +i \gamma  \eta +\eta ^2\right) e^{-\frac{i t \left(4 \beta  J-\eta ^3\right)}{2 \eta^2}}+\left(2 \beta -i \gamma  \eta +\eta ^2\right) e^{-\frac{i \alpha  t}{2 \eta ^2}}-(rr-i \gamma  \eta )e^{-\frac{i t \left(\alpha +2 \eta ^3+4 pp\right)}{2 \eta ^2}}+(rr+i \gamma  \eta ) e^{-\frac{i t \left(\alpha -4\eta ^3+4 pp\right)}{2 \eta ^2}}$, $h=3 i \gamma e^{\frac{2 i t \left(\eta ^3+pp\right)}{\eta ^2}}-3 i \gamma e^{\frac{i t \left(\eta ^3+2 pp\right)}{\eta ^2}}-i \gamma e^{3 i\eta t}+i\gamma-\eta  e^{\frac{2 i t \left(\eta ^3+pp\right)}{\eta ^2}}-\eta e^{\frac{i t \left(\eta ^3+2 pp\right)}{\eta ^2}}+\eta e^{3 i \eta
t}+\eta$, $q=\frac{2Jt(\gamma^2+2\Omega^2)}{\eta^2}$, $p=\frac{2Jt(\Omega^2+2\gamma^2)}{\eta^2}$.
		\subsection{C. The process of entanglement preparation for the truly $\mathcal{PT}$ symmetric system}
		In the analysis of entanglement preparation, we here examine the evolution dynamics of the system. In the absence of coupling between the qubits, i.e., $J=0~\rm{rad/\mu s}$, three qubits evolve independently, the state evolves successively through $|ggg\rangle\rightarrow\bigotimes_{n=1,2,3}1/\sqrt{2}(|g\rangle-i|e\rangle)_n\rightarrow|eee\rangle\rightarrow\bigotimes_{n=1,2,3}1/\sqrt{2}(|g\rangle+i|e\rangle)_n\rightarrow|ggg\rangle$. The evolution dynamics of states $|ggg\rangle$ and $|eee\rangle$ exhibit Rabi-like oscillation with the period of $\tau=2\pi/\eta\sim1.16~\rm{\mu s}$ (the duration required to complete a Rabi oscillation of the decoupled single qubit), wherein a $\pi$ phase is accumulated over a complete Rabi cycle. Considering the coupling between the qubits, i.e., $J=0.001~\rm{rad/\mu s}$ (Figs.~\ref{figs2}(e) and (f)), the evolution dynamics of states $|ggg\rangle$ and $|eee\rangle$ reveal deviations from perfect Rabi oscillation.  The weak inter-qubit coupling leads to imperfect dynamic evolution, preventing a complete return to the initial state within each periodic cycle. The residual tangle versus $\Omega$ and $t$ is in Fig.~\ref{figs2}(g), demonstrating how to determine optimal parameters for a given coupling strength $J$, we select $\Omega=6.58~\rm{rad/\mu s}$ here.
		\par To more clearly demonstrate the preparation of the maximal entanglement, we present the dynamic trajectory of reduced single qubit on the Bloch sphere in Fig.~\ref{figs2}(h). In the absence of coupling, the reduced single qubit consistently evolves along the surface of the Bloch sphere, indicative of the lack of entanglement among the three qubits. Conversely, the reduced single qubit gradually evolves into the inside of the Bloch sphere, demonstrating the system is entangled. Furthermore, it's notable that the evolution trajectory increasingly diverges from the state $1/\sqrt{2}(|g\rangle+i|e\rangle)$ with each cycle, making it possible to prepare maximal entanglement. After several cycles, the trajectory passes through the origin of the Bloch sphere, indicating the single qubit is in the fully mixing state, corresponding to the maximal entanglement of the system with the degree of entanglement $\tau_3=0.999$. For the normalized maximal entangled state $|\psi'_m\rangle$, the coefficients are $a_1=0.5656e^{i0.9\pi}$, $a_2=a_3=a_5=0.2354e^{-i0.243\pi}$, $a_4=a_6=a_7=0.2519e^{i1.248\pi}$, and $a_8=0.5687e^{i1.39\pi}$, the normalization factor of the pre-normalized state approaches 1, representing an exponential increase over the loss-only system. Moreover, the preparation time for maximal entanglement is about $11.6~\rm{\mu s}$, twice as rapid as the loss-only scenario and three orders of magnitude more efficient than the conventional Hermitian system.
		\begin{figure*}
			\centering
			\includegraphics[width=16cm,height=8cm]{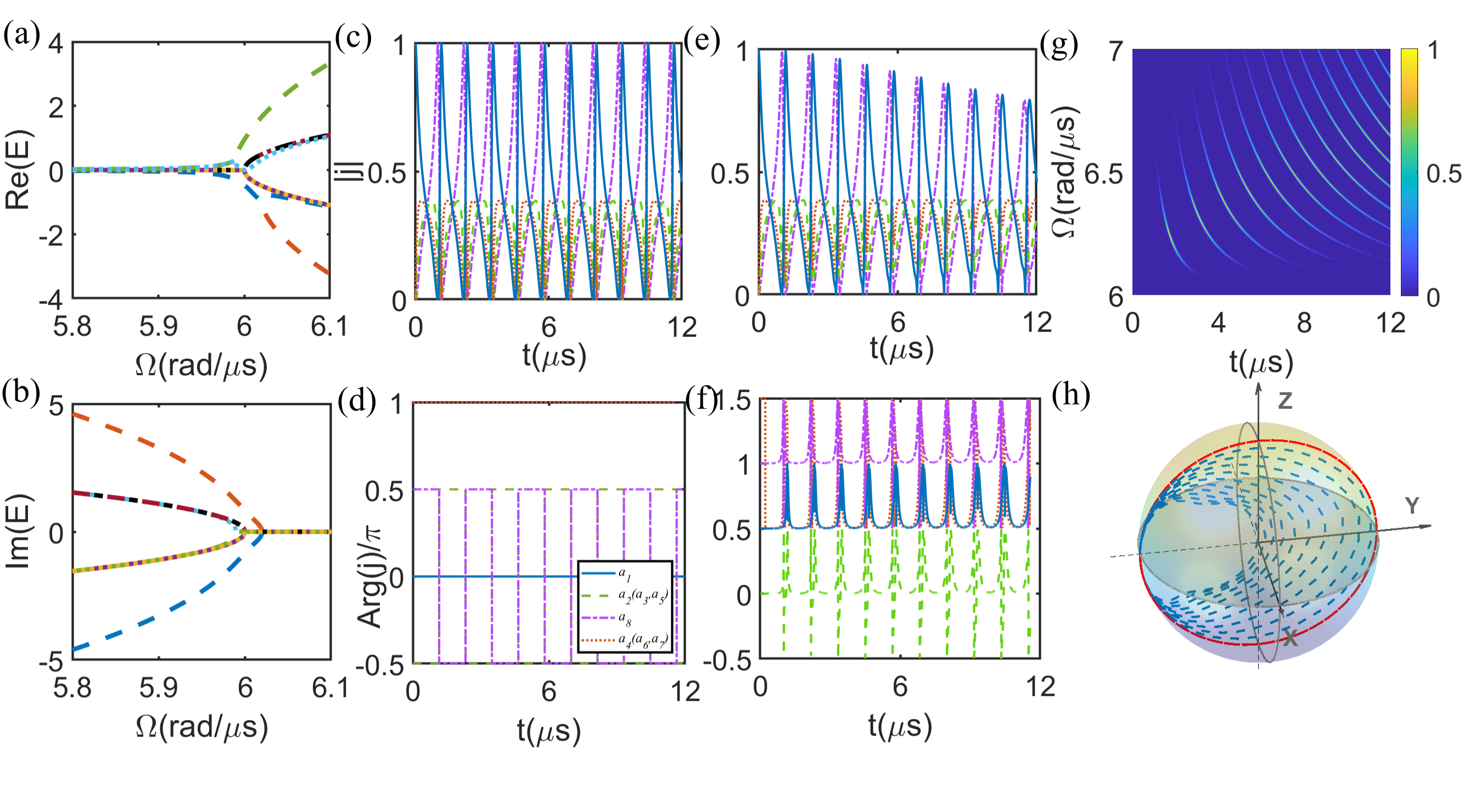}
			\caption{Real (a) and imaginary (b) parts of the eigenvalues for the active $\mathcal{PT}$ symmetric system. Time evolution of the modulus $|a_j|$ (c) and auxiliary angle Arg($a_j$) (d) for each of eight complex amplitudes of the triplet state with the coupling $J=0~\rm{rad/\mu s}$. The evolution of modulus $|a_j|$ (e) and auxiliary angle Arg($a_j$) (f) for the coupling $J=0.001~\rm{rad/\mu s}$. (g) Residual tangle as the function of the drive $\Omega$ and the evolution time. (h) Dynamic trajectory of reduced single qubit on the Bloch sphere with the coupling $J=0~\rm{rad/\mu s}$~(red) and $J=0.001~\rm{rad/\mu s}$~(blue).} \label{figs2}
		\end{figure*}
	\begin{figure*}
			\centering
			\includegraphics[width=13.5cm,height=5.3cm]{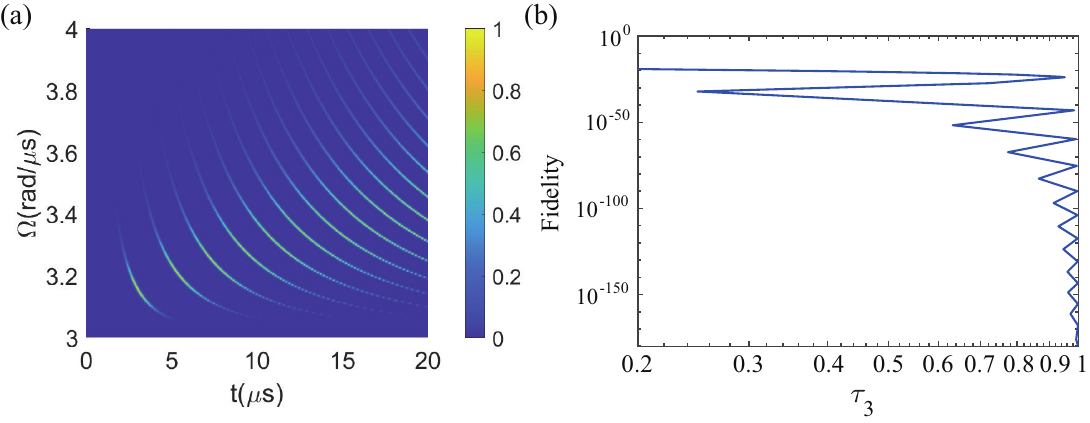}
			\caption{(a)  The residual tangle is the function of the drive and evolution time. (b) The fidelity under the corresponding maximal residual tangle, where $\gamma=12~\rm{\mu s^{-1}}$ and $J=0.001~\rm{rad/\mu s}$.}\label{figs3}
		\end{figure*}
		\begin{figure*}
			\centering
			\includegraphics[width=12cm,height=6cm]{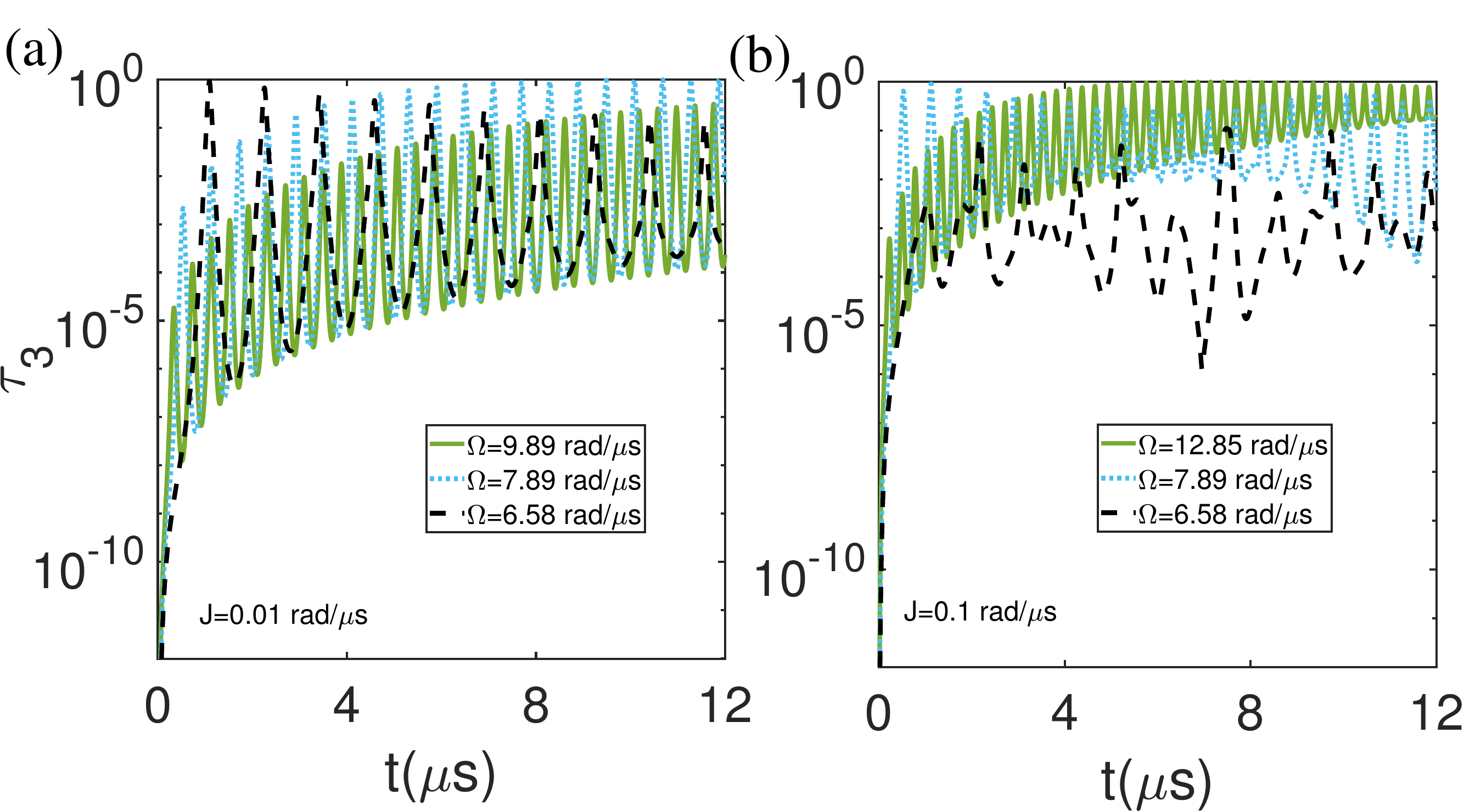}
			\caption{The evolution of the residual tangle for the coupling strength $J=0.01 ~\rm{rad/\mu s}$ (a) and $J=0.1 ~\rm{rad/\mu s}$ (b).}\label{figs4}
		\end{figure*}
  \begin{figure*}
			\centering
\includegraphics[width=6cm,height=4.5cm]{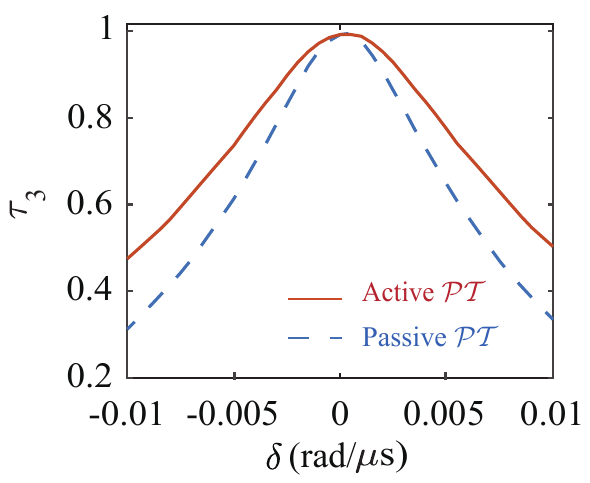}
    			\caption{The degree of entanglement with the non-resonant drive.}\label{figsss}
		\end{figure*}
		\begin{figure}
			\centering
			\includegraphics[width=13cm,height=4cm]{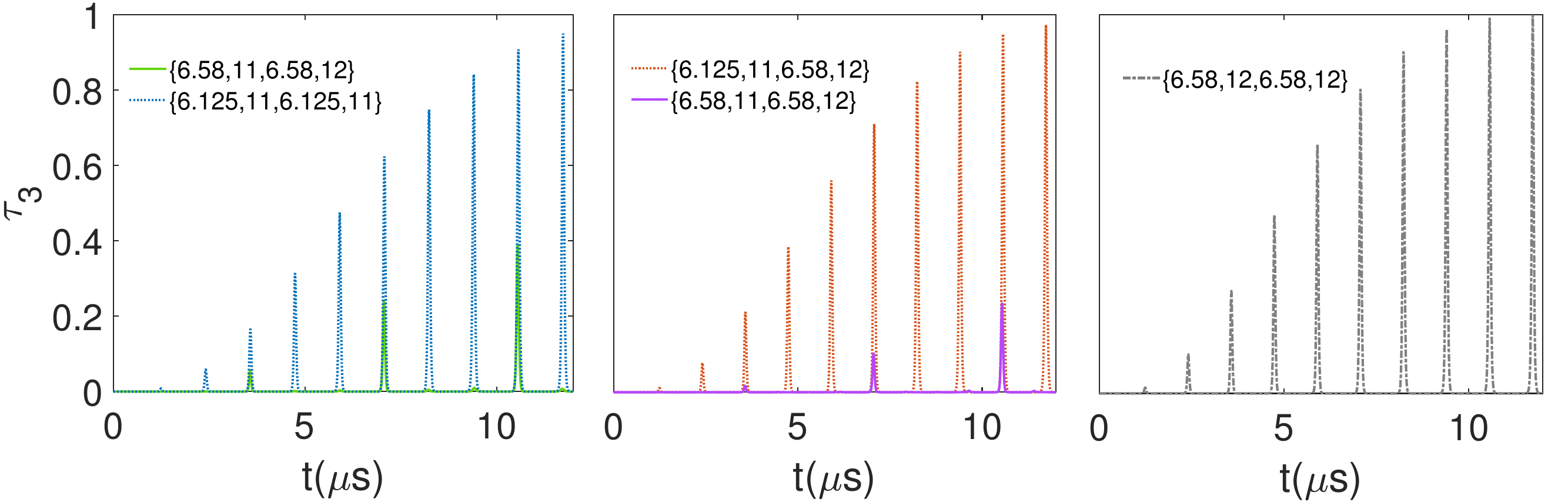}
			\caption{The evolution of residual tangle with different parameters \{$\Omega_2(\mathrm{rad}/\mu s)$, $\gamma_2(\mu s^{-1})$, $\Omega_3(\mathrm{rad}/\mu s)$, $\gamma_3(\mu s^{-1})$\}, where $\Omega_1=6.58~\mathrm{rad}/\mu s$ and $\gamma_1=12~\mu s^{-1}$ are fixed, the initial state and coupling between the qubits are $|ggg\rangle$ and $J=0.001~\mathrm{rad}/\mu s$, respectively.}\label{figs6}
		\end{figure}	\subsection{D. Passive $\mathcal{PT}$ symmetric system}
		\par The eigenvalue of the passive $\mathcal{PT}$ symmetric system always has an imaginary part, which means that the probability amplitude of the state decreases exponentially with time during the evolution of the system. From Fig.~\ref{figs3}(a), we can see the residual tangle reaches its maximal value after several oscillations over time after selecting the optimal $\Omega$ for preparing triplet entanglement. For the trade-off relation, the upper bound of the residual tangle oscillates with the different fidelity due to the different optimal drive amplitudes, as shown in Fig.~\ref{figs3}(b), where the optimal parameter \{$\Omega$,t\} is identified from the Fig.~\ref{figs3}(a). Here we point out that the ideal state for calculating the fidelity is $|\psi_i\rangle$ is the GHZ state.
		\subsection{E. Acceleration effect under strong coupling}
		Similarly, Figs.~\ref{figs4} (a) and (b) illustrate the residual tangle evolution for $J=0.01~\rm{rad/\mu s}$ and $J=0.1~\rm{rad/\mu s}$ under different $\Omega$ for the active $\mathcal{PT}$ symmetric system, which also shows an acceleration effect for the maximal entanglement preparation. 
		\subsection{F. Error consideration}
		
For the triplet system, there still exists stronger robustness against the detuning error, as shown in Fig.~\ref{figsss}. In addition, employing identical parameters for all three qubits poses experimental challenges. However, it is fortunate that as long as the same period is maintained among the three qubits, a high degree of entanglement can still be achieved in the prepared entangled state. As illustrated in Fig.~\ref{figs6}, it is evident that when the coupling strength between the qubits is altered while maintaining the coupling strength of the energy level, there is a significant decrease in residual tangle. Nevertheless, when condition $4\Omega_1^2-\gamma_1^2=4\Omega_2^2-\gamma_2^2=4\Omega_3^2-\gamma_3^2$ is satisfied, the high-degree of entanglement is still preserved.

[1] B. Simon, Large orders and summability of eigenvalue perturbation theory: A mathematical overview, International Journal of Quantum Chemistry 21, 3 (1982).

[2] Y. Chu and Yu. Liu and H. Liu and J. Cai, Quantum Sensing with a Single-Qubit Pseudo-Hermitian System, Phys. Rev. Lett. 124, 020501 (2020).

[3] \r{A}. Bj{\"O}rck, Numerics of Gram-Schmidt orthogonalization, Linear Algebra and its Applications 197-198, 297 (1994).
\end{widetext}
\end{document}